\documentclass[%
twocolumn,
showpacs,preprintnumbers,
 aps,
prstab,
superscriptaddress
]{revtex4-1}
\usepackage{graphics} % for pdf, bitmapped graphics files
\usepackage[pdftex]{graphicx}
\usepackage{epstopdf}
\usepackage[cmex10]{amsmath}
\usepackage{amssymb}  % assumes amsmath package installed
\usepackage[colorlinks,linkcolor=blue,pdfauthor=Alexander \ Scheinker,citecolor=red]{hyperref}
\usepackage{footnote}

\usepackage{xcolor}

\def\x{\mathbf{x}}
\def\p{\mathbf{p}}
\def\f{\mathbf{f}}
\def\norm{{\mathrm{nrm}}}

\newcommand{\mean}[1]
{\mbox{$\langle{#1}\rangle$}}

\begin{document}

% Use the \preprint command to place your local institutional report
% number in the upper righthand corner of the title page in preprint mode.
% Multiple \preprint commands are allowed.
% Use the 'preprintnumbers' class option to override journal defaults
% to display numbers if necessary
%\preprint{}

%Title of paper
%\title{Feedback for Optimization of Betatron Oscillation Minimization of a Time-Varying Lattice}

%\title{Suppressing of the emittance degradation in the Double Emittance Exchanger-based Bunch Compressor}

\title{Phase space exchange-based bunch compression with reduced CSR effects}

\author{Alexander~Malyzhenkov}
\email[]{malyzhenkov@lanl.gov}
\affiliation{Los Alamos National Laboratory, Los Alamos, NM, USA}
\affiliation{Northern Illinois Center for Accelerator \& Detector Development and Department of Physics, Northern Illinois University, DeKalb IL, USA}

\author{Alexander~Scheinker}
\email[]{ascheink@lanl.gov}
\affiliation{Los Alamos National Laboratory, Los Alamos, NM, USA}

\thanks{This research was supported by Los Alamos National Laboratory.}

\date{\today}

\begin{abstract}
	Through an interaction of theory and optimization-guided design, we have developed a novel bunch compression scheme based on an asymmetric double emittance exchange (DEEX) with greatly reduced coherent synchrotron radiation (CSR) effects. First, we discuss the benefits of the phase space exchange-based bunch compressor (BC) in comparison to the chicane-based counterpart in the approximation of linear single particle dynamics and with respect to microbunching instabilities driven by longitudinal space charge forces. We present simulation results for nonlinear dynamics in a symmetric DEEX BC which accounts for the CSR effects of the scheme with direct and mirrored longitudinal phase space at the exit of the beamline. The later scheme demonstrates self-compensation of CSR effects on the longitudinal dynamics. Next, we optimize the parameters of the beam elements with an adaptive feedback, resulting in an asymmetric configuration of the scheme with compensated CSR and nonlinear effects on the transverse dynamics, which we further improve by incorporating several sextupoles relying on an eigen-emittance analysis. Finally, we discuss how this method can be generalized for optimizing nonlinear beam dynamics with collective effects.
\end{abstract}

% insert suggested PACS numbers in braces on next line
% insert suggested keywords - APS authors don't need to do this
%\keywords{}

%\maketitle must follow title, authors, abstract, \pacs, and \keywords
\maketitle

%%%%%%%%%%%%%%%%%%%%%%%%%%%%%%%%%%%%%%%%%%%%%%%%%%%%%%%%%%%%%%%%%%%%%%%%%%%%%%%%%%%%%%%%
%%%%%%%%%%%%%%%%%%%%%%%%%%%%%%%%%%%%%%%%%%%%%%%%%%%%%%%%%%%%%%%%%%%%%%%%%%%%%%%%%%%%%%%%
\section{Introduction}
%%%%%%%%%%%%%%%%%%%%%%%%%%%%%%%%%%%%%%%%%%%%%%%%%%%%%%%%%%%%%%%%%%%%%%%%%%%%%%%%%%%%%%%%
%%%%%%%%%%%%%%%%%%%%%%%%%%%%%%%%%%%%%%%%%%%%%%%%%%%%%%%%%%%%%%%%%%%%%%%%%%%%%%%%%%%%%%%%

Extremely bright and short wavelength fourth generation free electron lasers (FELs), including the Linac Coherent Light Source (LCLS)~\cite{ref-lcls}, LCLS-II~\cite{LCLS2}, FLASH~\cite{FLASH}, SACLA~\cite{SACLA}, FERMI~\cite{FERMI}, and the recently completed European X-ray FEL~\cite{EXFEL} and SwissFEL~\cite{SWISSFEL,SWISSFEL1} allow scientists to probe matter at unprecedented time and length scales. The next generation of hard X-ray FELs (XFELs) has the goal of achieving even shorter wavelengths and higher brightness than existing FELs. Achieving shorter wavelengths faces increasingly greater challenges in terms of emittance limitations. For example, the XFEL planned for the Matter-Radiation Interactions in Extremes (MaRIE) facility at Los Alamos National Laboratory (LANL) has considered a goal of achieving $0.01$ nm wavelength X-ray pulses. The wavelength of the light produced by an FEL is approximately
\begin{equation}
\lambda_{\mathrm{x-ray}} \approx \frac{\lambda_{\mathrm{u}}}{2\gamma^2}\left (1 + \frac{K^2}{2} \right ), \label{simple_xray}
\end{equation}
where $K \approx$ 0.934$\cdot B_0$[T]$\cdot$$\lambda_{\mathrm{u}}$[cm] is the undulator parameter, % a measure of undulator strength, where  
$B_0$ is the undulator magnetic field amplitude, $\gamma$ is the Lorentz factor, and $\lambda_{\mathrm{u}}$ is the period of the undulator. According to Eq.~(\ref{simple_xray}), an electron beam at 12 GeV, $K\sim 1$, and an undulator period of 1 cm would provide $\lambda_{\mathrm{x-ray}}\sim 0.01$ nm X-rays. 
However, for a given wavelength there is also a constraint on the normalized transverse emittance, which is required for proper lasing:
\begin{equation}
\epsilon_{n_x,n_y} \leq \beta \gamma \frac{\lambda_{\mathrm{x-ray}} }{4 \pi}. \label{emit_constraint}
\end{equation}
Existing photo-injectors are capable of producing electron beams with transverse emittances that satisfy the requirement in Eq.~(\ref{emit_constraint}). However, beam quality is strongly affected by space charge forces limiting the bunch charge to $\sim100$~pC for the required small transverse normalized emittance. Such a low bunch charge requires an extremely short bunch length in order to achieve high peak current when the beam enters the undulator. However, longitudinal space charge forces at the injector and in the low energy stages of the acceleration require long bunches with as small as possible transverse emittances and energy-spread on the photo-cathode. Therefore, one or several bunch compressors must be used to shorten the beam and increase peak current once the beam becomes highly relativistic. Increasing the beam current typically comes at the price of increasing energy spread. The constraint on beam energy spread at the undulator which must be also satisfied for proper FEL lasing is
\begin{equation}\label{dE_FEL}
\sigma_{\gamma}\leq 0.5\; \rho_{FEL},
\end{equation}
where $\rho_{FEL}=\frac{\lambda_u}{4\sqrt{3}\pi L_g} $ is the FEL parameter (Pierce parameter) and $L_g$ is the effective gain length as defined in 1D FEL theory~\cite{FEL_book_Schmuser,KimHuang}. Because XFELs require ultra high current to reach high brightness, electron bunches must be compressed in length by several orders of magnitude by utilizing bunch compressors, while keeping the transverse emittances and energy spread as small as possible. 

The main idea behind a standard chicane-based bunch compressor, of translating energy offset along the bunch (imposed upstream of the chicane, typically via off-crest acceleration) into path length difference, is simple. In practice, the actual design and implementation is extremely challenging when transverse and longitudinal emittances are to by maintained, due to collective effects such as coherent synchrotron radiation (CSR) and microbunch instabilities driven predominantly by the longitudinal space charge (LSC) forces. In this work, we present the design and optimization of an asymmetric bunch compressor based on the double emittance exchange (DEEX). Such a design has the advantage that it does not require typical chicane-type optics and does not require initial/residual energy-phase adjustments, \emph{e.g.} imposing/removing the energy chirp along the bunch accompanying the chicane. Thus, the beam is expected to experience much smaller LSC-induced microbunching instabilities, since the chicane, the actual ``amplifier'' of the density modulation noise unavoidably present in any electron beam, is eliminated. 

In our design, CSR effects are expected to cause degradation of beam quality and their minimization is an important problem being solved in this paper. The CSR and nonlinear effects imposed during the first phase space exchange are partially compensated in the second phase space exchange if the parameters of the beam elements are properly tuned. We utilized a novel adaptive feedback technique known in the control community as extremum seeking (ES) \cite{ref-Sch-bounded-ES,ref-SchLANL,ref-SchSch,ref-SchBk}, which has been applied for the non-destructive prediction of the longitudinal phase space of electron beams in the FACET plasma wakefield accelerator \cite{ref-SchGess} and has been applied in-hardware to automatically tune the longitudinal phase space of the LCLS electron bunch \cite{ref-PRL-Sch}. In this work, we applied ES in simulation studies in order to minimize the impact of the nonlinear and CSR effects on the emittance deterioration while optimizing the design. By allowing the ES optimizer complete freedom in parameter choice within physical constraints, while preserving the exact double emittance exchange in the approximation of the linear single-particle dynamics, an asymmetric design with greatly reduced coherent synchrotron radiation effects was found. 
A general overview of the ES algorithm is given in Appendix \ref{apx:ES}.

%%%%%%%%%%%%%%%%%%%%%%%%%%%%%%%%%%%%%%%%%%%%%%%%%%%%%%%%%%%%%%%%%%%%%%%%%%%%%%%%%%%%%%%%
%%%%%%%%%%%%%%%%%%%%%%%%%%%%%%%%%%%%%%%%%%%%%%%%%%%%%%%%%%%%%%%%%%%%%%%%%%%%%%%%%%%%%%%%
\section{Bunch Compressors}\label{sec:bc}
%%%%%%%%%%%%%%%%%%%%%%%%%%%%%%%%%%%%%%%%%%%%%%%%%%%%%%%%%%%%%%%%%%%%%%%%%%%%%%%%%%%%%%%%
%%%%%%%%%%%%%%%%%%%%%%%%%%%%%%%%%%%%%%%%%%%%%%%%%%%%%%%%%%%%%%%%%%%%%%%%%%%%%%%%%%%%%%%%

\begin{figure*}[ht]
	\centerline{\includegraphics[width=0.9\textwidth]{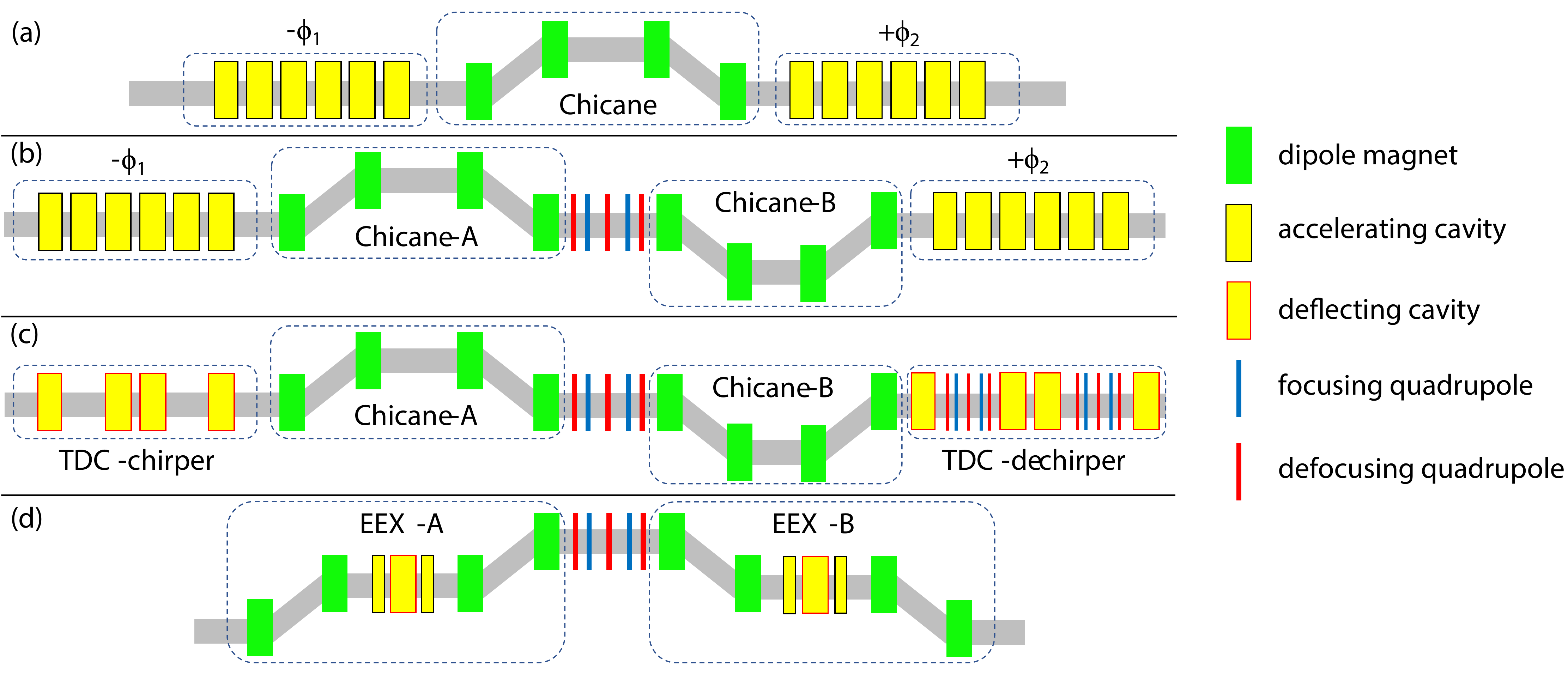}}
	\caption{Schemes of various bunch compressors: (a) Single and (b) double magnetic chicane surrounded by two off-crest acceleration sections for imposing/removing correlated energy spread onto the beam (c) Double magnetic chicane with the TDC-based chirper and dechirper working via transverse-longitudinal mixing (d) Two emittance exchangers separated by a demagnifying transverse-optics systems compressing the beam in the longitudinal direction.} 
	\label{fig:chicane}
\end{figure*}

There are several types of bunch compressors (BC), as shown in Fig.\ref{fig:chicane}, which are capable of compressing beams and increasing beam current, while increasing energy spread~\cite{Energy-spread}. Here, we divide BCs into two categories: chicane-based BCs and EEX-based BCs.

\subsection{Chicane-based BC}
The key element of a conventional bunch compressor~(Fig.~\ref{fig:chicane} (a-c)) is a magnetic chicane with four bending magnets (or C-chicane) with a nonzero $R_{56}$ element. The matrix elements of a chicane in terms of an electron's 6D phase space coordinates $\boldsymbol{\xi}=(x,\;x',\;y,\;y',\;z=s-ct,\;z'=\delta\gamma/\gamma_0)$ are expressed in terms of its structural components in Appendix~\ref{apx:elements}. The transverse ($x,\;x'$) and longitudinal ($z,\;z'$) beam dynamics are coupled within the elements of a chicane resulting in the following transfer matrix in 4-dimensional ($x,\;x',\;z,\;z'$) phase space:
\begin{equation}\label{eq:chicane-4D}
R_{chicane}=\begin{pmatrix}
1 & L_{x}& 0 &0  \\ 
0 & 1 & 0 &0\\ 
0 & 0 &1 &R_{56}\\
0 & 0 &0 &1\\
\end{pmatrix}
,
\end{equation}
where $R_{56}>0$, while transverse motion in the ($y,y'$) phase space is uncoupled. According to Eq.~(\ref{eq:chicane-4D}), the chicane acts on the longitudinal coordinate of a particle as:
\begin{equation}\label{eq:chicane_z}
z_f=z_i+R_{56}\cdot z'_i.
\end{equation}
Therefore particles having a smaller energy than the reference particle ($\boldsymbol{\xi}_r=0$) will travel over a longer path when passing through the chicane and vise versa. For an ensemble of particles, according to the 6-dimensional $\Sigma-$matrix: $\Sigma_f=\mean{\boldsymbol{\xi}_f\cdot\boldsymbol{\xi}_f^T}=\mean{R\cdot\boldsymbol{\xi}_i\cdot(R\cdot\boldsymbol{\xi}_i)^T}=R\cdot\mean{\boldsymbol{\xi}_i\cdot\boldsymbol{\xi}^T_i}\cdot R^T=R\cdot\Sigma_i\cdot R^T$, one can determine the evolution of the root-mean-square (rms) bunch length as:
\begin{equation}\label{eq:R_56_sigma}
\sigma_{z_f}^2=\sigma_{z_i}^2+R_{56}^2\sigma_{z'_i}^2+2R_{56}\mean{zz'}_i,
\end{equation}
where the conventional notations $\sigma_{z_i}=\sqrt{\mean{z^2}}$ and $\sigma_{z'_i}=\sqrt{\mean{z'^2}}$ are used.
Since $R_{56}>0$, equation~(\ref{eq:R_56_sigma}) suggests that if a bunch has a sufficiently negative correlated energy spread $\sigma_{z'_u}=\mean{zz'}_i/\sigma_{z_i}<0$, it will be compressed in the chicane. Therefore, compression strongly depends on the applied chirp $\mean{zz'}_i/\sigma^2_{z_i}$ prior to the chicane. In terms of the longitudinal emittance, $\epsilon_z =\sqrt{\sigma_z^2\sigma_{z'}^2- \mean{zz'}^2}$, the bunch length on the exit of the chicane becomes:
\begin{equation}\label{eq:R_56_1}
\sigma_{z_f}^2=\sigma_{z_i}^2+R_{56}^2\frac{\epsilon_z^2}{\sigma_{z_i}^2}+R_{56}^2\frac{\mean{zz'}_i^2}{\sigma_{z_i}^2}+2R_{56}\mean{zz'}_i.
\end{equation}
Analyzing the right-hand side of equation~(\ref{eq:R_56_1}), one can find that a minimal bunch length is achieved when
\begin{equation}\label{eq:R_56_cond}
\mean{zz'}_i=-\sigma_{z_i}^2/R_{56}
\end{equation}
and the longitudinal bunch size is equal to:
\begin{equation}\label{eq:R_56_min}
\min\left(\sigma_{z_f}^2\right)=R_{56}^2\frac{\epsilon_z^2}{\sigma_{z_i}^2}=-\frac{R_{56}}{\mean{zz'}_i}\epsilon_z^2=\frac{\sigma_{z_i}^2}{\mean{zz'}_i^2}\epsilon_z^2,
\end{equation}
under the assumption that the longitudinal emittance remains invariant, which is valid in the approximation of linear single particle dynamics. This results in an estimate of the maximum compression ratio which can be achieved in a single chicane:
\begin{equation}\label{eq:m_chicane}
m_{max}=\max\left(\frac{\sigma_{z_i}}{\sigma_{z_f}}\right)=\frac{|\mean{zz'}_i|}{\epsilon_z},
\end{equation}
and implies that one needs to maximize correlated energy spread to maximize compression. Equation~(\ref{eq:m_chicane}) also shows that it is easier to compress beams with smaller longitudinal emittances than beams with larger emittances in chicane-based BCs. 

The standard approach to impose and remove correlated energy spread onto the beam is via accelerating sections upstream and downstream of the chicane~(Fig.~\ref{fig:chicane} (a,b)). Particles experience a linearly-dependent acceleration gradient along the bunch, resulting in an energy chirp. Radio frequency (RF) cavity-based acceleration relies on synchronization of the electron beam's arrival and the phase of the electromagnetic wave at the moment of interaction. Since the dependence along the bunch is quadratic in the maximum of the RF field in the sinusoidal approximation (higher even orders are present), imposing a linear energy correlation along the bunch requires off-crest acceleration. One can find a scaling for the $R_{65}$ matrix element in the simple approximation of a uniform electromagnetic field across the bunch and ideal synchronization (the phase between the RF field and the reference particle of the beam is constant), given by:
\begin{equation}
R_{65}=-\frac{eE_0 k \sin\phi_0}{\gamma m_ec^2}s,
\end{equation} 
where $E_0$ and $k$ are respectively the amplitude and wave vector of the RF field, $s$ is the length of the accelerator structure, $m_e$ and $e$ are the mass and charge of the electron and $\gamma$ is the averaged Lorentz factor of the electron beam. This suggests that to impose a large energy chirp on the beam while still accelerating the beam, one needs to maintain a phase of approximately $\pi/4$. Therefore, this method leads to a significantly reduced acceleration gradient and consequently a longer length of accelerating sections is required to reach the same energy as compared to on-crest acceleration ($\phi_0=0$). An alternative configuration~(Fig.~\ref{fig:chicane} (c)), recently proposed and not yet practically implemented, relies on the transverse-longitudinal mixing imposed in a series of transverse deflecting cavities (TDC) operating in TM$_{110}$ mode and separated by free space propagation sections~\cite{TDC-conference-slides,TDC-paper}. Interestingly, such a scheme has topological similarities with a chicane. The chicane relies on the four bending magnets combined into two oppositely oriented doglegs providing with the transverse-longitudinal coupling of $x$ from $z'$ and $z$ from $x'$ and resulting in the nonzero $R_{56}$ at the end. The novel TDC-based scheme is based on the transverse-longitudinal coupling of $x'$ from $z$ and $z'$ from $x$ imposed in the deflecting cavities, resulting in the nonzero $R_{65}$ element and the absence of the residual transverse-longitudinal correlations at the exit of the scheme~\cite{TDC-paper}.

Regardless of the particular configuration, a beamline imposing/removing energy correlations along a bunch can be represented by the following general matrix in 4-dimensional ($x,\;x',\;z,\;z'$) phase space:
\begin{equation}\label{eq:chirper}
R_{chirp}=\begin{pmatrix}
1 & L_{c} &0 &0&  \\ 
0 & 1 &  0 &0\\ 
0 & 0 & 1 &0\\
0 & 0 & R_{65}&1\\
\end{pmatrix}
,
\end{equation}
where $R_{65}<0$ for a ``chirper'' and $R_{65}>0$ for a ``dechirper''. The chirper does not affect the longitudinal beam size $\sigma_{z_1}=\sigma_{z_0}$, while the correlated energy spread at its exit becomes:
\begin{equation}\label{eq:zz'_chirp}
\mean{zz'}_1=R_{65}\cdot\sigma_{z_0}^2,
\end{equation}
and the total energy spread is:
\begin{equation}\label{eq:z'2_chirp}
\sigma_{z'_1}^2=\sigma_{z'_0}^2+R_{65}^2\cdot\sigma_{z_0}^2,
\end{equation}
expressed in the terms of the beam parameters at the entrance of the chirper. Using Eq.~(\ref{eq:R_56_1}), the longitudinal beam size after the chicane becomes:
\begin{equation}\label{eq:z2_chirp}
\sigma_{z_2}^2=(1+R_{56}R_{65})^2\sigma_{z_0}^2+R_{56}^2\sigma_{z'_0}^2
\end{equation}
and the $\Sigma-$matrix element defining linear energy correlations along the beam can be found as:
\begin{equation}\label{eq:zz'_chirp_chicane}
\mean{zz'}_2=R_{65}(1+R_{56}R_{65})\sigma_{z_0}^2+ R_{56} \sigma_{z'_0}^2,
\end{equation}
while the energy spread remains unchanged $\sigma_{z'_2}=\sigma_{z'_1}$. Clearly, one needs to complete the scheme with an additional beamline element in order to remove the correlated energy spread $\mean{zz'}_2/\sigma_{z_2}^2$ after the chicane. Up to now matrix elements $R_{56}$ of the chicane and $R_{65}$ of the chirper could have been independent of the beam parameters unless ones aims to maximize the compression ratio according to equation~(\ref{eq:R_56_cond}). However, due to the fact that the longitudinal beam size is changing in the chicane (as intended), a proper matrix element $R_{65}'$ of the dechirper added to compensate the previously imposed chirp is:
\begin{equation}\label{eq:R65_dechirp}
R_{65}'=-\frac{\mean{zz'}_2}{\sigma_{z_2}^2}=-\frac{R_{65}(1+R_{56}R_{65})\sigma_{z_0}^2+ R_{56} \sigma_{z'_0}^2}{(1+R_{56}R_{65})^2\sigma_{z_0}^2+R_{56}^2\sigma_{z'_0}^2},
\end{equation}
and depends not only on the parameters of the chirper $R_{65}$ and the chicane $R_{56}$, but also on the initial beam length $\sigma_{z_0}$ and energy spread $\sigma_{z'_0}$. Consequently, the final matrix of the chicane-based bunch compressor is specific for each set of longitudinal beam parameters through the matrix elements $R''_{65}=\frac{R_{56}\sigma_{z'_0}^2}{(1+R_{56}R_{65})^2\sigma_{z_0}^2+R_{56}^2\sigma_{z'_0}^2}$, and $R''_{66}=\frac{(1+R_{56}R_{65})\sigma_{z_0}^2}{(1+R_{56}R_{65})^2\sigma_{z_0}^2+R_{56}^2\sigma_{z'_0}^2}$ and can be found as:
\begin{equation}\label{eq:chicane_BC}
R_{BC}=\begin{pmatrix}
1 & L_{BC_x} & 0 &0  \\ 
0 & 1 &  0 &0\\ 
0&0&1+R_{56}R_{65} &R_{56}\\
0&0&R''_{65}&R''_{66}\\
\end{pmatrix}
,
\end{equation}
where the overall effective length in the transverse direction is $L_{BC_{x}}=L_{c}+L_{x}+L_{d}$~\cite{universal-chicane-BC}.

\emph{\bf LSC-driven microbunch instabilities.}
Chicane-based schemes suffer from LSC-induced microbunch instabilities. Saldin \emph{et al.}~\cite{Saldin_Klystron} predicted that the noise present in the density distribution along a bunch will be amplified by LSC space charge forces in a chicane-based BC, comparing this effect with a working principles of a klystron amplifier. Saldin's elegant model assumes that the DC electron beam upstream of the bunch compressor has a current modulation of the form
\begin{equation}\label{eq:current_mod}
	I_1(z) = I_0 \left ( 1 + \rho_i \cos(kz) \right),
\end{equation}
where $k=\omega/c$, and $\rho_i$ and $\omega$ are respectively the modulation density, and frequency. As the bunch passes through a long drift and/or accelerator section, the LSC forces amplify noise resulting in strong energy modulations along the beam  $\Delta\gamma_m=\frac{|Z(k)|}{Z_0}\frac{I_0}{I_A}\rho_i$, where $I_A=17$~kA is the Alfven current, $Z_0=377\;\Omega$ is the vacuum impedance, and $Z(k)$ is the longitudinal impedance characterizing the action of longitudinal wakefields. Due to the nonzero $R_{56}$ matrix element in the chicane, these energy modulations result in modulations of the current at the exit of the chicane, which after linearization can be found as (from \cite{Saldin_Klystron}):
\begin{equation}\label{eq:current_mod_chic_appr}
I_2(z)=CI_0\left(1+\rho_{ind}\cos(Ckz)\right),
\end{equation}
where $\rho_{ind}=CkR_{56}\frac{\Delta\gamma_m}{\gamma_0}{\exp}\left ( -\frac{C^2k^2R_{56}^2\sigma_{z'_1}^2}{2} \right )$ is the induced density modulation in the chicane, $C=(1+hR_{56})^{-1}$ is the compression factor, and $h$ is the linear energy chirp in the approximation of a continuous beam in the longitudinal direction. If $\sigma_{z'_1}^2\gg\Delta\gamma_m/\gamma_0$ analysis of the second term on the right-hand side of equation~(\ref{eq:current_mod_chic_appr}) shows that the harmonic component becomes negligible. Otherwise, the chicane can significantly amplify initial density and energy modulations along the bunch at the exit of the BC, which can eventually lead to the possible loss of FEL performance~\cite{microbunching-Borland,microbunching-SLAC-report,microbunching_LaserHeater2}.

There are several methods to suppress microbunching instabilities in chicane-based bunch compressors based on increasing the uncorrelated energy spread within the FEL tolerance described in Eq.~(\ref{dE_FEL})~\cite{microbunching-SLAC-report,microbunching_LaserHeater2,microbunching-TESLA-report,microbunching_LaserHeater1,Huang_Microbunch_Inst}. For example, at the LCLS it was initially proposed that the local energy spread could be enlarged with quantum fluctuations originated by induced incoherent synchrotron radiation (ISR) in a superconducting wiggler placed directly before the second BC at 4.54 GeV beam energy \cite{microbunching-SLAC-report}. This method effectively works at high beam energies where ISR can be significant. However, this does not suppress the energy modulations at the high-frequency imposed upstream~\cite{microbunching_LaserHeater2}. An alternative solution, relying on a laser heater and thus effectively working at low beam energies, ($\sim100-200$~MeV) was initially proposed for the TESLA Test Facility linac at DESY~\cite{microbunching-TESLA-report,microbunching_LaserHeater1} and was later designed~\cite{microbunching_LaserHeater2} and implemented~\cite{Huang_Microbunch_Inst} for the LCLS linac. In the original approach~\cite{microbunching_LaserHeater1}, a laser light induces rapid FEL-type energy modulations at the optical frequency in an undulator, which wash out the short wave modulations in the chicane of the downstream bunch compressor effectively ``heating'' the beam. In the modified design~\cite{microbunching_LaserHeater2} the undulator is placed in the middle of an additional small chicane in order to simplify optical access and initiate an additional temporal smearing effect of the energy modulations imposed by the laser. Using a SC wiggler or a laser heater in various modifications result in effective suppression of the growing harmonic current proportional to $\Delta\gamma_m$ for the large uncorrelated energy spread of the beam $\sigma_{z'_1}$, as calculated by the exponent in Eq.~(\ref{eq:current_mod_chic_appr}). Although these methods practically make the beam ``immune'' to LSC-induced microbunching instabilities, they come at the price of an enlarged energy spread which can be critical for the advanced FEL techniques such as seeding, etc.~\cite{microbunching_DCavity}.

Alternative approaches aimed at mitigating this issue rely on transverse to longitudinal mixing of the beam motion~\cite{microbunching1,microbunching_Chicane,microbunching_DCavity,microbunching_Bends}. Initial estimates of the emittance-initiated smearing of the microbunch instabilities using the 2-dimensional Vlasov solver~\cite{microbunching1-V} were reported in~\cite{microbunching1}. An interesting approach relying on transverse to longitudinal coupling in the magnetic chicane was proposed in~\cite{microbunching_Chicane}. In \cite{microbunching_Chicane}, the chicane of the second bunch compressor was modified from compressing the beam (the correlated energy spread at the entrance was removed) to suppressing the energy modulations along the bunch imposed in the first BC via rotation of the longitudinal phase space. Furthermore, reversible beam heating relying on transverse-longitudinal coupling in two TDCs is an elegant solution capable of eliminating the laser heating from the scheme~\cite{microbunching_DCavity}. Beam energy spread is increased in the first TDC downstream of the chicane by imposed correlations between the transverse vertical ($y,\;y'$) and longitudinal ($z,\;z'$) motions, while the beam dynamics through the elements of the chicane are coupled in 4D phase space ($x,\;x',\;z,\;z'$). Following the chicane, the uncorrelated energy spread is returned close to its initial value in the second TDC by uncoupling the transverse and longitudinal dynamics. The downside of the method is a degradation of the vertical emittance in the tail and front of the bunch caused by CSR effects in the chicane, which also prevents the complete compensation of the energy spread. 

Finally, suppression of current modulation growth in the chicane was proposed via longitudinal mixing taking advantage of the naturally present transverse beam spreads $\sigma_x$ and $\sigma_{x'}$, which is possible with the respectively nonzero $R_{51}$ and $R_{52}$ matrix elements of the two bending magnets accompanying the chicane~\cite{microbunching_Bends}. While this method demonstrated promising performance in the linear single particle dynamics approximation, its feasibility for nonlinear dynamics accounting for CSR effects requires further investigation. Indeed, the collective effects can result in the large degradation of the emittances and consequently ``destroy'' the compensation mechanism removing the correlations in the final bending magnet.  

Adding extra elements to a chicane to initiate transverse-longitudinal coupling in order to suppress LCS-induced microbunching instabilities intuitively suggests a search for an alternative compression scheme. The potential candidate is a bunch compressor based on a double emittance exchange separated by a demagnifying transverse-optics system~\cite{Zholents}.

\subsection{EEX-based BC} \label{subsec:EEX}

The first emittance exchange (EEX) scheme was proposed in 2002~\cite{EMMA}. It was realized as a chicane with an inserted transverse deflecting cavity between its doglegs. Although such a scheme highly couples transverse and longitudinal phase spaces, the exchange  between the corresponding emittances is not exact and only works if the following condition is satisfied:
\begin{equation}
	4 \left < x'^2 \right > \left < z'^2 \right > \eta^2\ll\epsilon_{x,z}^2.
\end{equation}
The exact, to first order, EEX is realized as two doglegs oriented in the same direction separated by a transverse deflecting cavity (Fig.~\ref{fig:chicane} (d))~\cite{EMMA2006}. The working principles of the exact EEX scheme can be explained by analyzing the transfer matrices of its structural components (Appendix~\ref{apx:elements}). Briefly, the transverse coordinate $x$ is affected by the energy spread $z'$ proportionally to dogleg dispersion $\eta$. The longitudinal coordinate $z$ similarly depends on the initial divergence $x'$. The upstream TDC provides a back action: the initial coordinates $z$ and $x$ impact the final momentums of the motions in the opposite degree of freedom, respectively $x'$ and $z'$. A combination of one dogleg and a TDC is not yet enough to cross-couple transverse and longitudinal phase spaces and exchange the emittances. Completing the scheme with another dogleg (identical to the first and oriented in the same direction) results in the overall transfer matrix delivering exact emittance exchange:
\begin{equation}\label{eq:EEX-direct}
R_{EEX}=\begin{pmatrix}
0 & 0 & -\frac{L}{\eta} &\eta- \frac{L\psi}{\eta}   \\ 
0 & 0 & -\frac{1}{\eta} & -\frac{\psi}{\eta}  \\
-\frac{\psi}{\eta} & \eta- \frac{L\psi}{\eta} & 0 & 0   \\
-\frac{1}{\eta} & -\frac{L}{\eta} & 0 & 0   
\end{pmatrix}
,
\end{equation}
where the condition on the geometric strength of the TDC, $\kappa=-1/\eta$, has to be satisfied. The introduced parameters are defined according to the relations:
\begin{align}\label{eq:EEXparameters}
L&=\left( D+D\cos 2\theta+S_1\right) \sec^3\theta+S_2,\\
\psi&=-2D\;\theta\;\csc\theta+\sec\theta\left(2D+S_1\tan^2\theta\right),\\\label{eq:EEXparameters-l}
\eta&=-\sec^3\theta\;\tan\left(\frac{\theta}{2}\right)\left(S_1+S_1\cos\theta+D+D\cos 2\theta \right),
\end{align} 
where $D=r\sin\theta$ is the length of the bending magnet along the $z-$axis in the laboratory frame and $\theta$ is the bending angle, $S_1$ and $S_2$ are respectively separations between two magnets and magnet -- deflecting cavity.
The matrix elements in the main diagonal blocks completely vanished justifying that transverse and longitudinal phase spaces at the exit of the scheme are completely independent from their corresponding initial values, and emittances are exchanged.

The idea to use a single EEX as a bunch compressor was proposed by Carlsten~\emph{et al.}~\cite{singleEEX_Bruce}. The main advantage of this scheme is that the final longitudinal phase space is decoupled from the initial longitudinal phase space and does not require any initial and residual energy-phase compensation as a chicane-based scheme. Consequently, such a scheme  should be much less affected by LSC-induced microbunching instabilities. This compressor configuration requires all initial emittances to be of the same order ($\epsilon_{n_x}\sim\epsilon_{n_y}\sim\epsilon_{n_z}$) because an emittance exchanger only swaps the projections of eigen emittances between the transverse and longitudinal phase spaces~\cite{Yampolsky_arxiv,Malyzhenkov_AAC}. Since transverse emittances from photo-cathode injectors are typically of the same order and significantly different than the longitudinal emittance, the idea of a single EEX based bunch compressor eventually transformed into a double EEX~\cite{Zholents}. First, transverse ($x,x'$) phase space is exchanged with the longitudinal phase space. Then, a ``telescope'' consisting of transverse optics elements, quadrupole magnets and drifts, is applied to compress the beam in the new ($x,x'$) phase space, which is the original ($z,z'$) phase space. Finally, a second EEX switches the transverse and longitudinal phase spaces back, returning a beam that is compressed in the $z-$direction with enlarged energy spread. 

Zholents~\emph{et al.}~\cite{Zholents} proposed using the chicane-type EEX modules with focusing and defocussing quadrupole magnets inserted along the beamline with respect to ($x,x'$) phase space to adjust the transverse beam parameters in the original design of the DEEX-based BC. In contrast, we discuss the exact EEX modules with unidirectional doglegs in this paper. The direction of the second EEX module of the bunch compressor is flipped so the electron beam can travel back to its original path for simple integration within existing linacs (Fig.~\ref{fig:chicane} (c)). For this configuration, the transverse phase space should be properly reshaped in the telescope if one needs to keep the original orientation of the longitudinal phase space at the exit of the BC. For the sake of compensation of CSR effects it might be useful to reshape and mirror the original orientation of the transverse phase space in the telescope resulting in the mirrored beam in the longitudinal phase space at the exit of the BC (see Sections~\ref{sec:results} and~\ref{sec:ES-Optimization} for details). %~\cite{EEX-orientation}. 
Furthermore, we generally ``allow'' unique configuration of the structural components for the emittance exchangers A and B: $R_{EEX_{A,B}}=\pm R_{EEX}(L_{A,B},\eta_{A,B},\psi_{A,B})$ using Eq.~(\ref{eq:EEX-direct}). 
Conversely, the parameters of each EEX are defined according to Equations~(\ref{eq:EEXparameters}-\ref{eq:EEXparameters-l}) as following:
\begin{align}\label{eq:EEXparameters-AB}
L_{A,B}&=\L(D_{A,B},\theta_{A,B},S_{1_{A,B}},S_{2_{A,B}}),\\
\psi_{A,B}&=\psi(D_{A,B},\theta_{A,B},S_{1_{A,B}}),\\
\eta_{A,B}&=\eta(D_{A,B},\theta_{A,B},S_{1_{A,B}}),
\end{align}
and cavity strengths are $\kappa_{A,B}=\mp1/\eta_{A,B}$.
Such a asymmetry is compensated in the transverse-optics between two EEX modules:
\begin{equation}\label{eq:telescope}
R_{T_{1,2}} = \begin{pmatrix}
\pm R_{11} & \pm R_{12} & 0 & 0  \\ 
\pm R_{21} & \pm R_{22} & 0 & 0  \\
0 &0 & 1 & 0   \\
0 & 0 & 0 & 1   
\end{pmatrix},
\end{equation}
where the transverse matrix elements are defined through the parameters of emittance exchangers:
\begin{align}\label{eq:TELEparameters-AB}
R_{11}&=\frac{-L_B (\psi_A + m^2 \psi_B) + 
	m^2 \eta_B^2}{m\eta_A\eta_B},\\
R_{21}&=\frac{\psi_A + m^2 \psi_B}{m \eta_A \eta_B},\\
R_{22}&=\frac{-L_A (\psi_A+ 
	m^2\psi_B) + \eta_A^2}{m\eta_A\eta_B},\\\label{eq:TELEparameters-AB_F}
R_{12}&=\frac{R_{11}R_{22}-1}{R_{21}},
\end{align}
where $m$ is the compression factor. The transverse optics, $R_{T_{1}}$, preserve the orientation of the final longitudinal phase space (from now on referred to as direct or standard) and $R_{T_{2}}$ turns it around (from now on referred as mirrored).
Finally, the transfer matrix of the DEEX BC derived as a multiplication product of its structural components, $R_{DEEX}=R_{EEX_A}\cdot R_{T_{1,2}}\cdot R_{EEX_B}$, can be found in ($x,x',z,z'$) 4D phase space:
\begin{align}\label{DEEX-matrix}
R'_{DEEX}=\begin{pmatrix}
R'_{11} & R'_{12} & 0 & 0  \\ 
R'_{21} & R'_{22} & 0 & 0  \\
0 & 0 & \pm 1/m & 0   \\
0 & 0 & 0 & \pm m   
\end{pmatrix},
\end{align}
where the transverse matrix elements are:
\begin{align}\label{eq:DEEX-matrix_parameters}
R'_{11}&=-\frac{L_B (\psi_A + \psi_B) - 
	\eta_B^2}{\eta_A\eta_B},\\
R'_{21}&=-\frac{\psi_A + \psi_B}{ \eta_A \eta_B},\\
R'_{22}&=-\frac{L_A (\psi_A+ 
	\psi_B) - \eta_A^2}{\eta_A\eta_B},\\\label{eq:DEEX-matrix_parametersF}
R'_{12}&=\frac{R'_{11}R'_{22}-1}{R'_{21}}.
\end{align}
The motion in ($y,y'$) phase space is uncoupled and described in Appendix~\ref{apx:elements}. In contrast to the chicane-based BC in Eq.~(\ref{eq:chicane_BC}), the final matrix of the DEEX-based BC in Eq.~(\ref{DEEX-matrix}) is independent of the beam parameters, and the compression ratio $m$ is theoretically unlimited in the approximation of linear single particle dynamics and can be easily adjusted by tuning the transverse-optics system between two EEXs.

\subsection{CSR effects} \label{subsec:CSR}
Chicane-based bunch compressors experience emittance degradation due to CSR effects. Particles at the front (end) of the bunch gain (lose) energy in bending magnets due to the interaction with the self-produced CSR radiation, which results in the appearance of a CSR wake in the longitudinal phase space of the beam~\cite{Bruce_Tor,Derbenev_CSR,Saldin_CSR,Borland_CSR,Stupakov_CSR}. In the dispersive element such as a bending magnet, this distribution in energy unavoidably affects the transverse phase space resulting in a degradation of the transverse emittance. According to the theoretical calculations in~\cite{Bruce_Tor}, the impact of the CSR effects on the transverse emittance enlargement (added in quadrature to the initial emittance $\epsilon_{n_f}^2=\epsilon_{n_i}^2+\Delta \epsilon_n^2$) in the bending plane of a dipole magnet can be quantified as:
\begin{equation}\label{eq:Emitt-growth_scale}
\Delta \epsilon_n=0.38\theta^2\frac{I_p}{I_A}\ln\left(\frac{\rho_0}{a}\right)\frac{\sigma_r^2}{\sigma_z}\;\;\;,
\end{equation}
where $I_p$ is the beam peak current, $I_A$ is the Alfven current, $\sigma_r$ is the radial beam size in the approximation of a cylindrical beam, $\sigma_z$ is the bunch length, $\rho_0$ is the radius of the beam pipe and $\theta$ is the bending angle. The $R_{16}$ and $R_{26}$ matrix elements of the chicane are absent in the approximation of the linear single-particle dynamics, however they become effectively nonzero for particles in the head and tail of the bunch due to the CSR wake~\cite{Derbenev_CSR,Saldin_CSR,Double-chicane}. This results in even bigger emittance deterioration in the chicane than in a single dipole magnet. 

The standard approach to preserve beam quality after the chicane is to minimize the emittance degradation in each element of the scheme. CSR radiation can exit the bending magnet and keep interacting with a beam in the downstream sections of the beamline~\cite{Saldin_CSR,Borland_CSR,Stupakov_CSR}. According to Eq.~(\ref{eq:Emitt-growth_scale}), emittance enlargement quadratically depends on beam peak current at a constant bunch charge $\Delta \epsilon_n\sim I_p/\sigma_z\sim I_p^2$, so the effect becomes larger for shorter bunches. Quadratic scaling of emittance growth with transverse beam size and bending angle suggests minimizing these parameters for suppressing the CSR effects in each dipole magnet. 

A more sophisticated and, perhaps, more effective approach relies on mutual compensation of the CSR effects imposed in various sections of the bunch compressor. For instance, this is realized in a double-chicane scheme~\cite{Double-chicane}. The orientation of the second chicane is flipped with respect to the first chicane. This results in the opposite sign of the dispersion $\eta$ of the doglegs in the second chicane which partially reverses the transverse phase-space displacement caused by the CSR effects in the first chicane and eventually reduces the emittance deterioration at the exit of the compressor. The compensation of the CSR effects can be optimized by adjusting the beam elements of each chicane, the input Twiss parameters of the beam, and the transverse-optics in between of two chicanes. An interesting approach to minimize the impact of the CSR effects on emittance deterioration via longitudinal bunch shaping is discussed in~\cite{Mitchell_CSR}.
Alternatively, minimizing the CSR-induced phase space degradation can be accomplished by blocking the CSR radiation with metal plates along the beam path in a bending magnet as was demonstrated in~\cite{PRL-block-CSR}.

The microbunching instabilities driven by CSR effects are typically much smaller than the LSC-initiated counterpart~\cite{Saldin_Klystron,ZHuang_2,DRatner}, while originally they were considered to be the major source of the magnified noise in the chicane-type BCs~\cite{microbunching_CSR,Stupakov}. The LSC-induced instabilities are expected to be much smaller in the DEEX-based BC, predominantly because of the absence of the chicane-type optics and constant transverse-longitudinal mixing of the beam motion. However, the CSR effects in the dipole magnets in the phase space exchange schemes are still expected to be critical, and might be even more problematic than in chicane-based bunch compressors.

%%%%%%%%%%%%%%%%%%%%%%%%%%%%%%%%%%%%%%%%%%%%%%%%%%%%%%%%%%%%%%%%%%%%%%%%%%%%%%%%%%%%%%%%
%%%%%%%%%%%%%%%%%%%%%%%%%%%%%%%%%%%%%%%%%%%%%%%%%%%%%%%%%%%%%%%%%%%%%%%%%%%%%%%%%%%%%%%%
\section{Symmetric DEEX}\label{sec:results}
%%%%%%%%%%%%%%%%%%%%%%%%%%%%%%%%%%%%%%%%%%%%%%%%%%%%%%%%%%%%%%%%%%%%%%%%%%%%%%%%%%%%%%%%
%%%%%%%%%%%%%%%%%%%%%%%%%%%%%%%%%%%%%%%%%%%%%%%%%%%%%%%%%%%%%%%%%%%%%%%%%%%%%%%%%%%%%%%%

Assuming similar beam parameters to those used for the second bunch compressor at 1.6 GeV in the conceptual design of LCLS-II~\cite{LCLS2-conceptual_design} (summarized in Table~\ref{Tbl:LCLS2-beam}), we consider a symmetric configuration of the DEEX BC. The parameters for beamline elements, chosen with realistic electro-magnetic fields in bends and transverse deflecting cavities, are summarized in Table~\ref{Tbl:LCLS2-beamline} and satisfy conditions of the corresponding matrix formalism presented in Section~\ref{sec:bc}.
\begin{table}[ht]
	\begin{ruledtabular}
		\caption{LCLS-II~\cite{LCLS2-conceptual_design} beam parameters on the entrance of the bunch compressor used in the simulations.
		}
		\begin{tabular}{lll}\label{Tbl:LCLS2-beam}
			{\bf Electron beam }& LCLS-II& Units\\ 
			Normalized emittance $\epsilon_{n_x,n_y}$& 0.45&$\mu$m\\
			Energy spread (rms)&  47 & keV\\
			Bunch length (rms)& 153 & $\mu$m\\
			Energy chirp (rms)& 0  & m$^{-1}$\\
			Beam energy&  1.6& GeV\\
			Bunch charge& 100  &pC\\
			Required compression ratio & 17  & \\	
		\end{tabular}
	\end{ruledtabular}
\end{table}

\begin{table}[ht]
	\begin{ruledtabular}
		\caption{Parameters of the symmetrical DEEX and input Twiss parameters of the beam for the unoptimized design with standard and mirrored telescope.}
		\begin{tabular}{lll}	\label{Tbl:LCLS2-beamline}
			{\bf Parameters }& Values &Units  \\ 
			Bend angle, $\theta_{A}=\theta_{B}$ & 2.5&deg  \\
			Bend radius, $r_{A}=r_{B}$ & 1.14& m   \\
			Drift, $S_{1A}=S_{1A}$ & 4.8& m   \\
			Drift, $S_{2A}=S_{2B}$ &1.6& m \\
			Cavity strength, $\kappa_{A}=-\kappa_{B}$&-4.71&$\rm m^{-1}$\\
			Telescope matrix element, $R_{11}$ & $\mp$5.768&\\
			Telescope matrix element, $R_{12}$ & $\pm$37.186&m\\
			Telescope matrix element, $R_{21}$& $\pm$3.495&m$^{-1}$ \\
			Telescope matrix element, $R_{22}$ &$\mp$22.709&\\
			Twiss parameter, $\beta_x$  &100& m \\
			Twiss parameter, $\alpha_x$  &0& \\
		\end{tabular}
	\end{ruledtabular}
\end{table}

\begin{figure*}[ht]
	\includegraphics[width=0.470\textwidth]{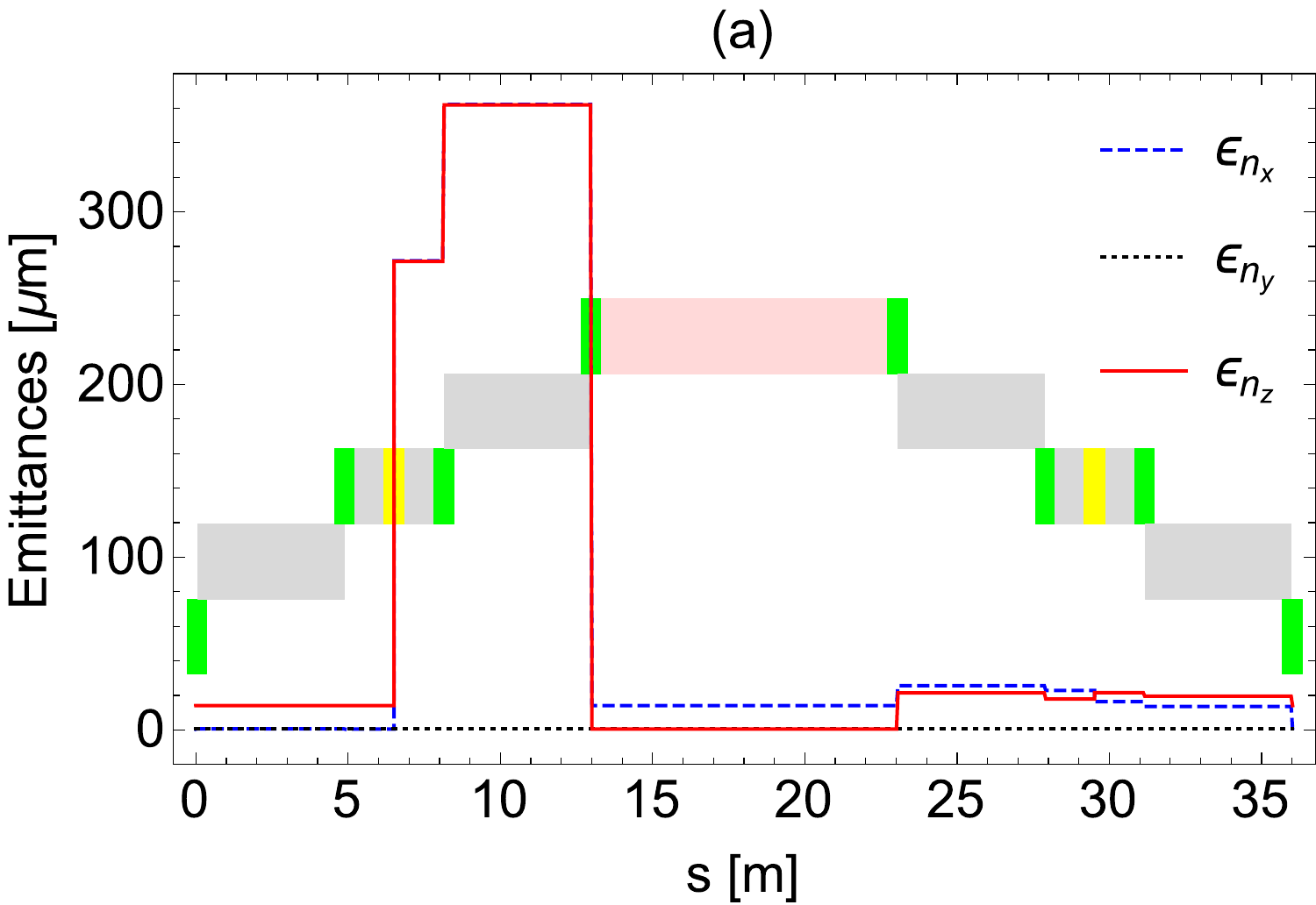}
	\includegraphics[width=0.470\textwidth]{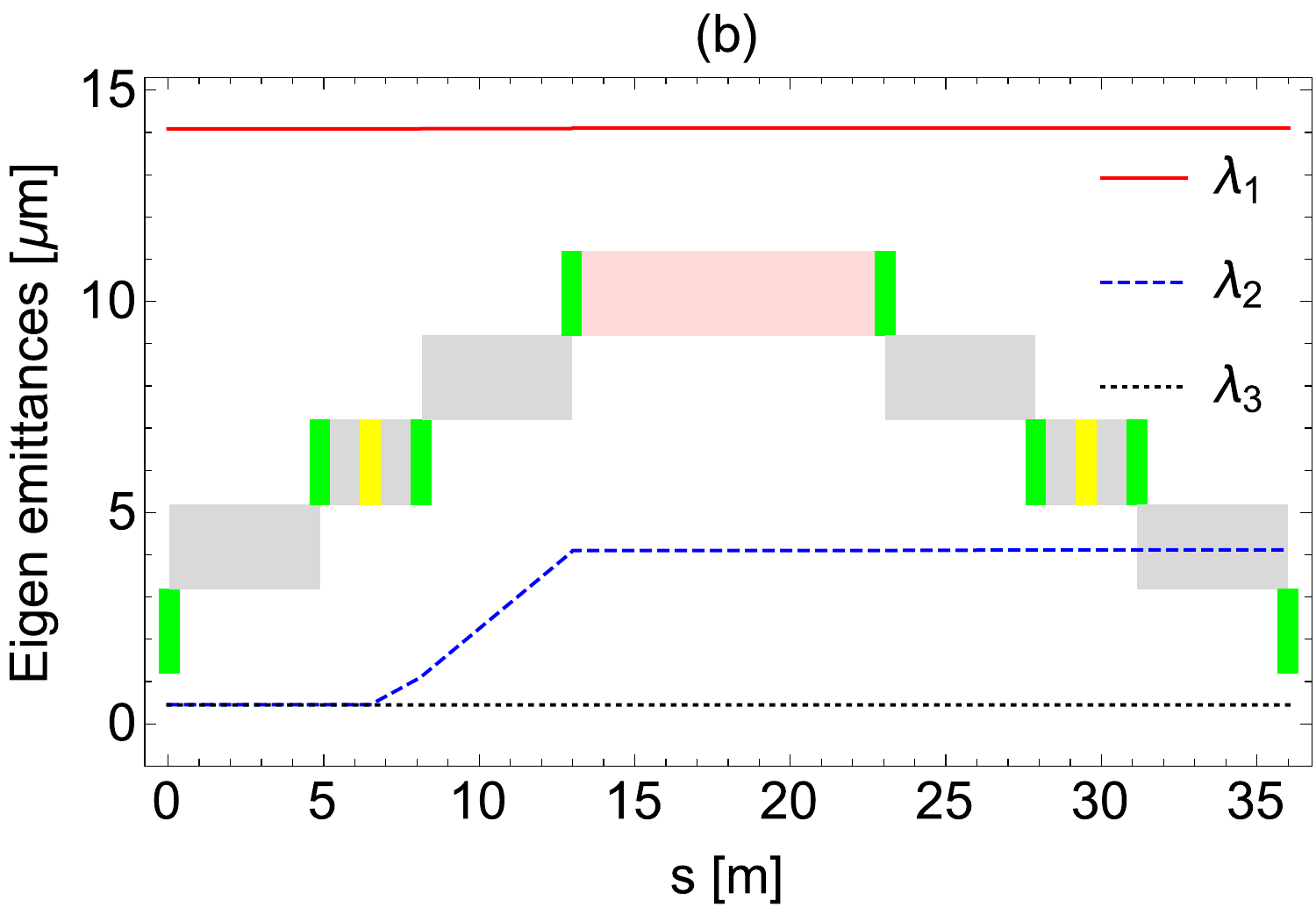}
	\includegraphics[width=0.470\textwidth]{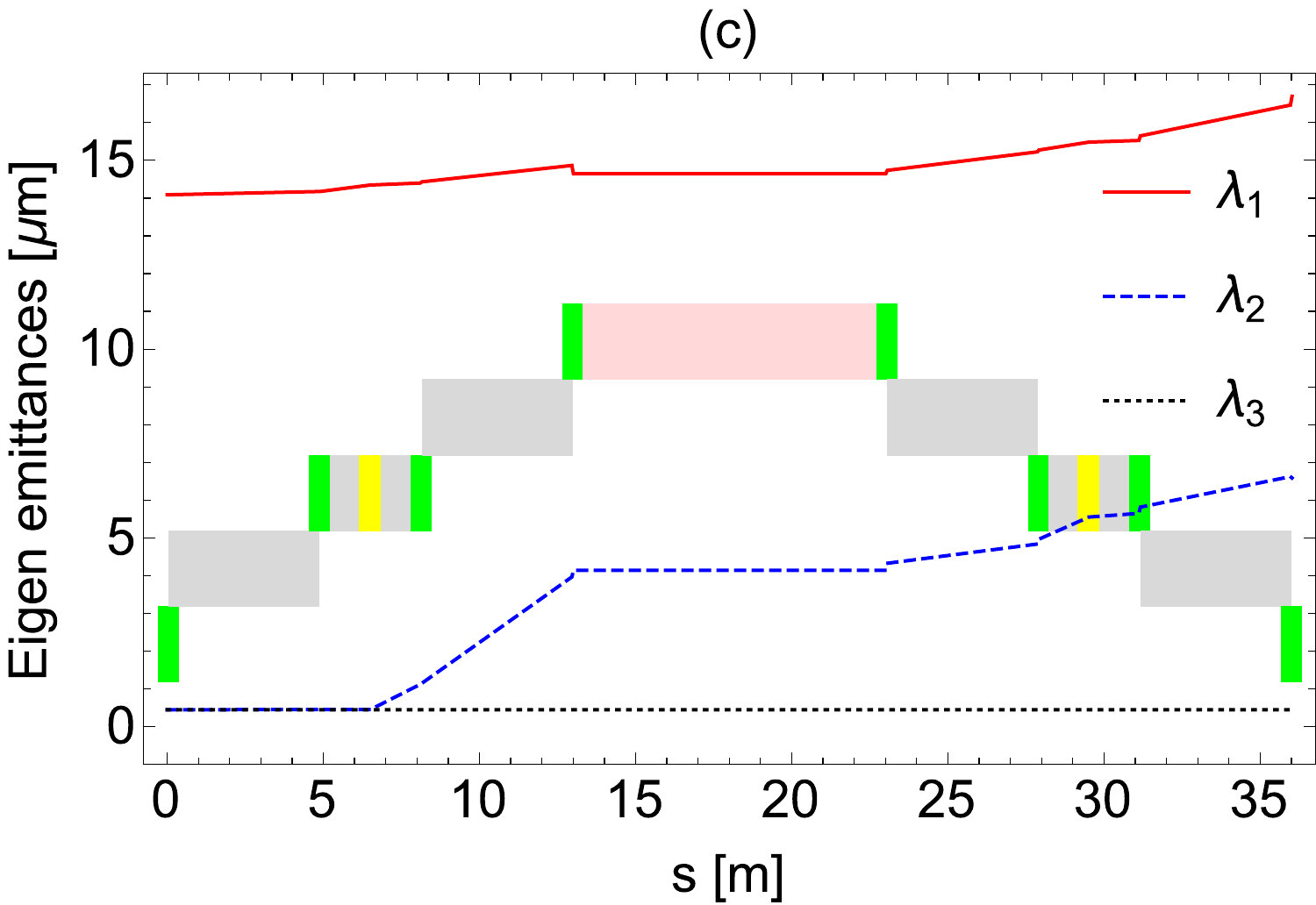}
	\includegraphics[width=0.470\textwidth]{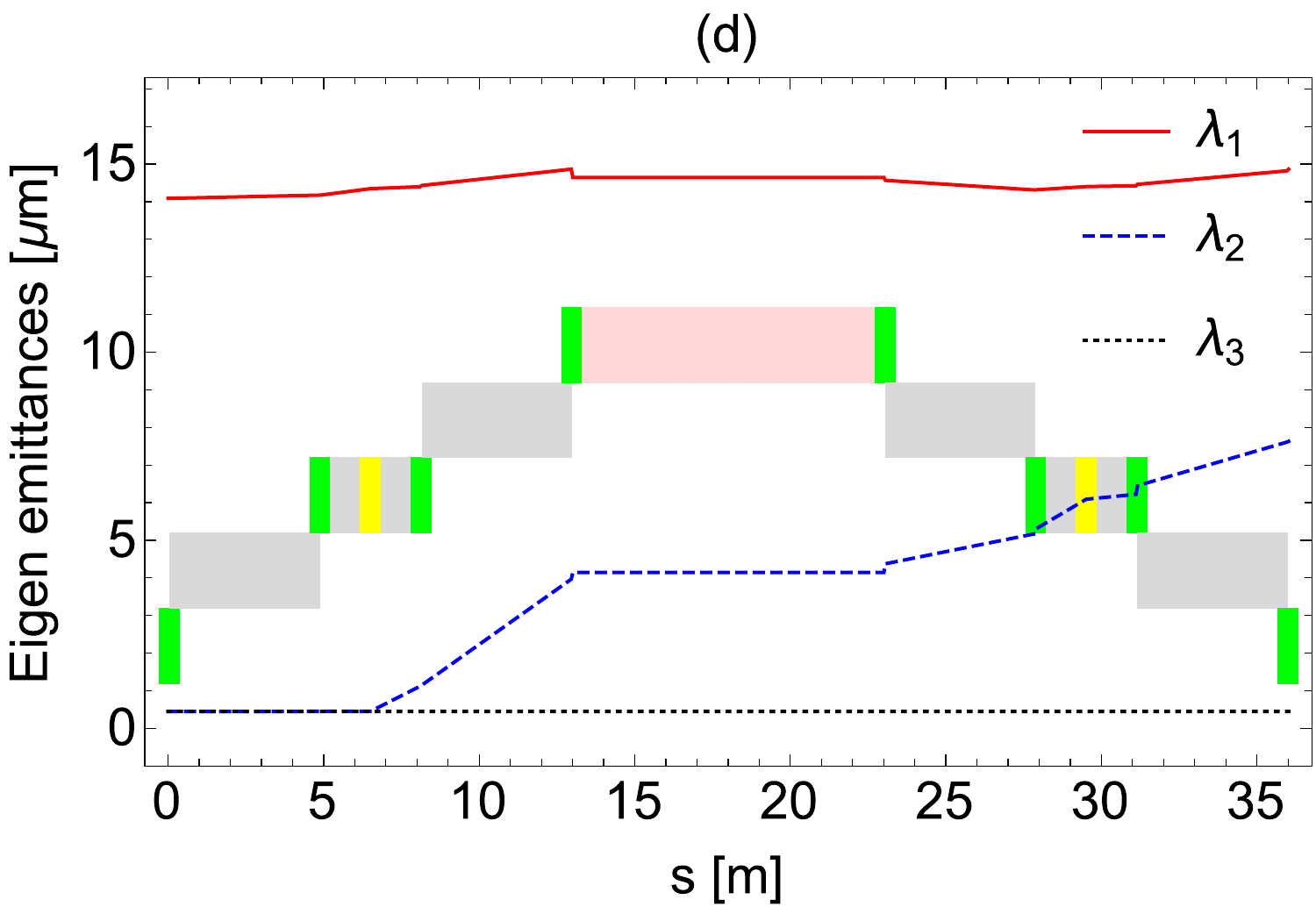}
	\caption{Evolution of the beam parameters along the beamline in different regimes: normalized emittances in linear regime (a); normalized eigen emittances in nonlinear regime (b);  normalized eigen emittances in CSR regime for direct (c) and mirrored (d) telescope configuration.
	} 
	\label{fig:Emittance_evol}
\end{figure*}
The scheme delivers exact direct and inverse exchanges in each EEX module and overall results in compressing the bunch length be a factor of 17 in the approximation of linear single particle dynamics. Simulation studies discussed in this paper are performed using tracking code {\sc elegant}~\cite{Borland} in the corresponding regimes of the linear and nonlinear single-particle dynamics (referred further as linear and nonlinear regimes) and of the multi-particle dynamics accounting for CSR effects in bending magnets and drifts (referred further as CSR regime).  
The transverse and longitudinal emittances grow within the first EEX because beam dynamics are highly coupled between various degrees of freedom as depicted in Fig.~\ref{fig:Emittance_evol}~(a). The emittances are exchanged after the first EEX and precisely correspond to the initial values of each other justifying the absence of transverse-longitudinal correlations. In the second EEX, transverse and longitudinal dynamics become coupled again resulting in enlargement of the corresponding emittances, which at the exit of the compressor return to their initial values. In contrast, the invariants of linear dynamics, eigen emittances, remain unchanged along the beamline in the linear regime as expected. The eigen-emittance formalism is reviewed in Appendix~\ref{apx:beam}.

It is convenient to introduce the coefficient $\mu$ as a ratio between the product of normalized eigen emittances and the product of normalized emittances:
\begin{equation}\label{eq:mu} \mu=\frac{\epsilon_{n_x}\cdot\epsilon_{n_y}\cdot\epsilon_{n_z}}{\lambda_1\cdot\lambda_2\cdot\lambda_3}=\frac{\epsilon_{n_x}\cdot\epsilon_{n_y}\cdot\epsilon_{n_z}}{\gamma^3\beta^3\sqrt{\det\Sigma}}=\frac{\epsilon_{n_x}\cdot\epsilon_{n_y}\cdot\epsilon_{n_z}}{\epsilon_{n_{6D}}}\;\;\;,
\end{equation}
which can be expressed in terms of the normalized 6-dimensional emittance $\epsilon_{n_{6D}}$.
The coefficient $\mu$ characterizes if an emittance exchange is complete, \emph{e.g.} if there are no remaining transverse-longitudinal correlations, it is exactly equal to 1 in the linear regime. 

\begin{figure*}[p]
	\centering
	\includegraphics[width=0.3\textwidth]{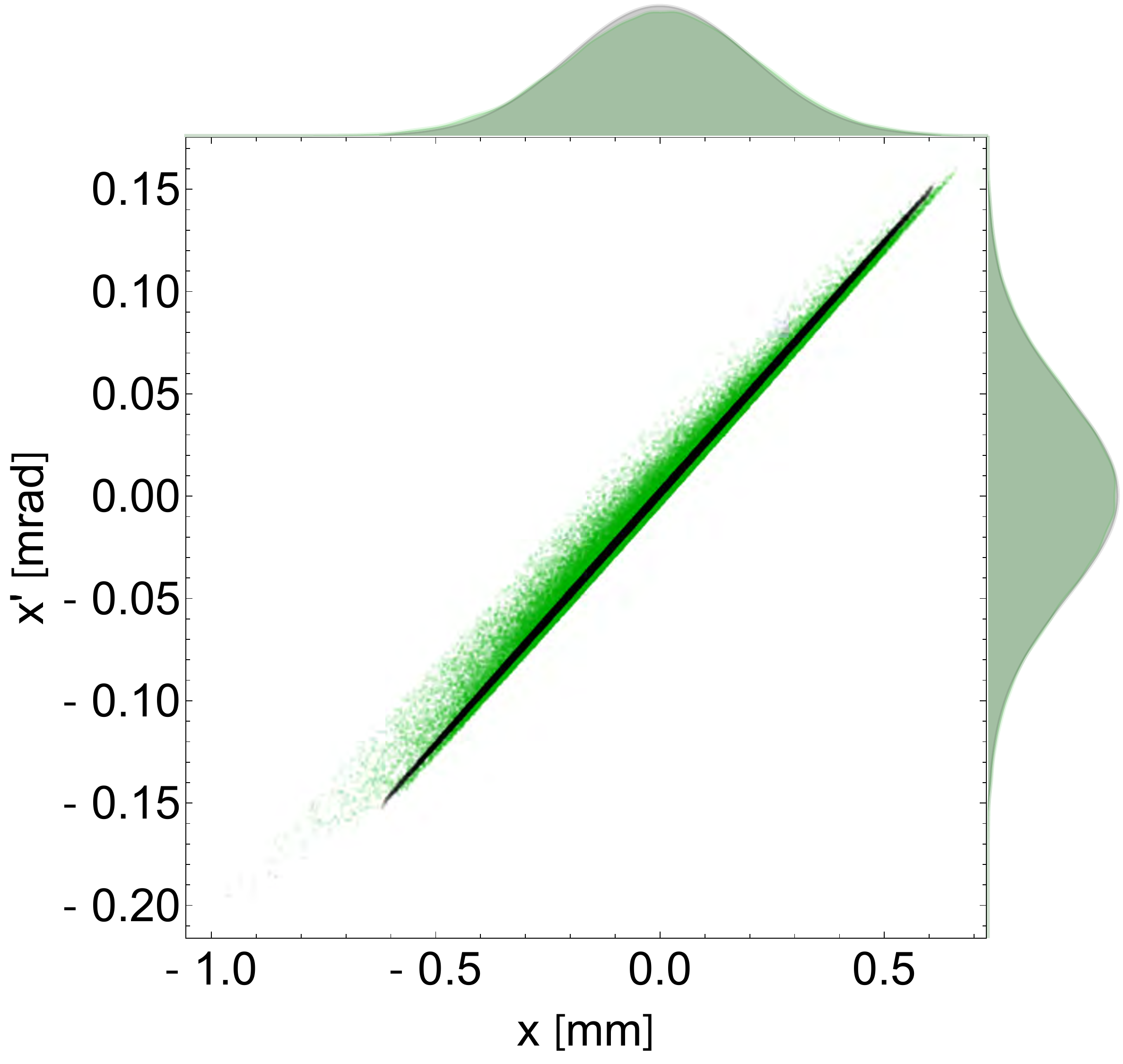} \
	\includegraphics[width=0.32\textwidth]{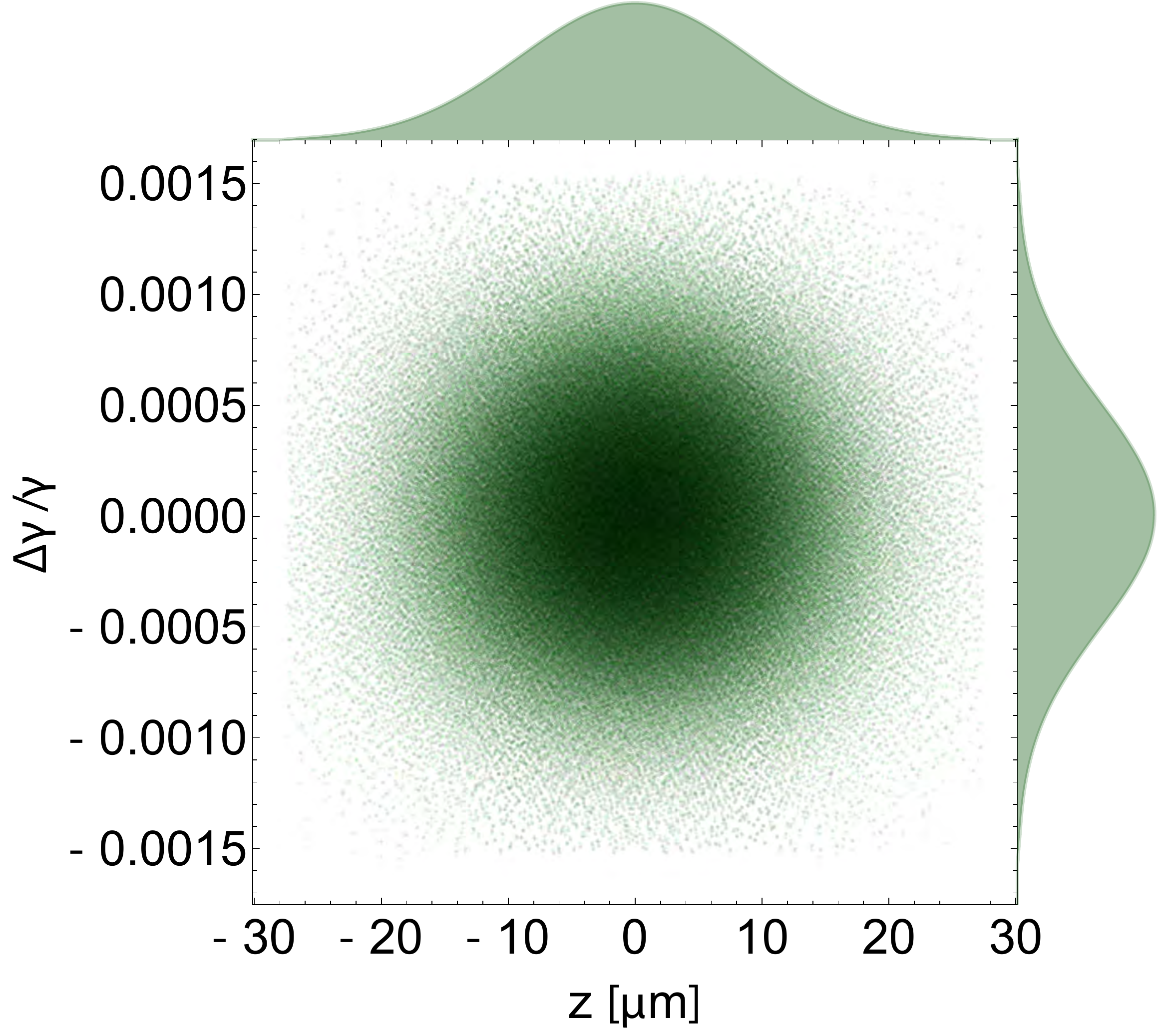} \
	\includegraphics[width=0.31\textwidth]{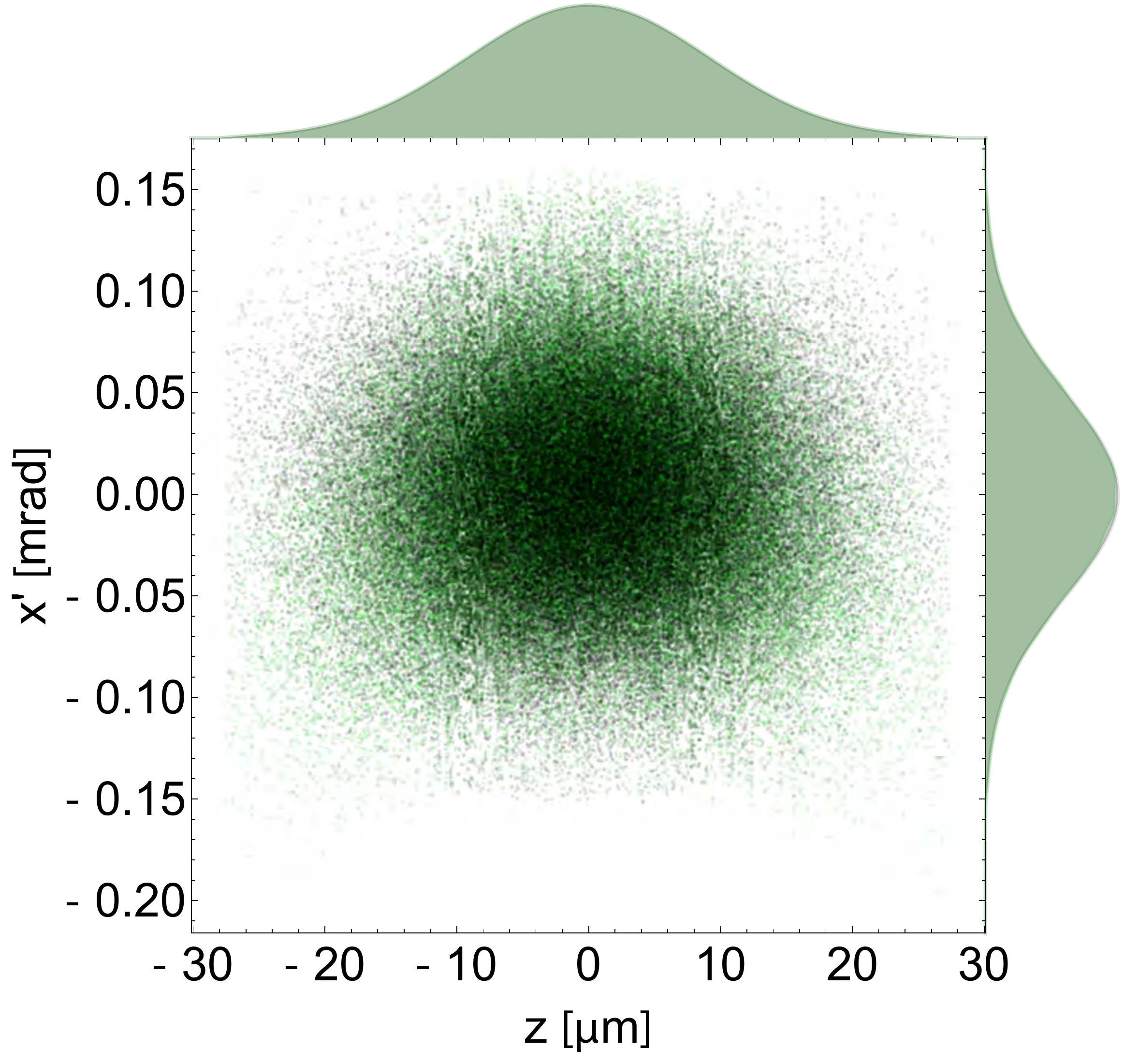} 
	\includegraphics[width=0.32\textwidth]{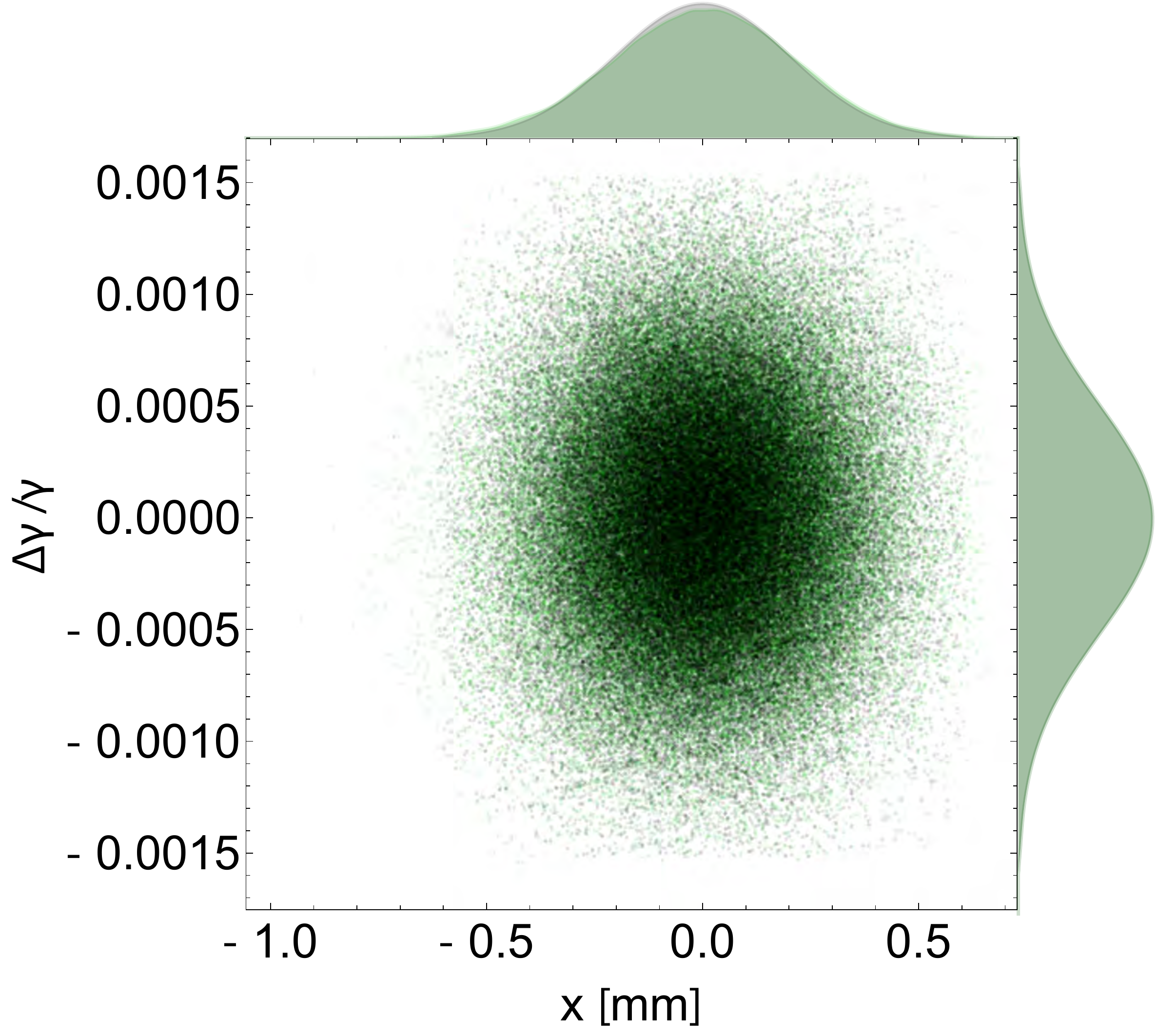} \
	\includegraphics[width=0.3\textwidth]{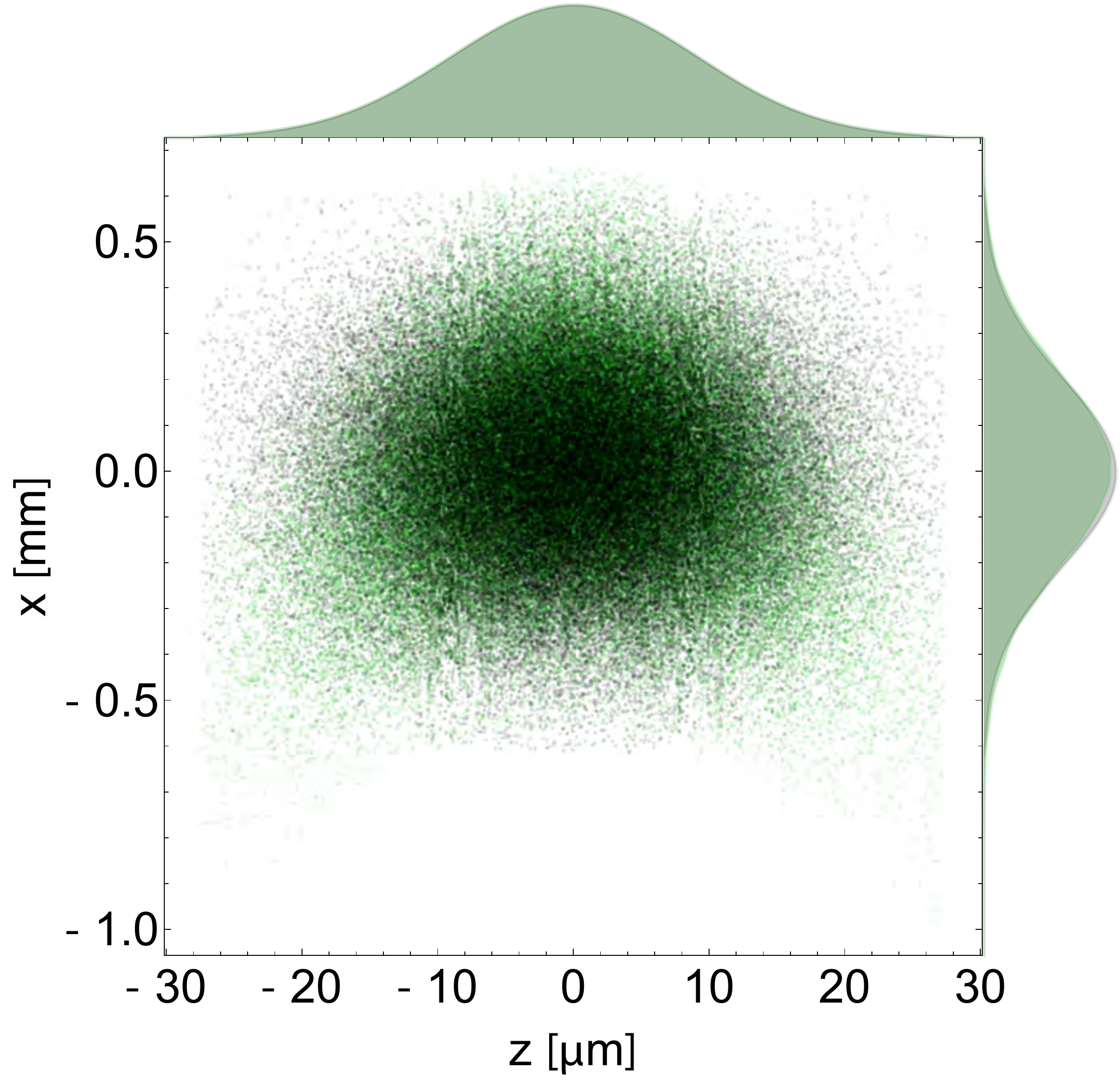} \
	\includegraphics[width=0.32\textwidth]{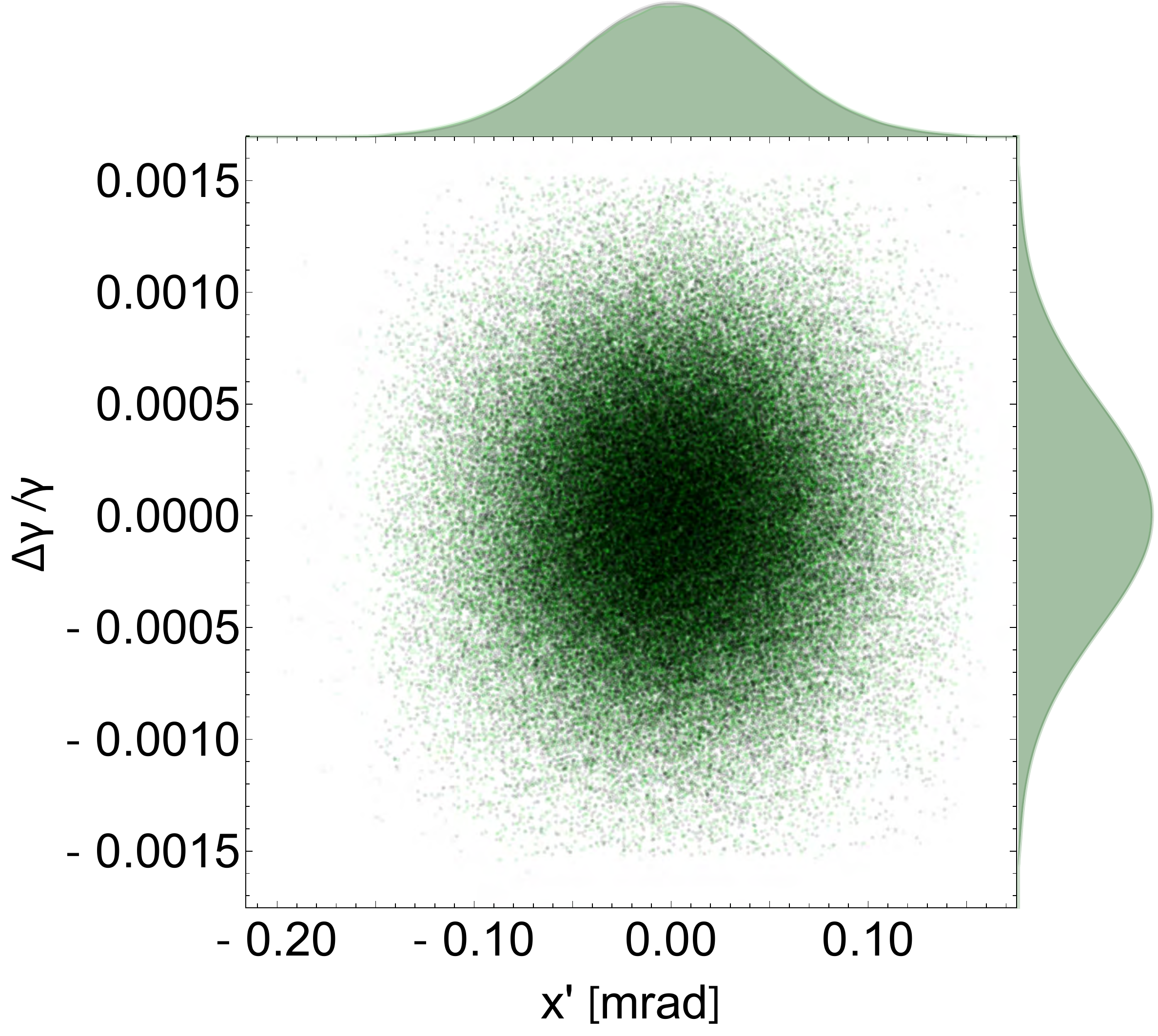} 
		\caption{Output phase spaces in linear (black) and nonlinear (green) regimes. In both regimes phase spaces are completely identical for the direct and mirrored telescope configuration hence only the former is depicted.
	} 
	\label{fig:LCLS2 phase spacesL-NL}
		\centering
	\includegraphics[width=0.3\textwidth]{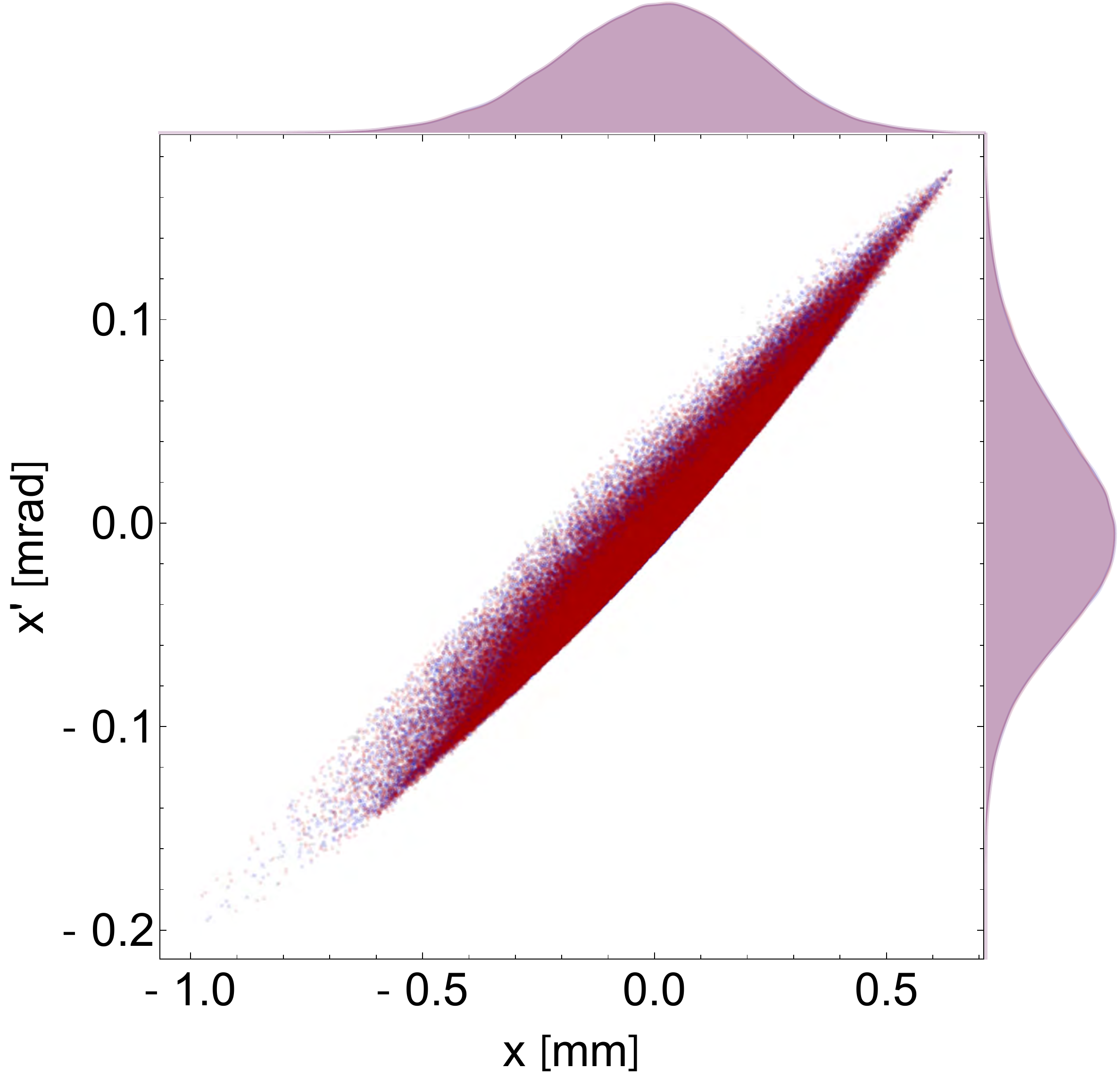} \
	\includegraphics[width=0.32\textwidth]{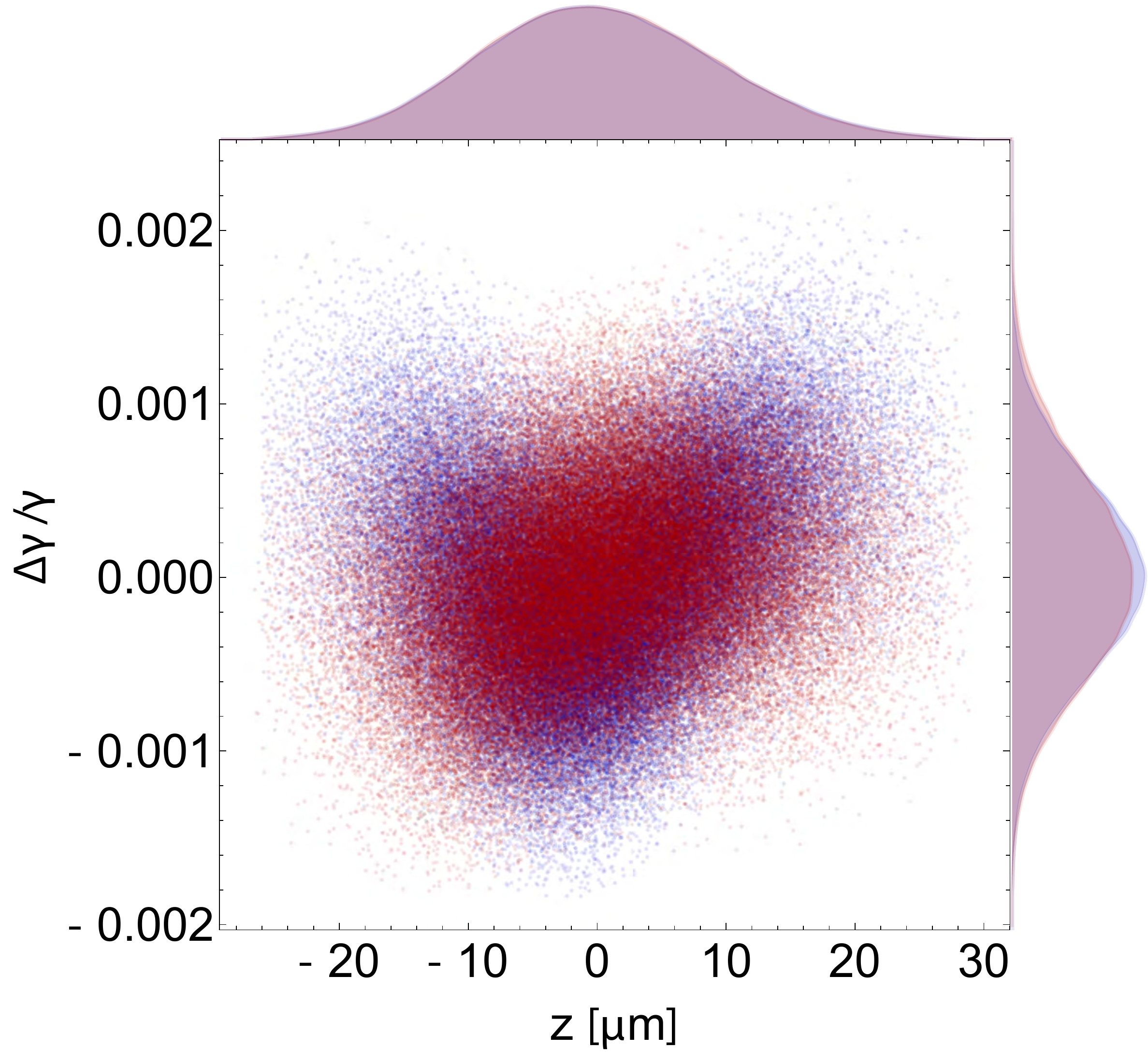} \
	\includegraphics[width=0.31\textwidth]{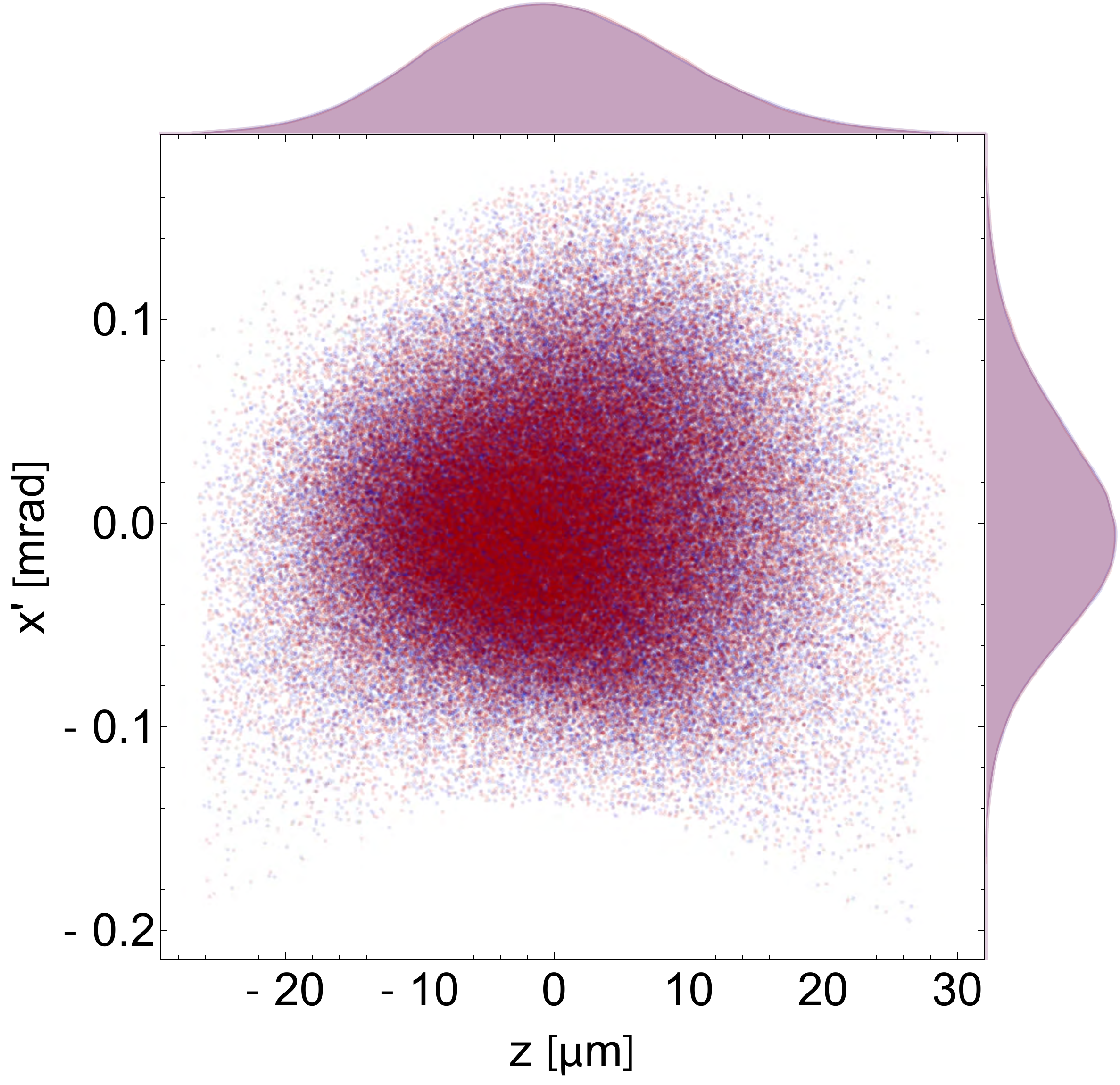} 
	\includegraphics[width=0.32\textwidth]{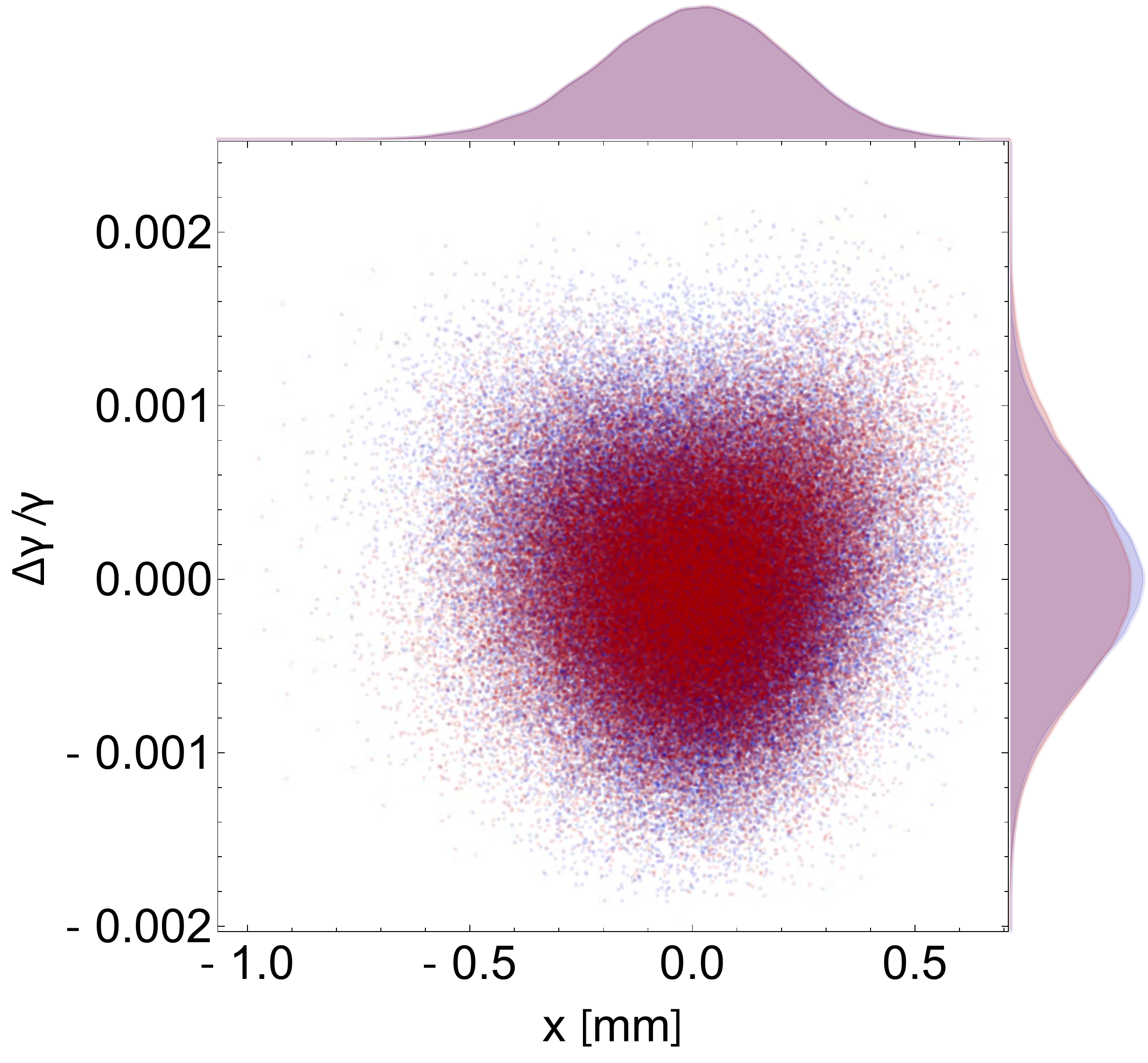} \
	\includegraphics[width=0.3\textwidth]{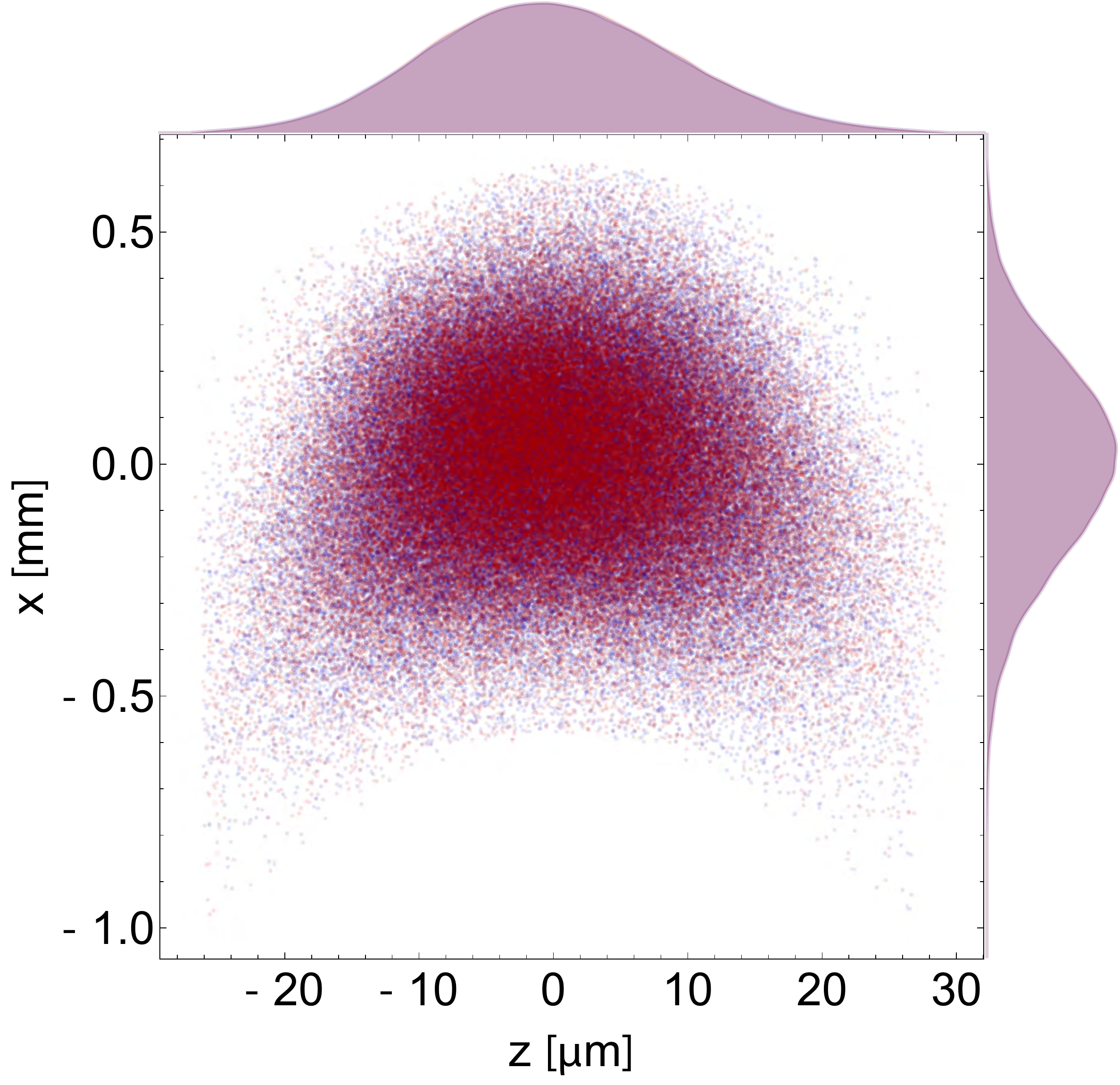} \
	\includegraphics[width=0.32\textwidth]{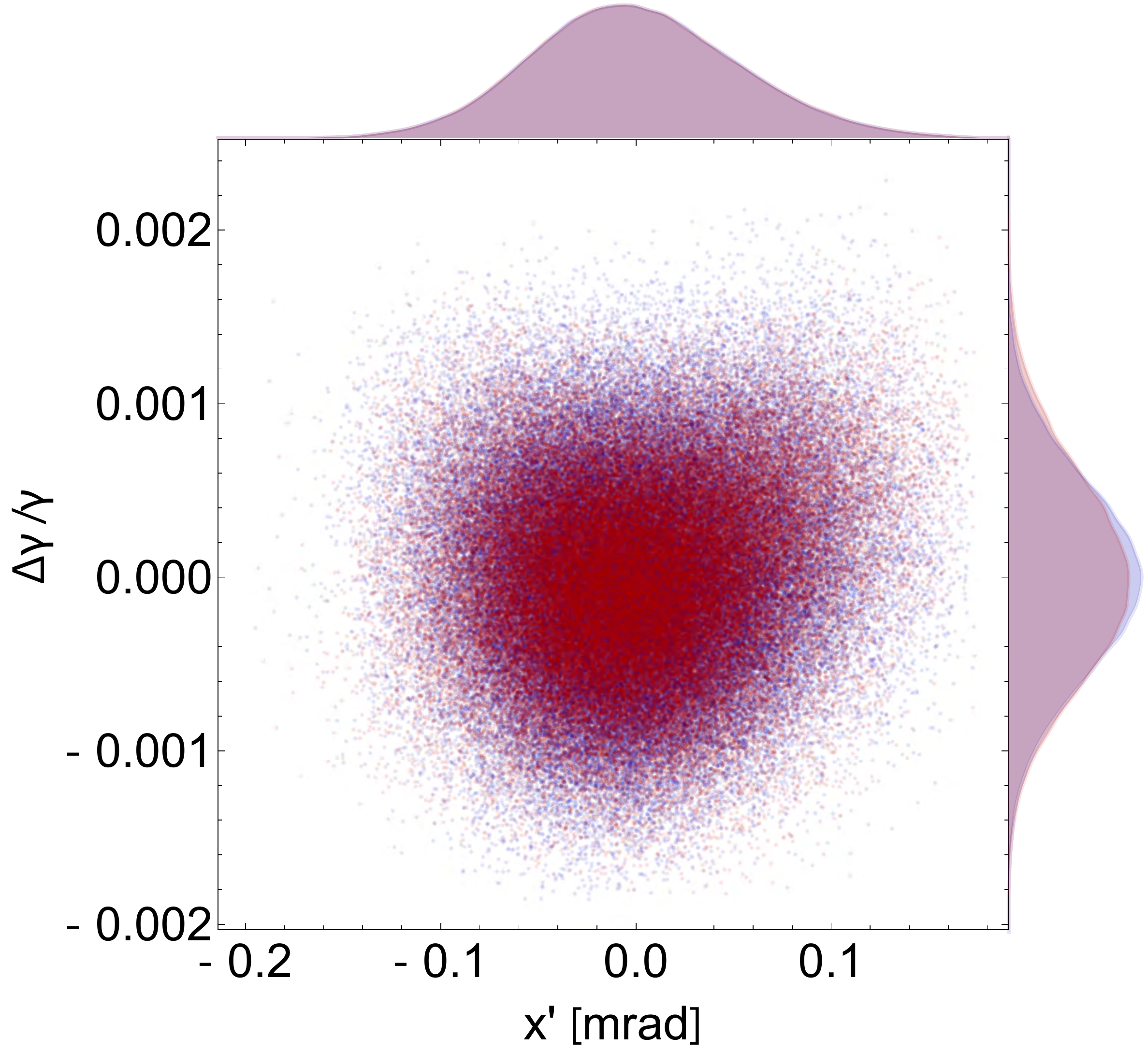} 
	\caption{Output phase spaces in CSR regime for the scheme with direct (blue) and mirrored (red) telescope.}
	\label{fig:Up-down-CSR-TRICKvsNOTRICK} 
\end{figure*}

The output beam has small transverse-longitudinal correlations in ($x,\;z$) and ($x',\;z$) phase spaces in the nonlinear regime (green) as depicted in Fig.~\ref{fig:LCLS2 phase spacesL-NL}~and justified by $\mu=1.0025$ in contrast to the linear regime (black). Our description of the propagation in the telescope and deflecting cavities has relied on the thin-lens approximations via first order matrices for simplicity. The nonlinear effects associated with TDCs can be suppressed using methods discussed in~\cite{TDC-paper,Aberration_Sext_MIT_2015}. The dynamics in linear and nonlinear regimes are practically identical between the schemes with direct and mirrored telescopes, hence only the former is presented in Fig.~\ref{fig:Emittance_evol} (a, b) and Fig.~\ref{fig:LCLS2 phase spacesL-NL}. In the nonlinear regime, the eigen emittances $\lambda_1$ and $\lambda_3$, corresponding respectively to the longitudinal phase space and ($y,\;y'$) transverse phase space, still remain invariant, while eigen emittance corresponding to ($x,\;x'$) transverse phase space $\lambda_2$ grows along the beamline as demonstrated in Fig.~\ref{fig:Emittance_evol}~(b). At the exit of the scheme, this results in a significant 9.2-fold increase of emittance $\epsilon_{n_x}$, while growth of $\epsilon_{n_z}$ is almost invisible (0.1\%). 

Accounting for CSR effects in bends and drifts results in a similar enlargement of eigen emittance $\lambda_2$ and various evolutions of eigen emittance $\lambda_1$ along the beamline for the distinct telescope configurations as compared in Fig.~\ref{fig:Emittance_evol} (c) and (d). The phase space dilution initiated by CSR effects in the first EEX is partially compensated by turning the beam around in the second EEX following a mirrored telescope and provides only a 5.3\% enlargement of the longitudinal emittance $\epsilon_{n_z}$ versus 14.6\% for the scheme with a standard telescope. The residual transverse-longitudinal correlations in the scheme with CSR-compensation are significantly smaller and quantified by $\mu=1.04$ versus $\mu=1.28$ for its counterpart. The 17.3-fold increase of transverse emittance $\epsilon_{n_x}$ at the exit of the scheme with mirrored telescope is slightly bigger than for the scheme with standard telescope demonstrating 17.1-fold increase. The final emittances and output phase spaces in the CSR regime are compared between the two schemes (red and blue) in Fig.~\ref{fig:Up-down-CSR-TRICKvsNOTRICK} and highly compromise the applicability of the DEEX bunch compressor in front of a LCLS-II type FEL, however the final compression ratio ($m=17$) is in absolute agreement with its designed value.

\begin{figure*}[ht]
	\includegraphics[width=1.0\textwidth]{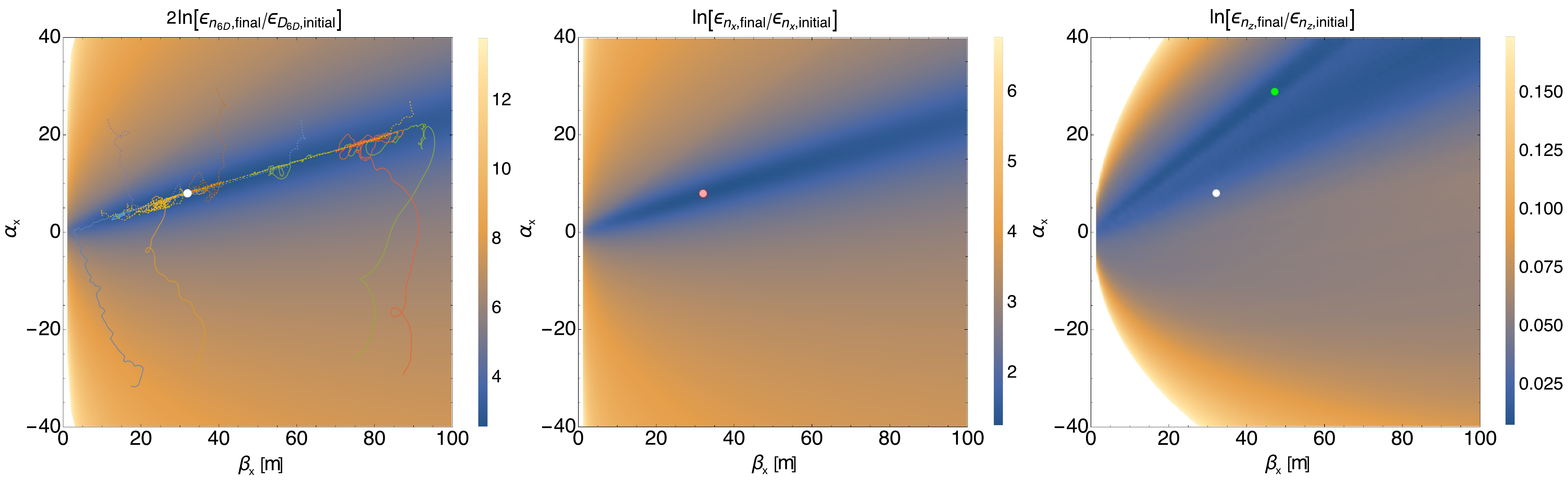}	
	\caption{ 2D density plots for the scheme with mirrored telescope from the input Twiss parameters $\beta_x$ and $\alpha_x$ of: (a) six-dimensional emittance $\epsilon_{n_{6D}}$, (b) transverse emittance $\epsilon_{n_x}$ and (c) longitudinal emittance $\epsilon_{n_z}$.} 
	\label{fig:landscape-LCLS2-d}
\end{figure*}
\begin{figure*}[ht]
	\includegraphics[width=0.470\textwidth]{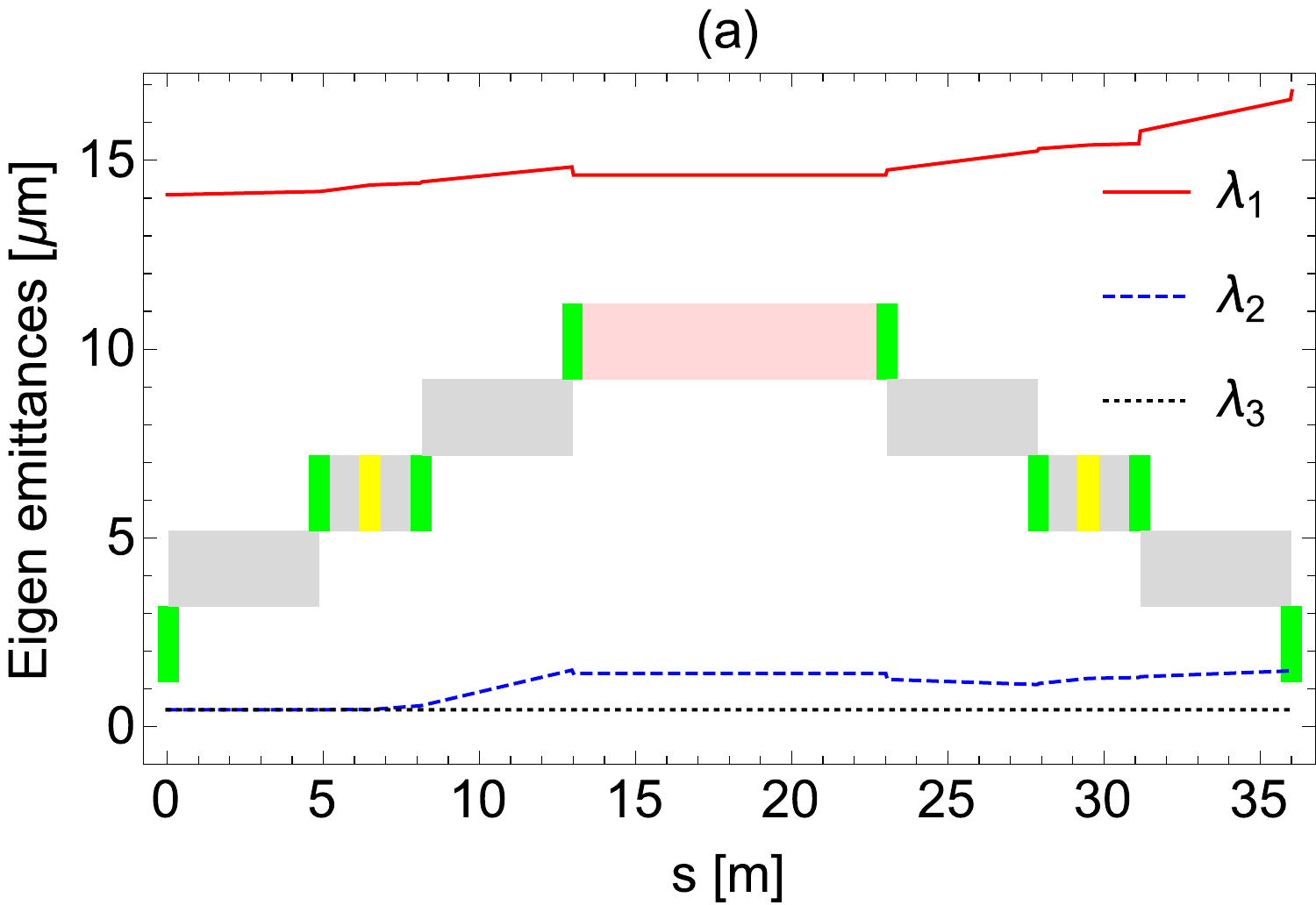}
	\includegraphics[width=0.470\textwidth]{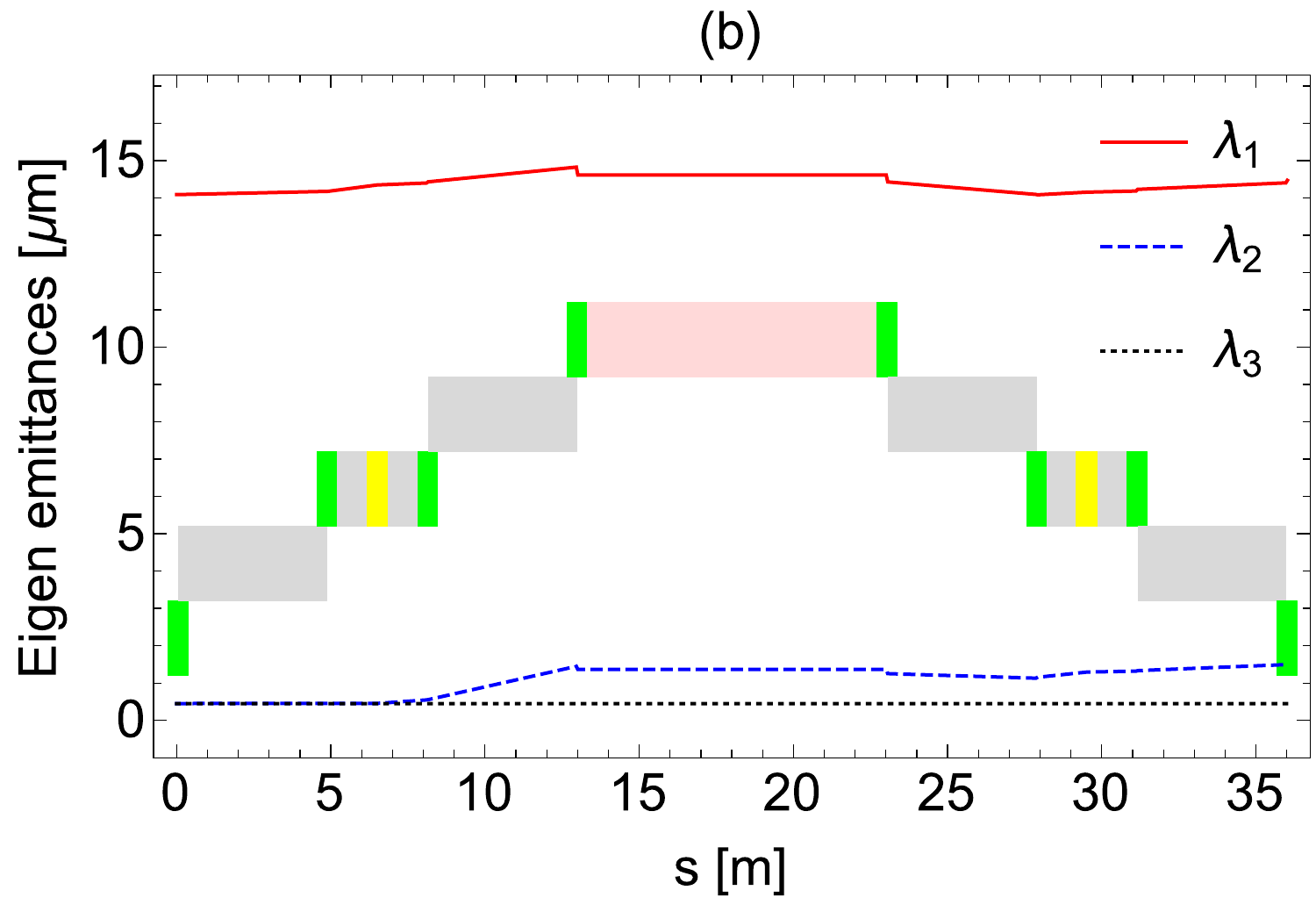}
	\caption{Evolution of normalized eigen emittances in  CSR regime for the scheme with direct (a) and mirrored (b) telescope.
	} 
	\label{fig:Emittance_evol_opt_Twiss}
\end{figure*}

\emph{\bf Optimization of input Twiss parameters}.
Nonlinear and CSR effects strongly depend on transverse and longitudinal beam parameters along the beamline. Since transverse and longitudinal dynamics are highly coupled in each EEX, both can be optimized by proper adjustment of the input Twiss parameters $\beta_x$ and $\alpha_x$, while the fold-increase of the normalized 6-dimensional emittance, characterized by the determinant of the $\Sigma-$matrix according to Eq.~(\ref{eq:mu}), is minimized. Figure~\ref{fig:landscape-LCLS2-d} demonstrates the dependence of the fold-increase of the 6-dimensional emittance (a), transverse (b) and longitudinal (c) emittances from the input Twiss parameters for the scheme with mirrored telescope.  A ``minimum-valley'' is clearly seen, characterized by the line $\alpha_x\approx0.2\cdot\beta_x$[m]. The white dot in Fig.~\ref{fig:landscape-LCLS2-d} (a-c) demonstrates the absolute minimum of the 6-dimensional emittance, while the ES searches depicted in Fig.~\ref{fig:landscape-LCLS2-d} (a) converge to the global minimum starting from the various initial conditions. The green dot in Fig.~\ref{fig:landscape-LCLS2-d} (c) shows the location of the absolute minimum of longitudinal emittance, while the absolute minimum of transverse emittance (red dot) in Fig.~\ref{fig:landscape-LCLS2-d} (b) is perfectly aligned with the absolute minimum of $\epsilon_{n_{6D}}$. 

\begin{figure*}[p]
	\centering
	\includegraphics[width=0.3\textwidth]{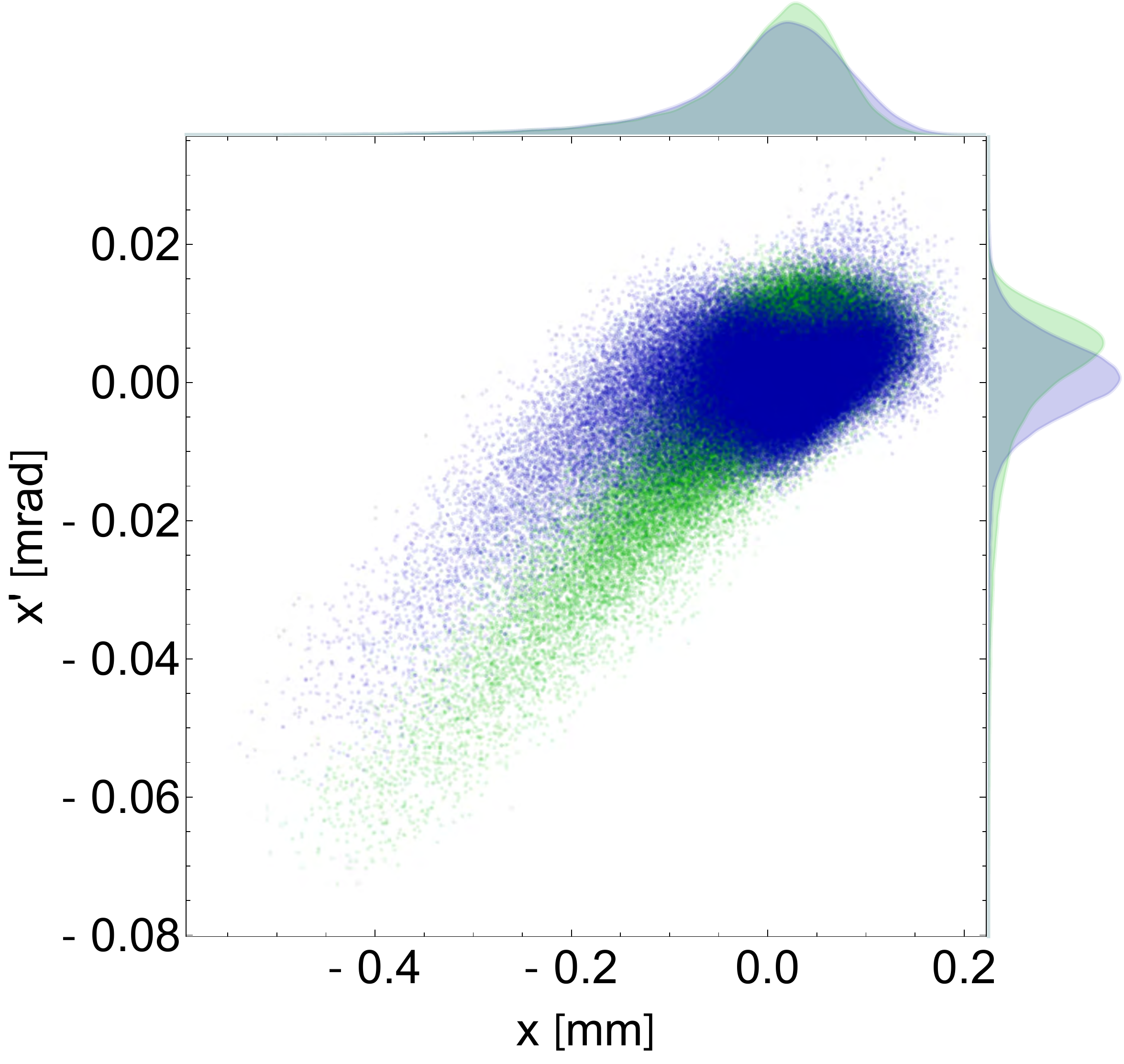} \
	\includegraphics[width=0.32\textwidth]{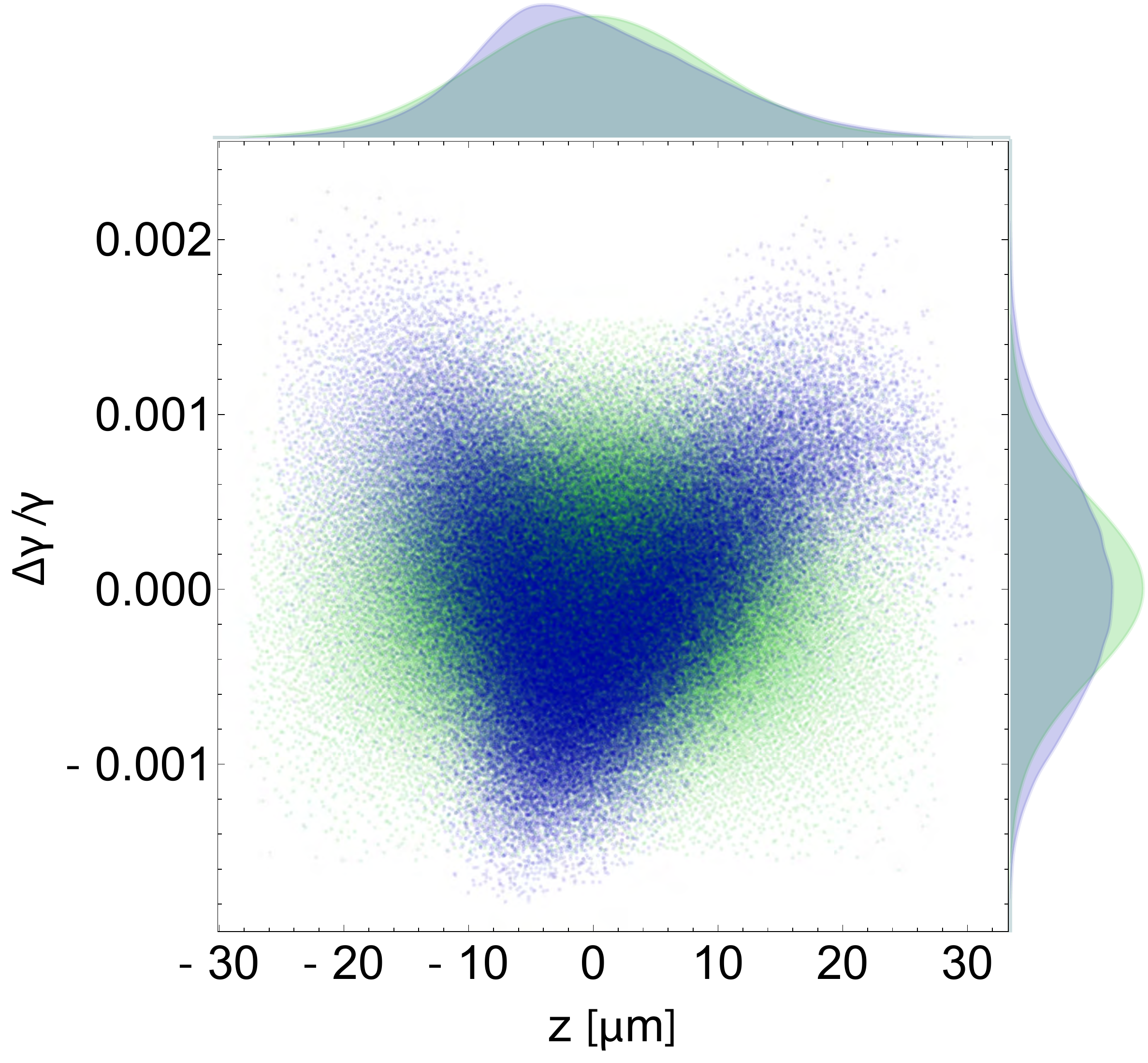} \
	\includegraphics[width=0.31\textwidth]{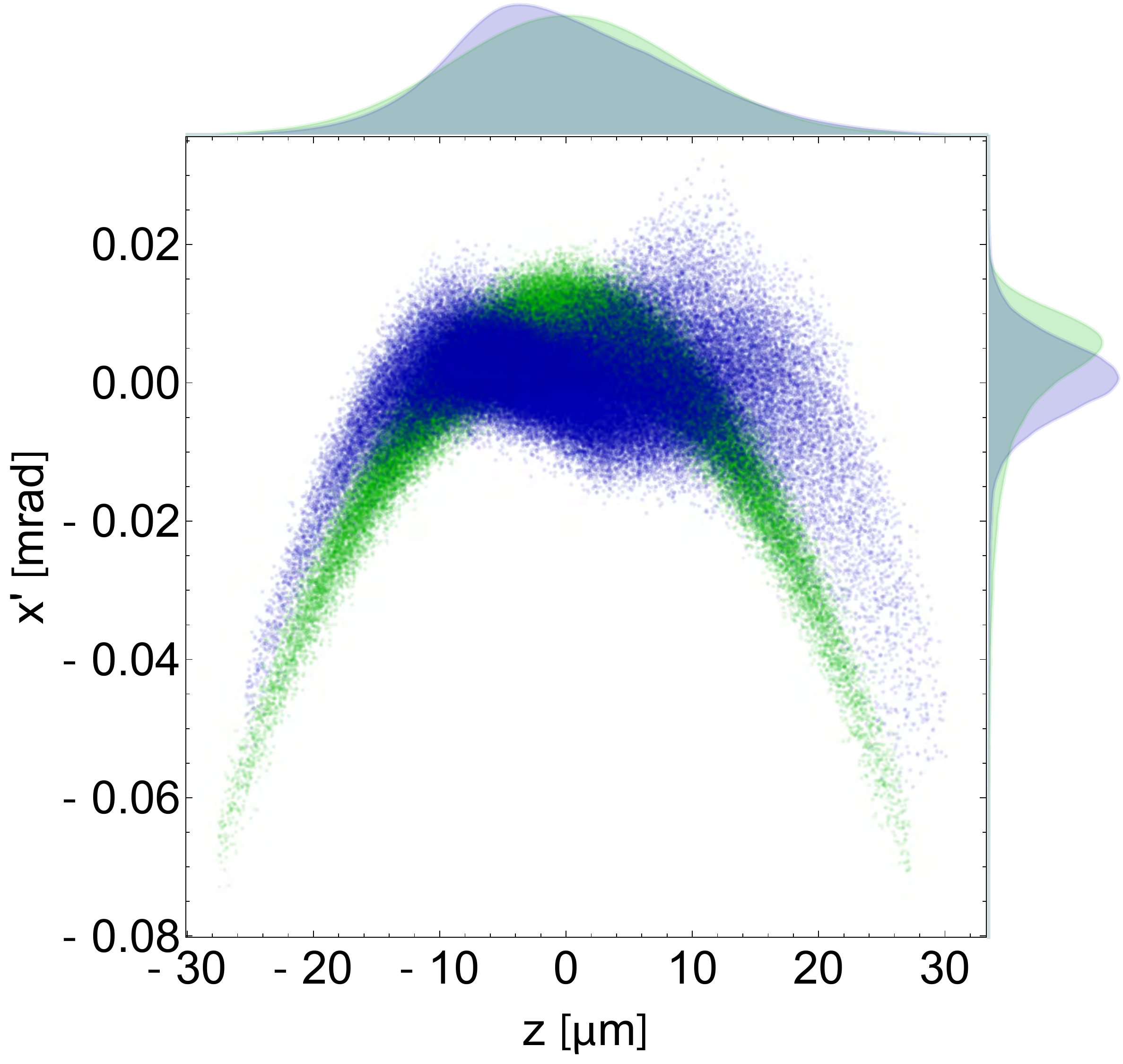} 
	\includegraphics[width=0.32\textwidth]{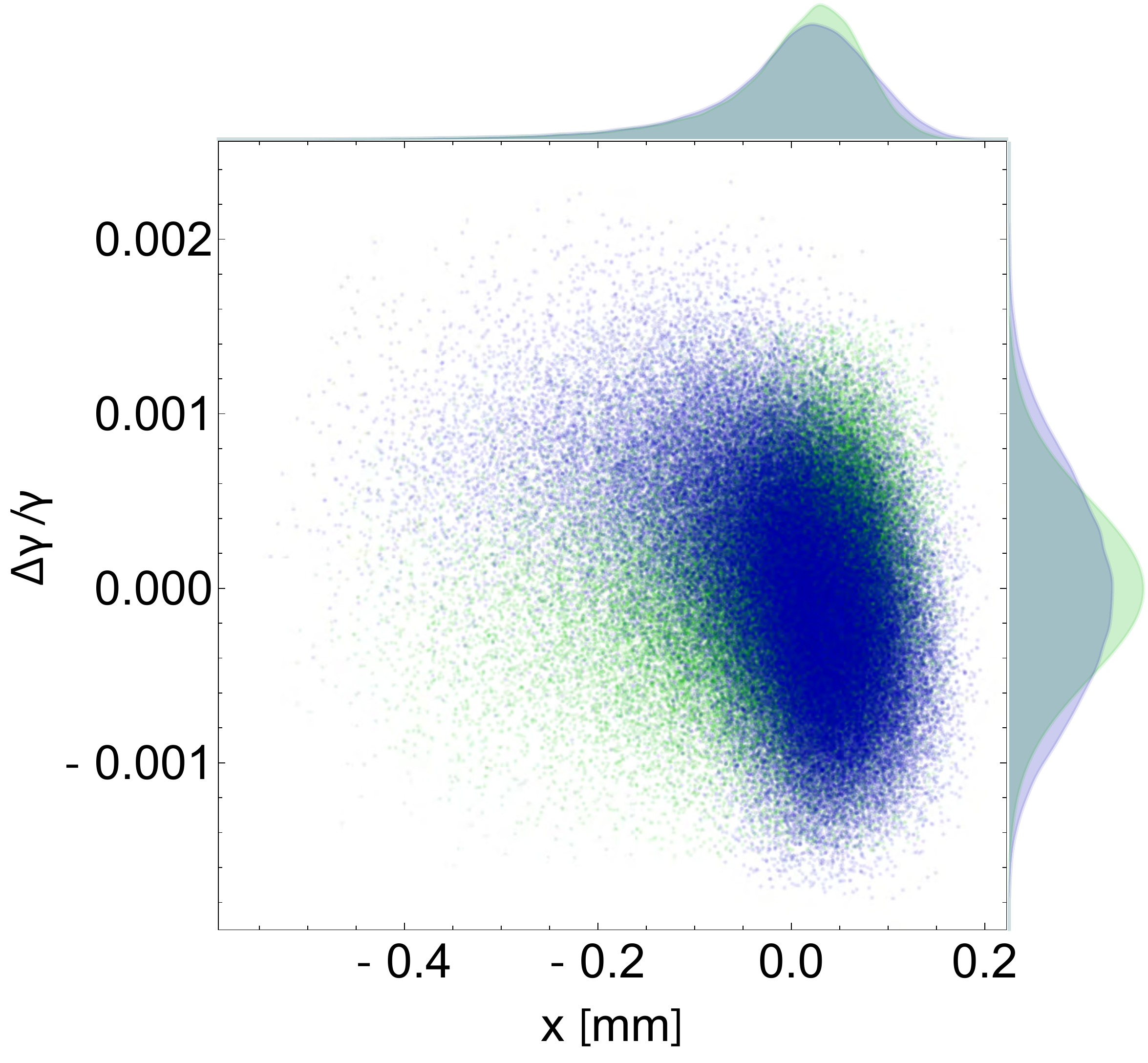} \
	\includegraphics[width=0.3\textwidth]{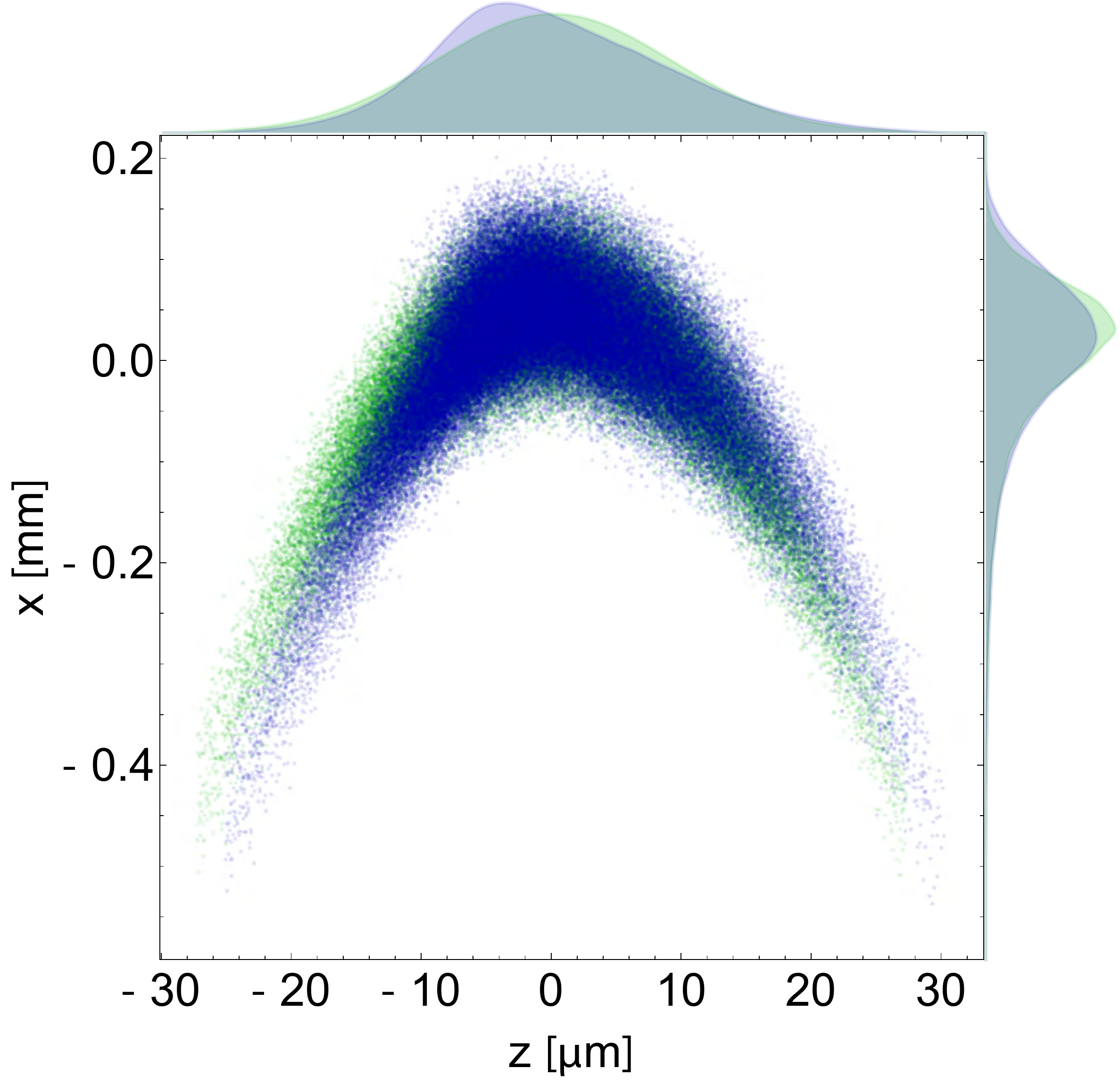} \
	\includegraphics[width=0.32\textwidth]{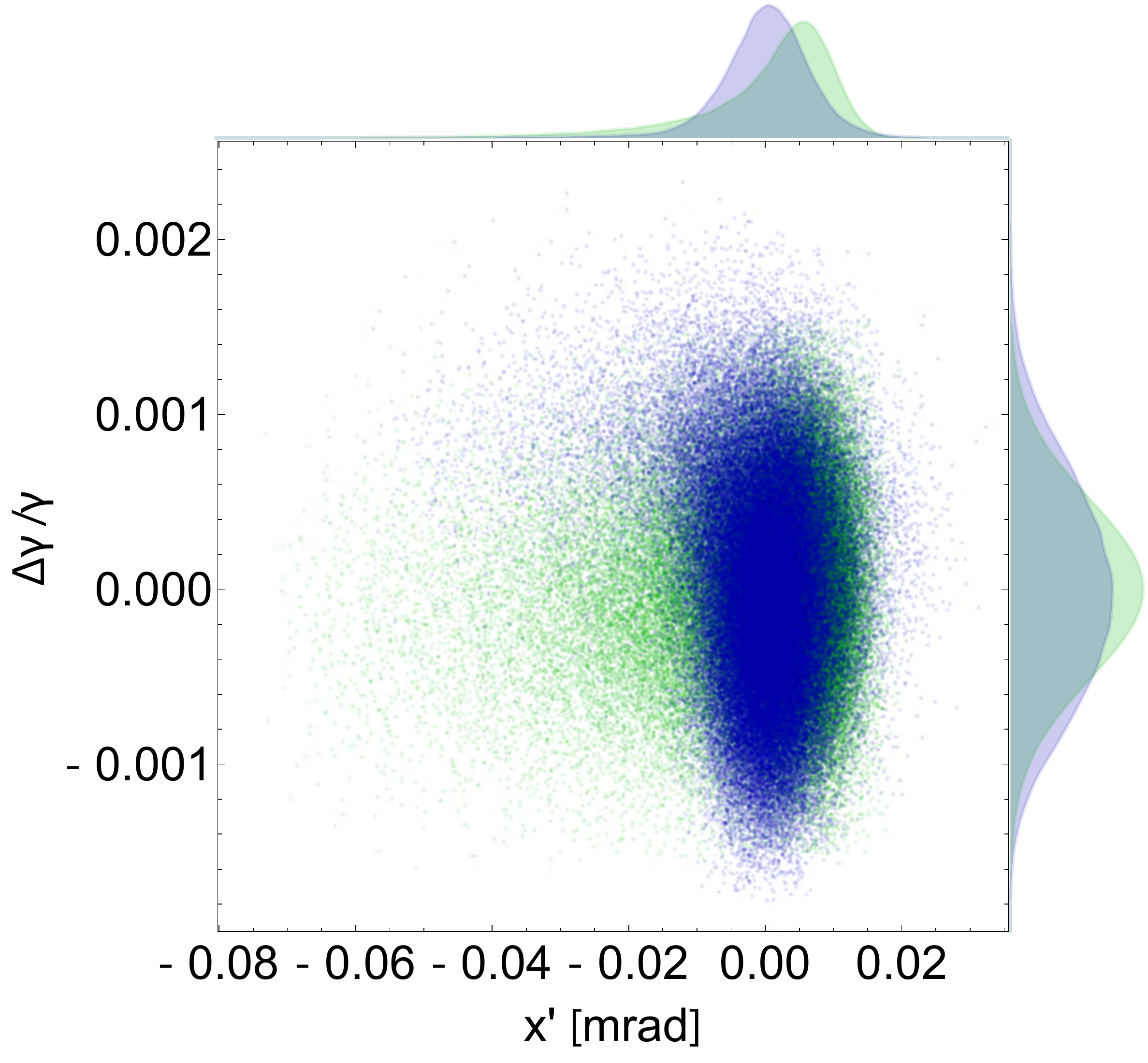} 
	\caption{Output phase spaces simulated in nonlinear (green) and CSR (blue) regime for the symmetric scheme with standard telescope for the beam with Twiss parameters optimized in CSR regime.} 
	\label{fig:OptTwiss-notrick_NLvsCSR}
	\includegraphics[width=0.3\textwidth]{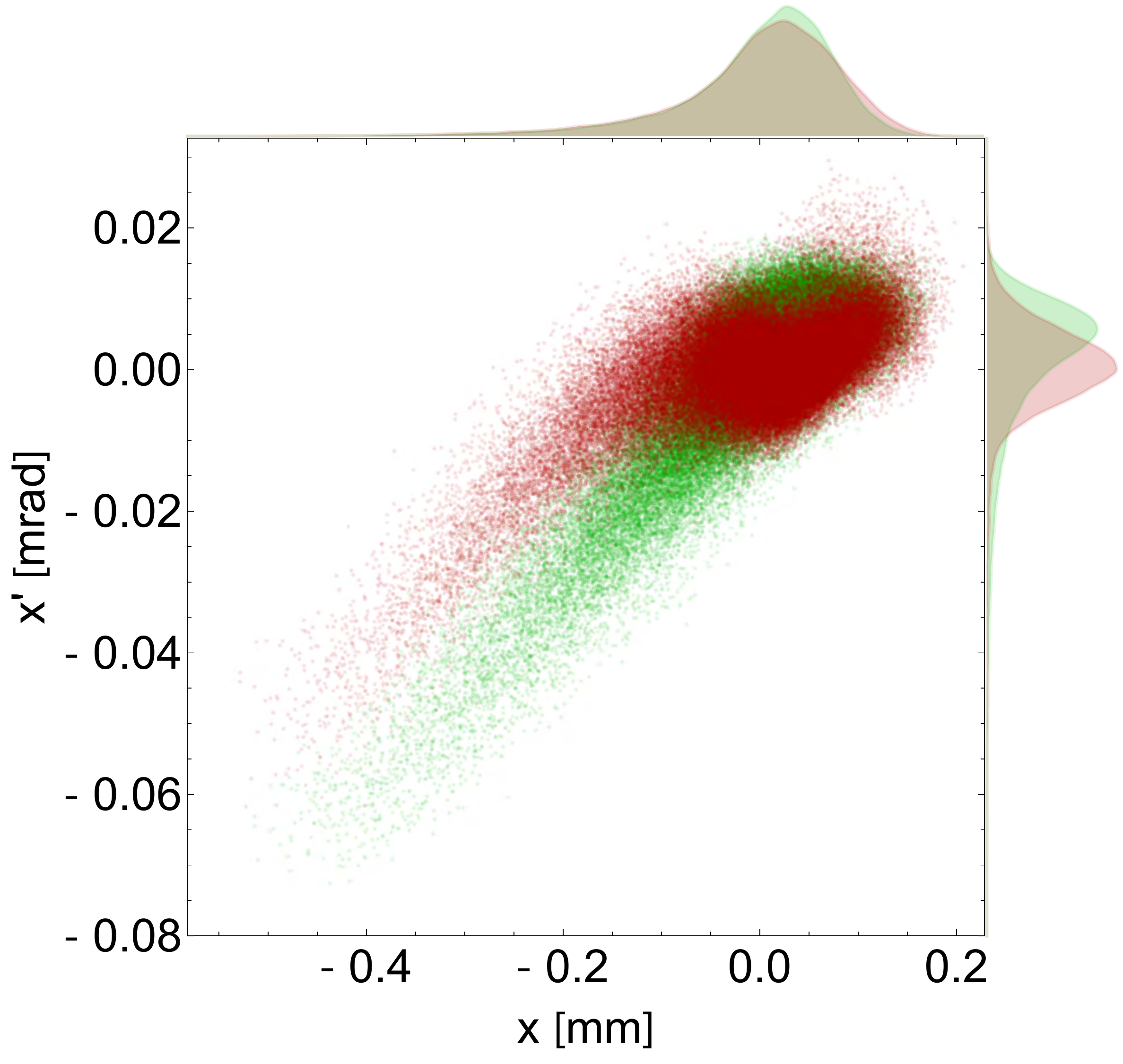} \
	\includegraphics[width=0.32\textwidth]{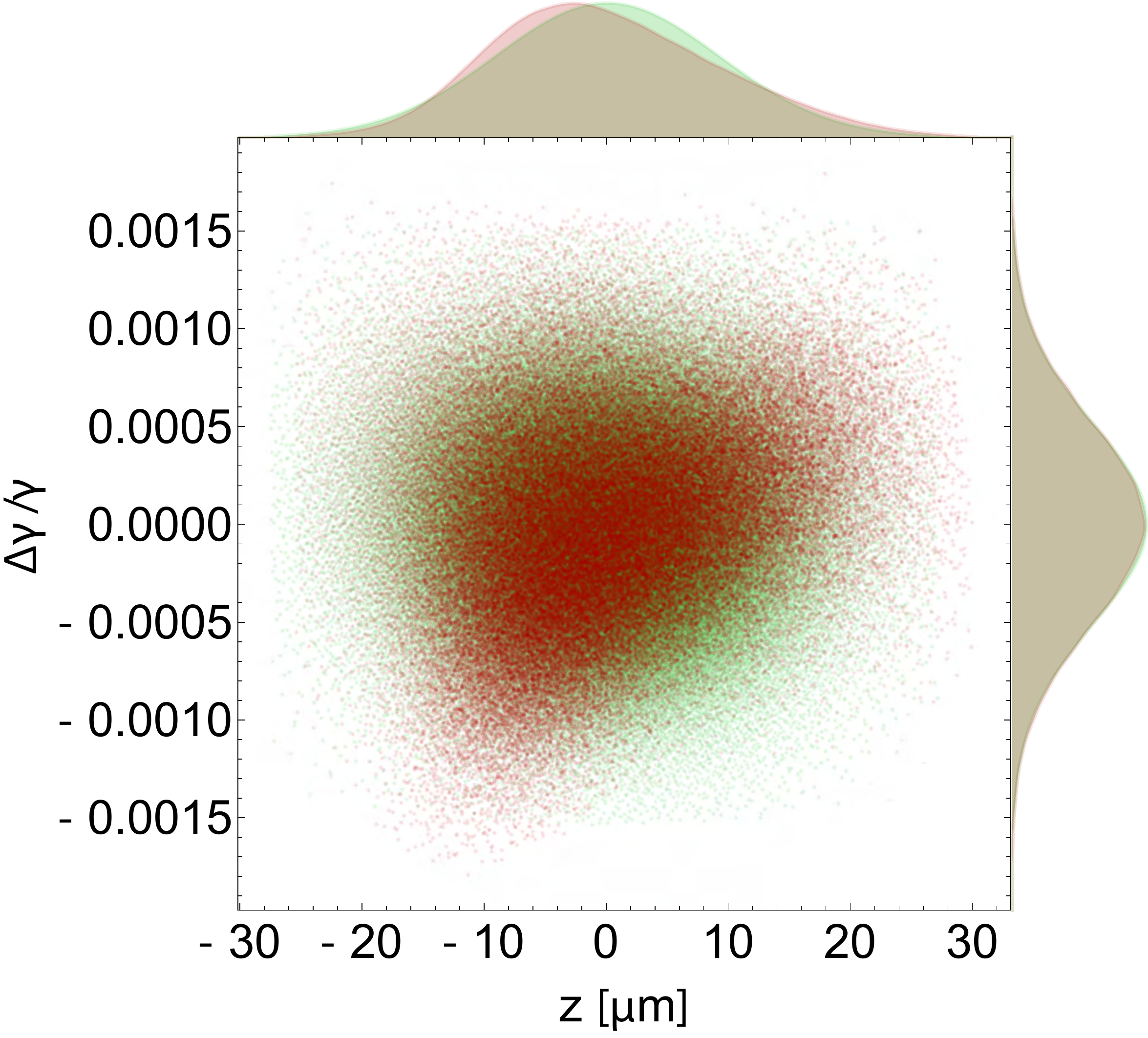} \
	\includegraphics[width=0.31\textwidth]{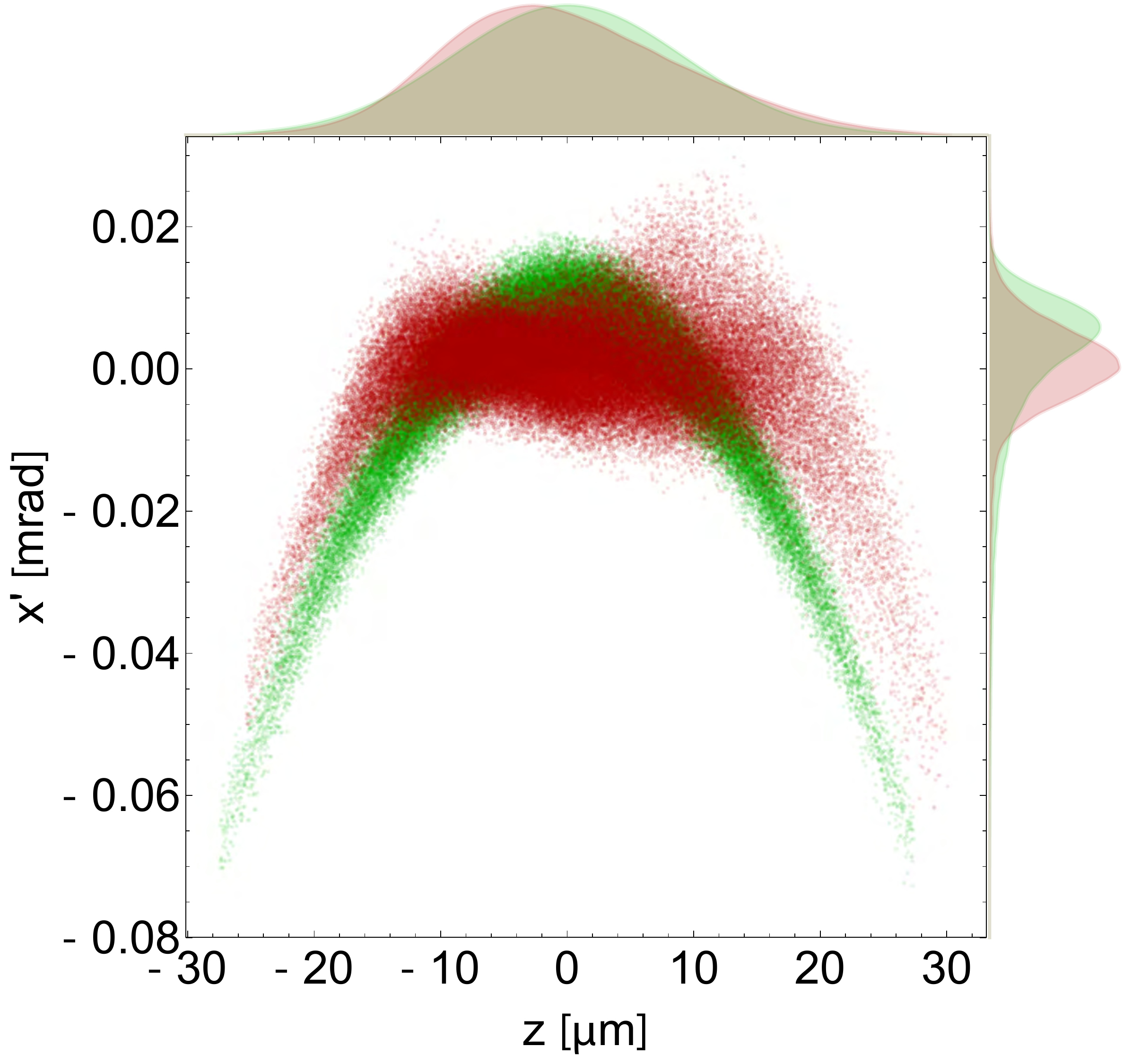} 
	\includegraphics[width=0.32\textwidth]{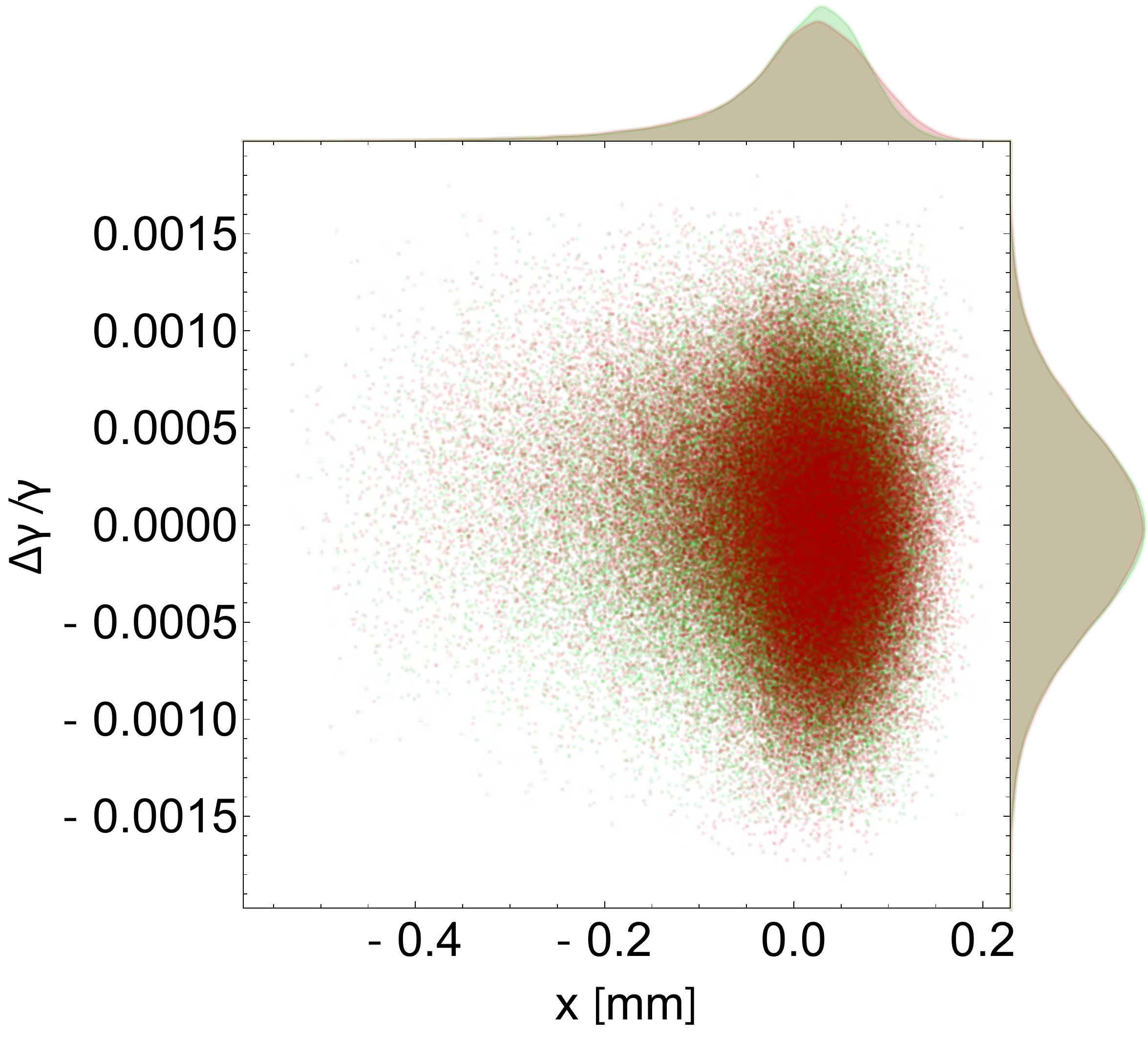} \
	\includegraphics[width=0.3\textwidth]{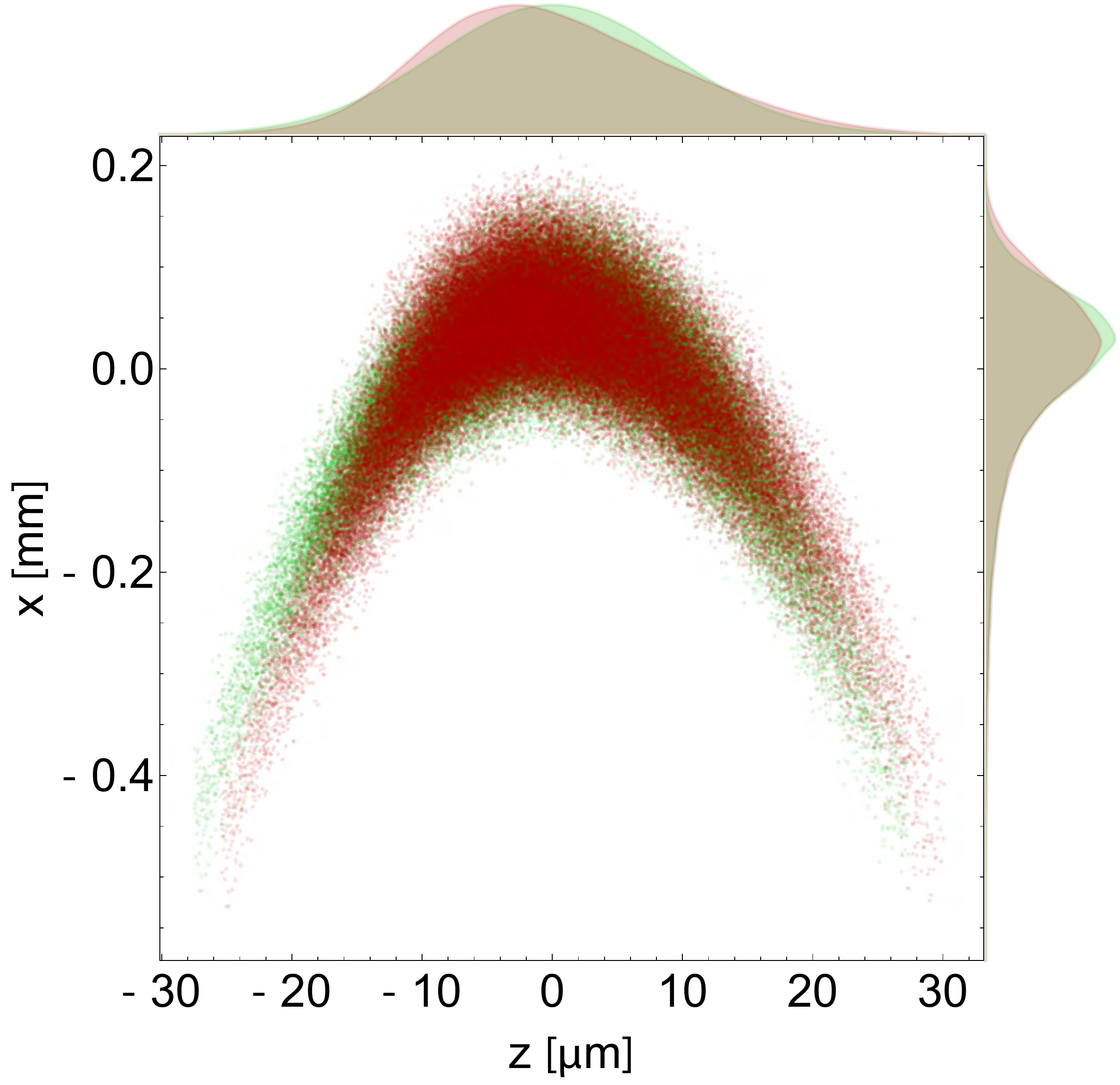} \
	\includegraphics[width=0.32\textwidth]{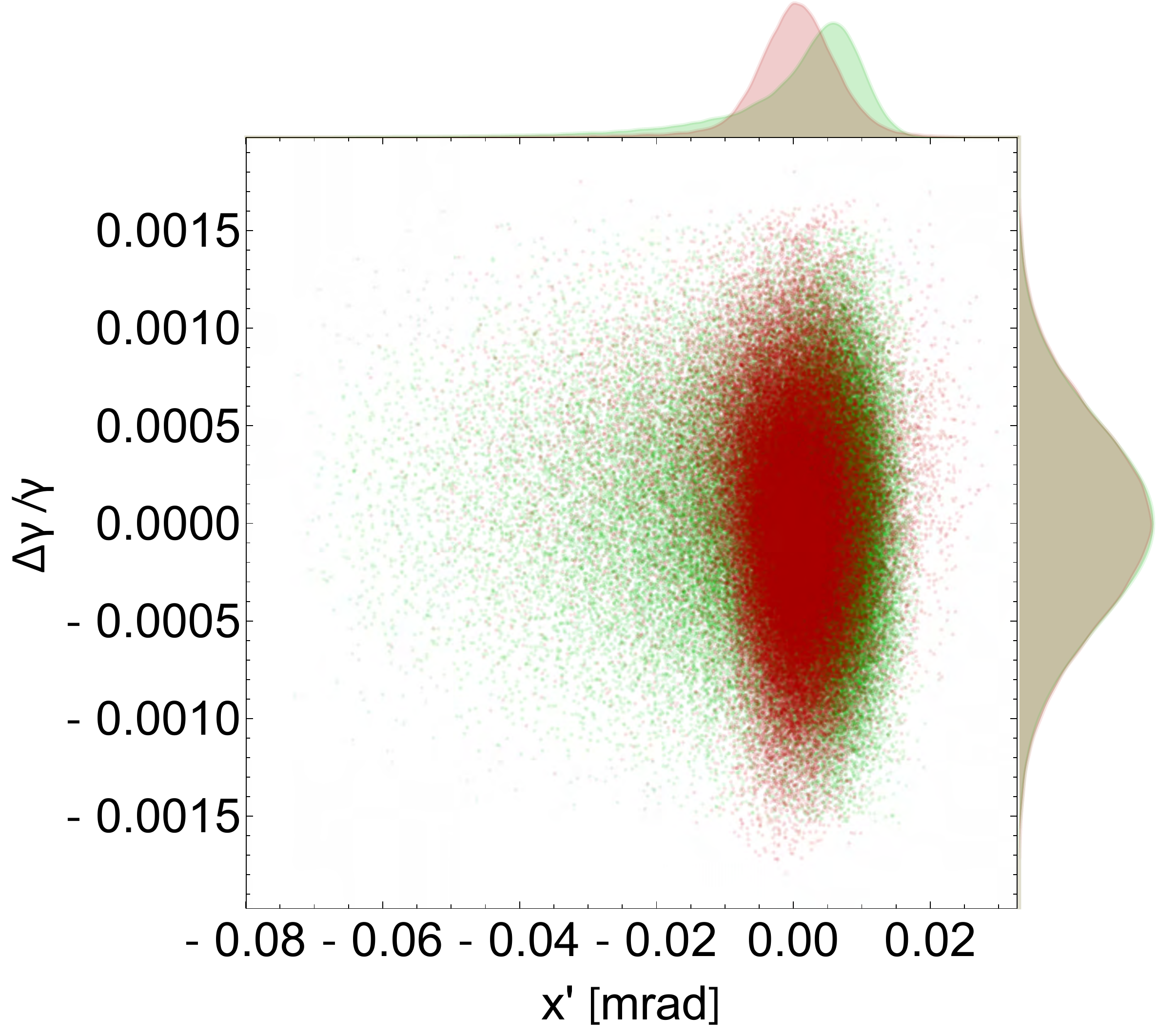} 
	\caption{Output phase spaces simulated in nonlinear (green) and CSR (red) regime for the symmetric scheme with mirrored telescope for the beam with Twiss parameters optimized in CSR regime.} 
	\label{fig:OptTwiss-trick_NLvsCSR}
\end{figure*}

Nonlinear and CSR effects partially compensate each other resulting in 3.7- and 3.4-fold increase of transverse emittance $\epsilon_{n_x}$ in the CSR regime for the scheme with direct and mirrored telescopes,  respectively. Compensation of CSR effects in the scheme with a mirrored telescope is depicted in Fig.~\ref{fig:Emittance_evol_opt_Twiss} (b), representing the evolution of eigen emittance along the beamline, and results in much smaller degradation of the longitudinal phase space quality which is quantified by emittance enlargement of 2.7\% versus 19.8\% for the scheme with direct telescope (Fig.~\ref{fig:Emittance_evol_opt_Twiss} (a)). Conversely, 3.4-fold increase of the 6-dimensional emittance and $\mu=1.03$ for the scheme with CSR-compensation are smaller than 3.85-fold increase of $\epsilon_{n_{6D}}$ and $\mu=1.3$ for its counterpart. 
Output phase spaces for the dynamics in nonlinear and CSR regimes for the scheme with standard telescope are compared in Fig.~\ref{fig:OptTwiss-notrick_NLvsCSR} and for the scheme with mirrored telescope in Fig.~\ref{fig:OptTwiss-trick_NLvsCSR}. 

Summarizing the above, the scheme with a mirrored telescope between identical (apart from orientation) EEX modules demonstrates significantly better quality of the longitudinal phase space due to compensation of CSR effects along the beamline, while the dilution of the transverse emittance $\epsilon_{n_x}$ is predominated by nonlinear effects showing slightly better results for the mirrored telescope configuration.

%%%%%%%%%%%%%%%%%%%%%%%%%%%%%%%%%%%%%%%%%%%%%%%%%%%%%%%%%%%%%%%%%%%%%%%%%%%%%%%%%%%%%%%%
%%%%%%%%%%%%%%%%%%%%%%%%%%%%%%%%%%%%%%%%%%%%%%%%%%%%%%%%%%%%%%%%%%%%%%%%%%%%%%%%%%%%%%%%
\section{Asymmetric DEEX}\label{sec:ES-Optimization}
%%%%%%%%%%%%%%%%%%%%%%%%%%%%%%%%%%%%%%%%%%%%%%%%%%%%%%%%%%%%%%%%%%%%%%%%%%%%%%%%%%%%%%%%
%%%%%%%%%%%%%%%%%%%%%%%%%%%%%%%%%%%%%%%%%%%%%%%%%%%%%%%%%%%%%%%%%%%%%%%%%%%%%%%%%%%%%%%%

\begin{figure*}[p]
	\includegraphics[width=0.3\textwidth]{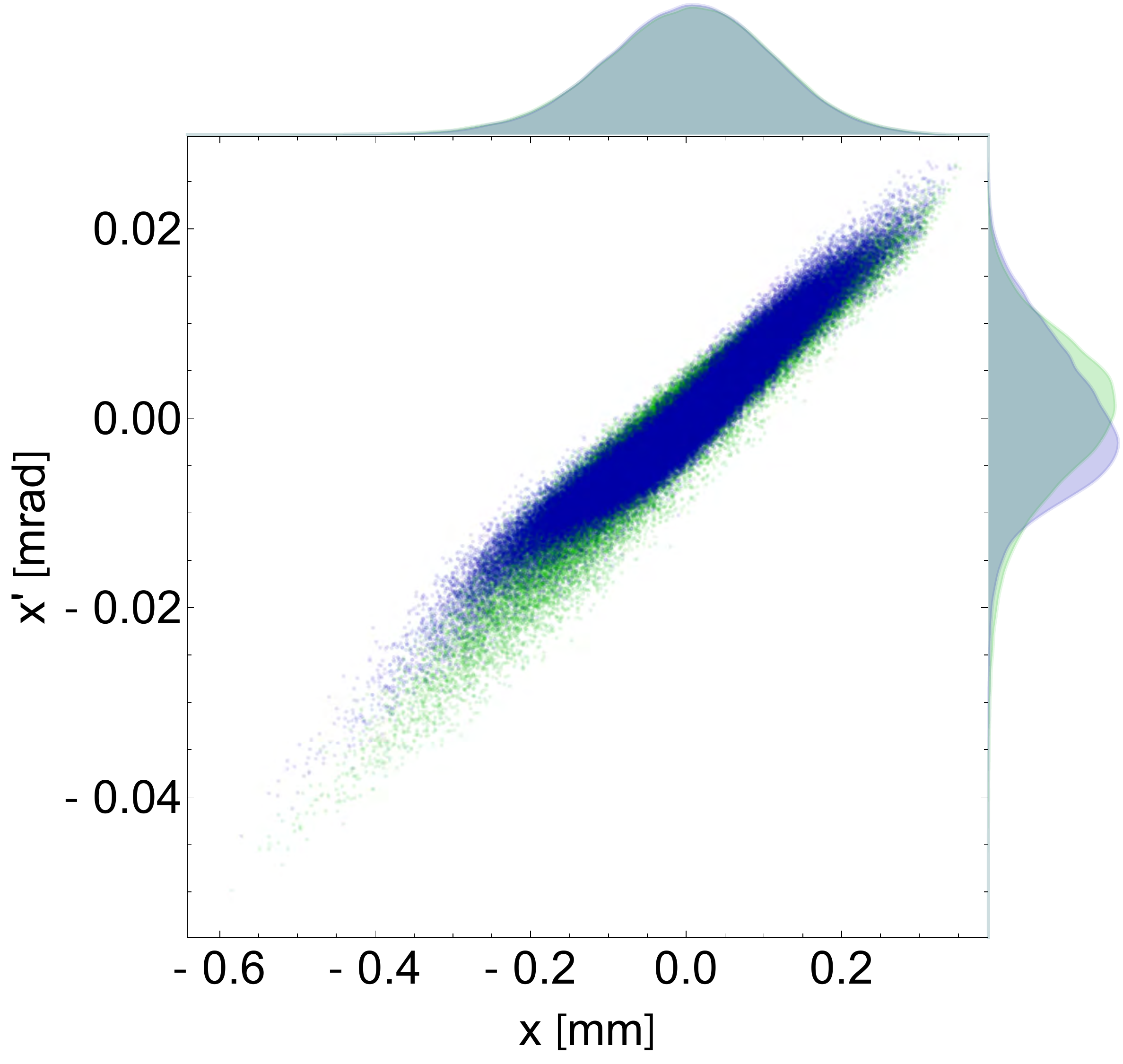} \
	\includegraphics[width=0.32\textwidth]{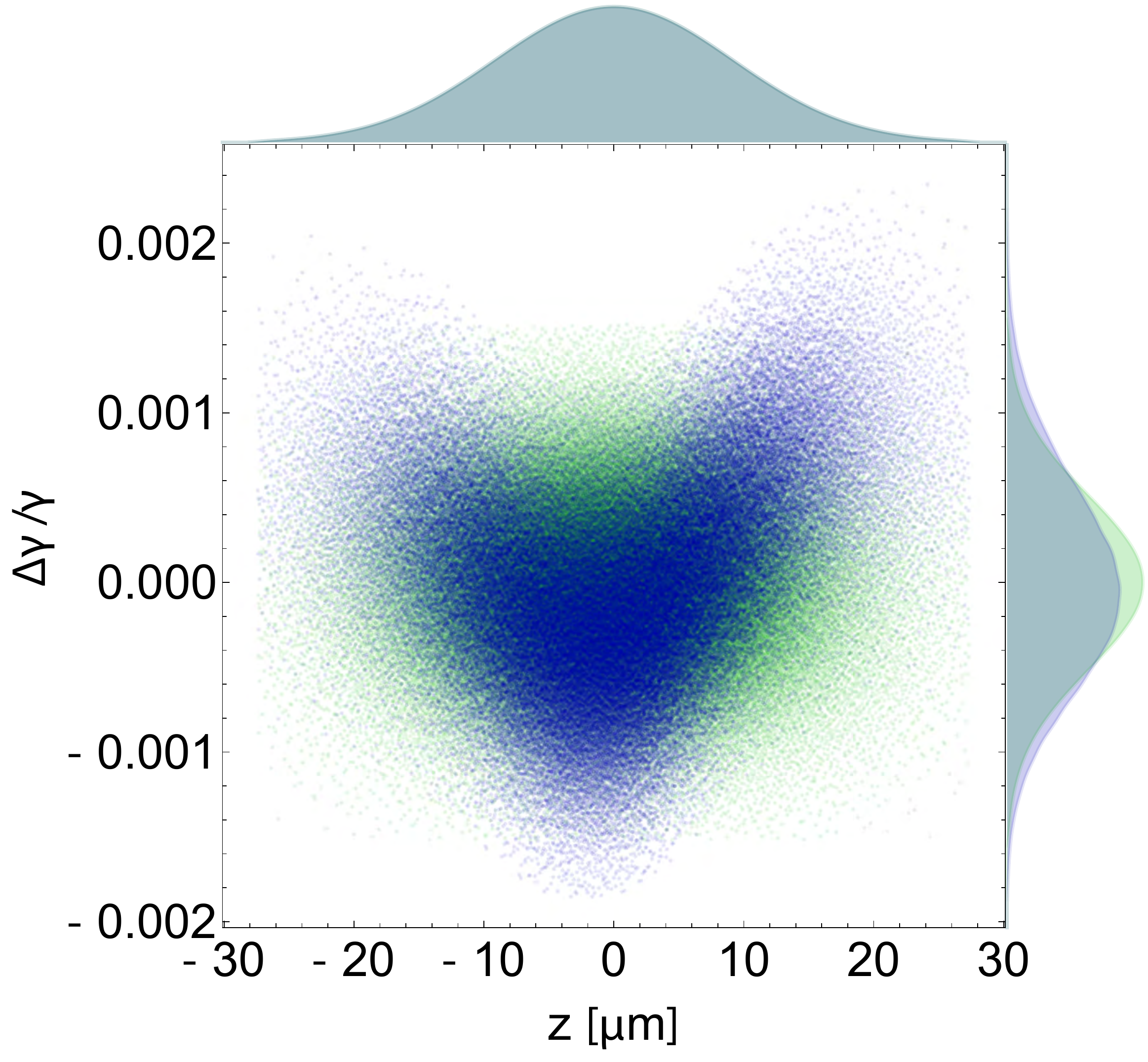} \
	\includegraphics[width=0.31\textwidth]{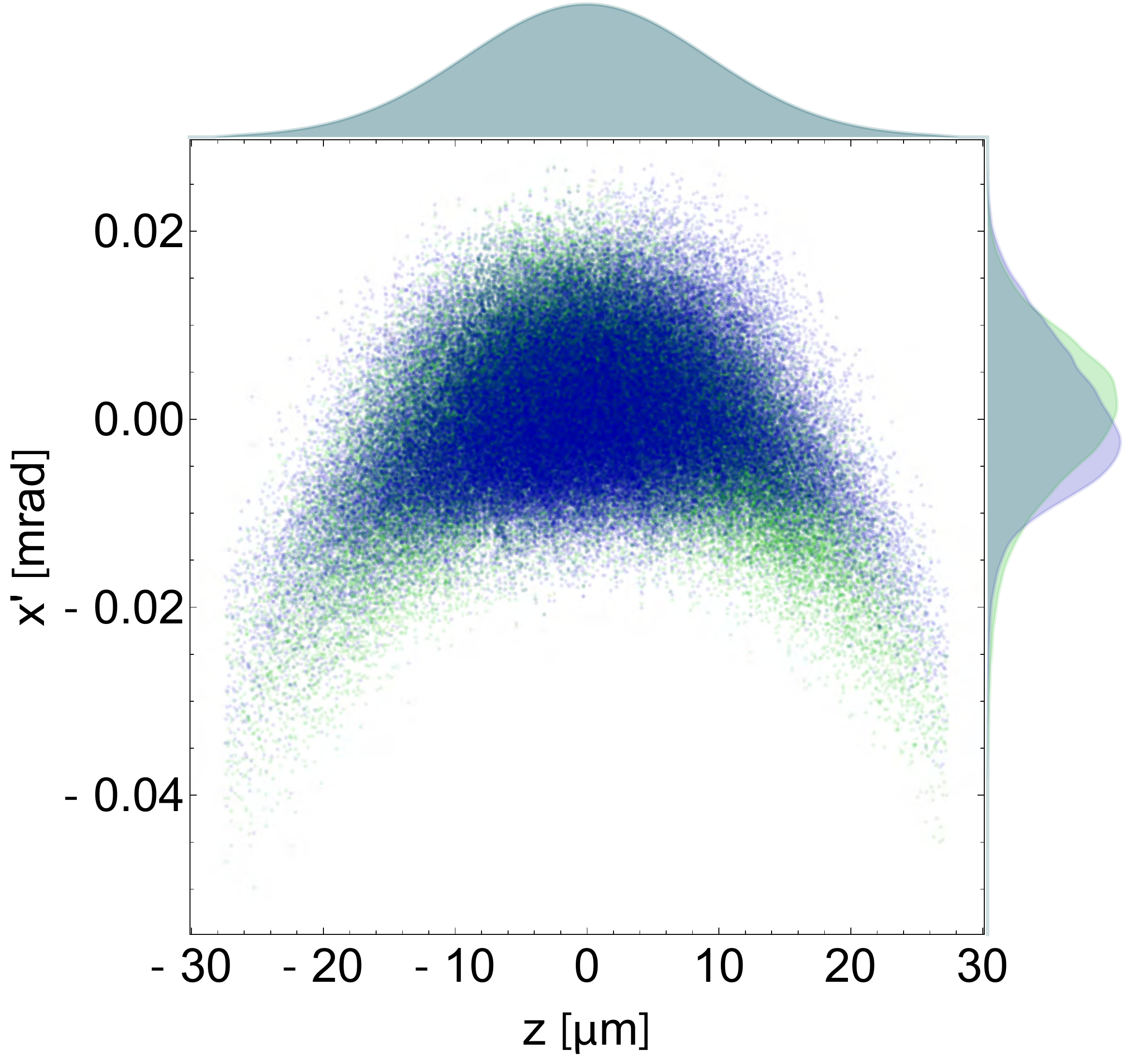} 
	\includegraphics[width=0.32\textwidth]{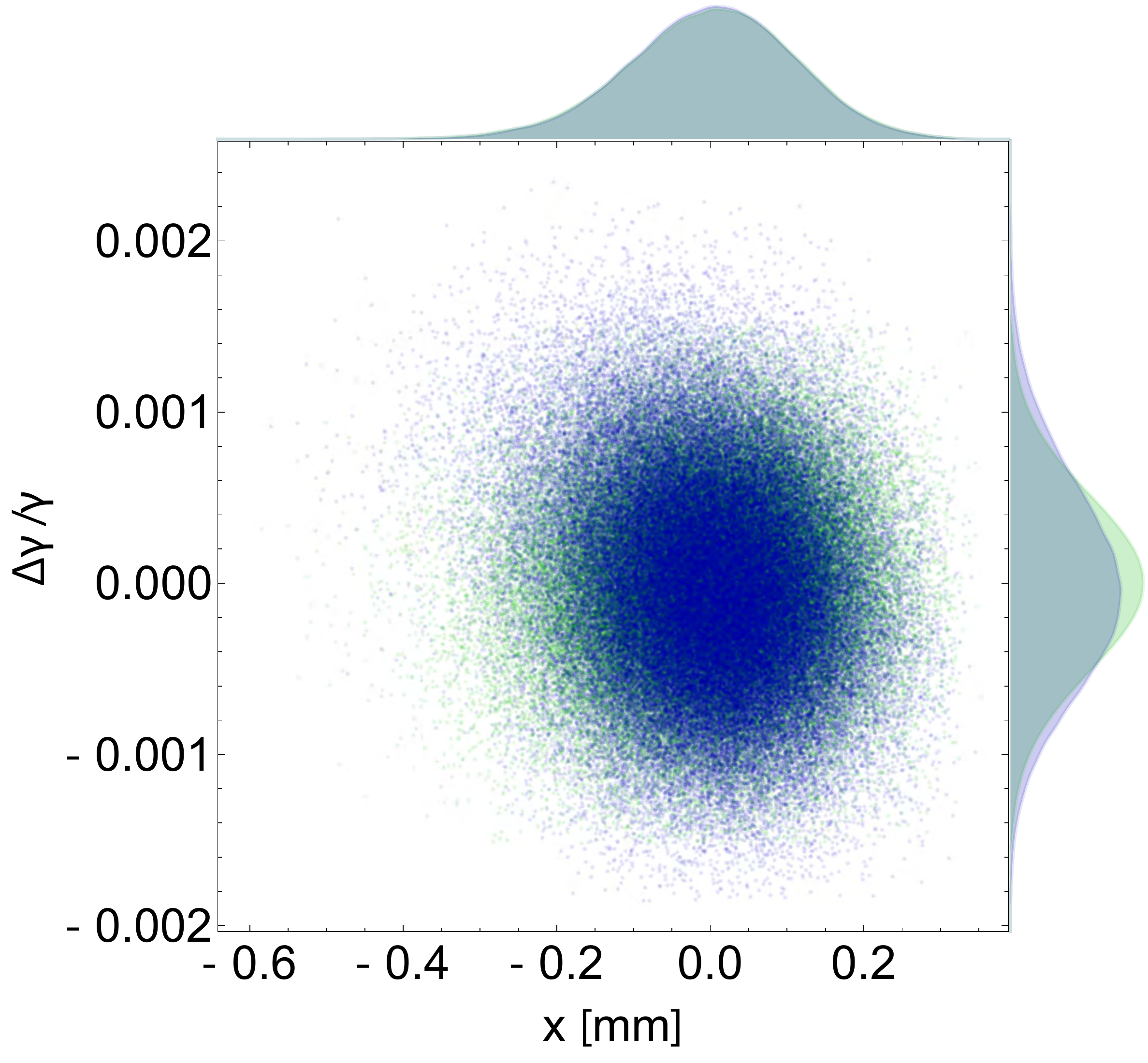} \
	\includegraphics[width=0.3\textwidth]{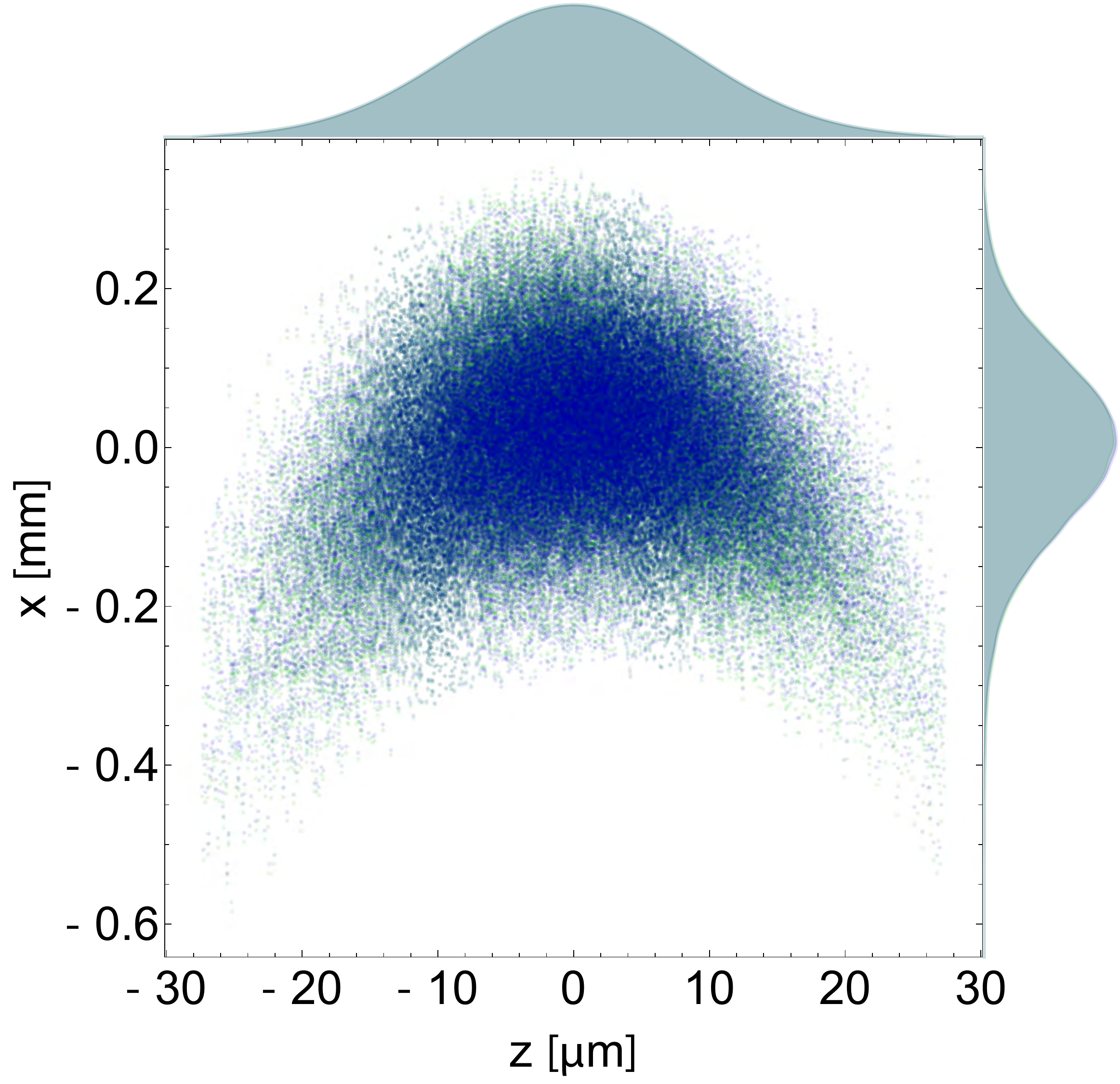} \
	\includegraphics[width=0.32\textwidth]{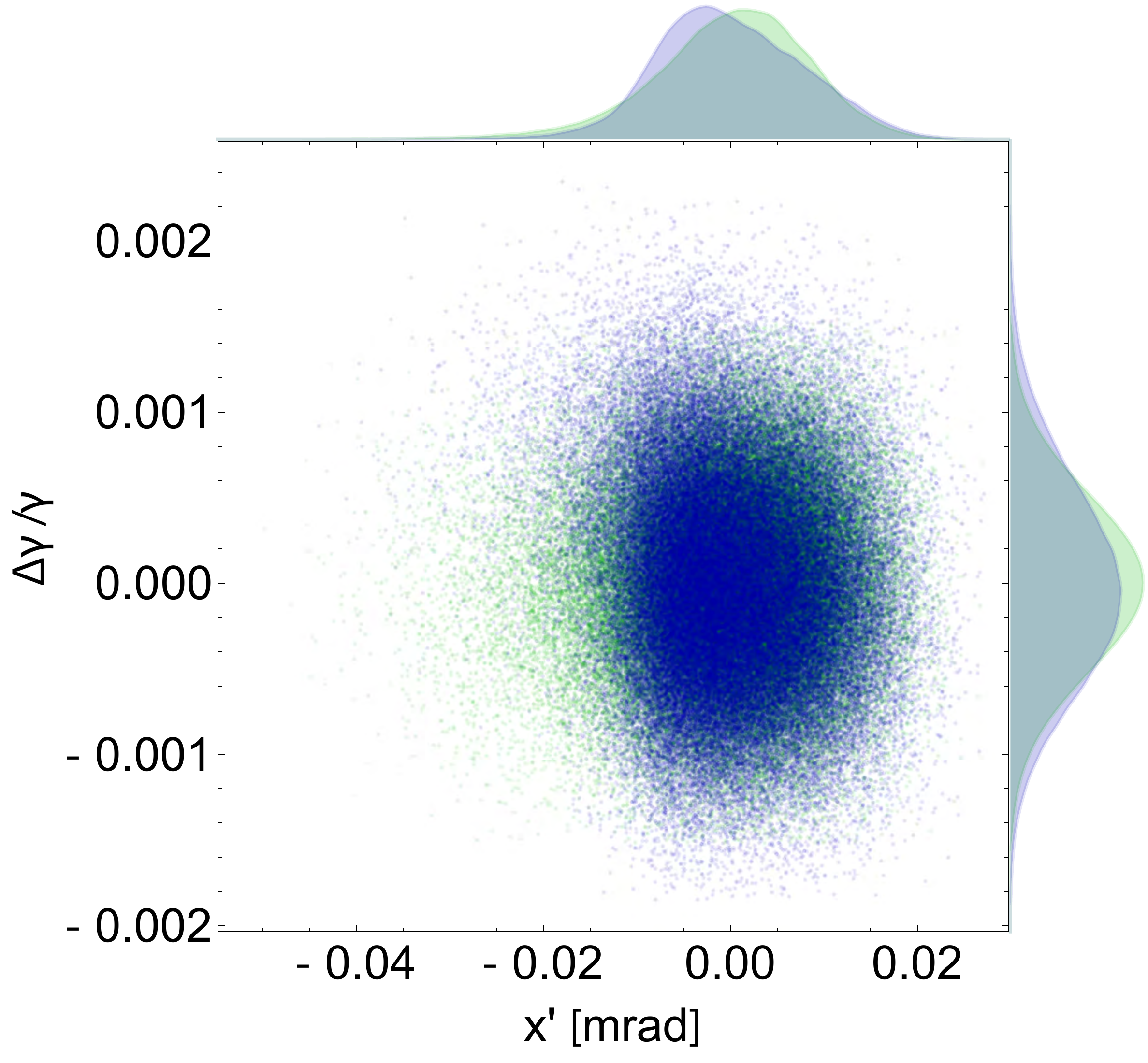} 
	\caption{Phase spaces at the exit of asymmetrical DEEX with direct telescope optimized in CSR regime and simulated in nonlinear (green) and CSR (blue) regime.} 
	\label{fig:OptScheme-notrick_NLvsCSR}
	\includegraphics[width=0.3\textwidth]{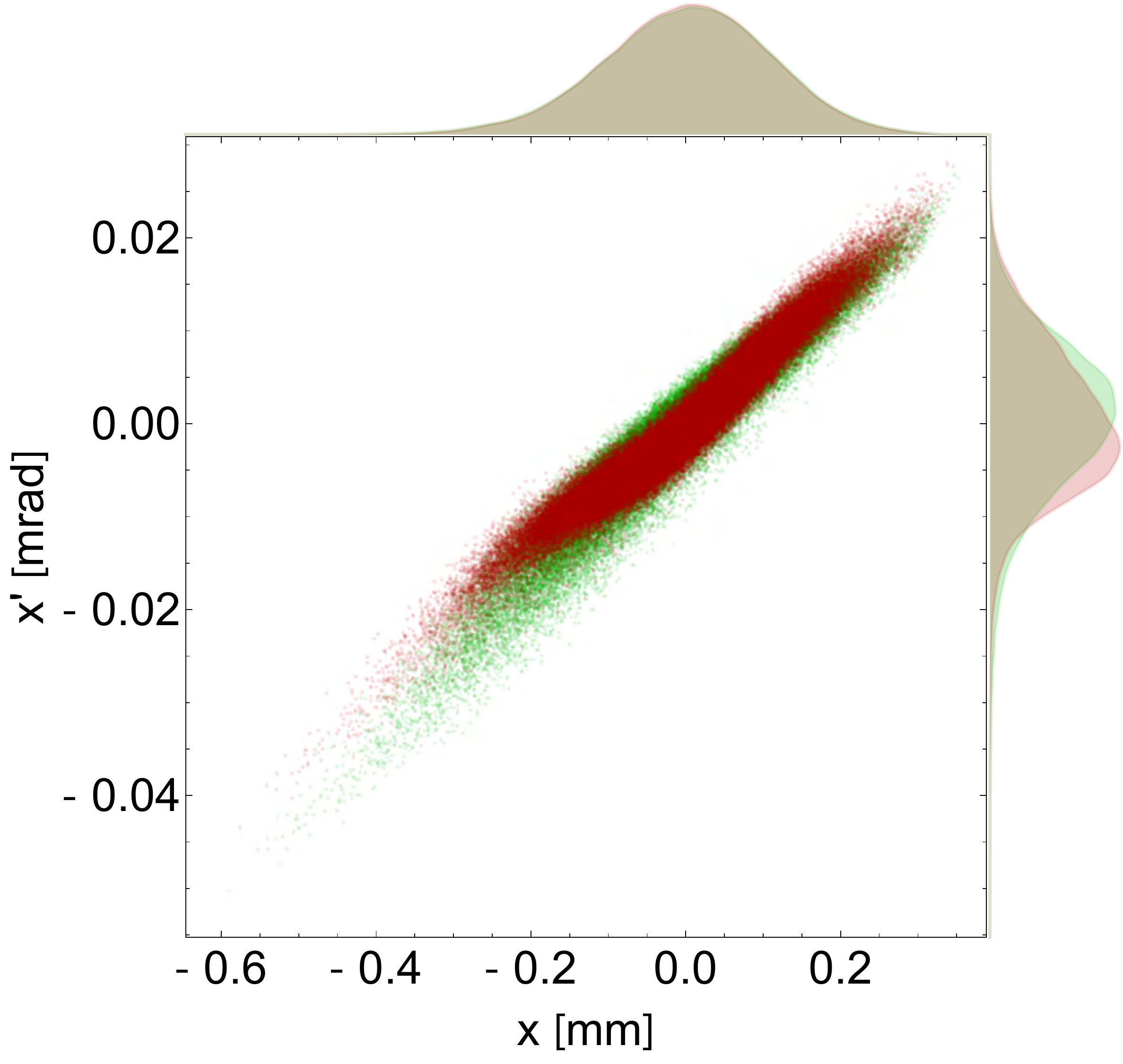} \
	\includegraphics[width=0.32\textwidth]{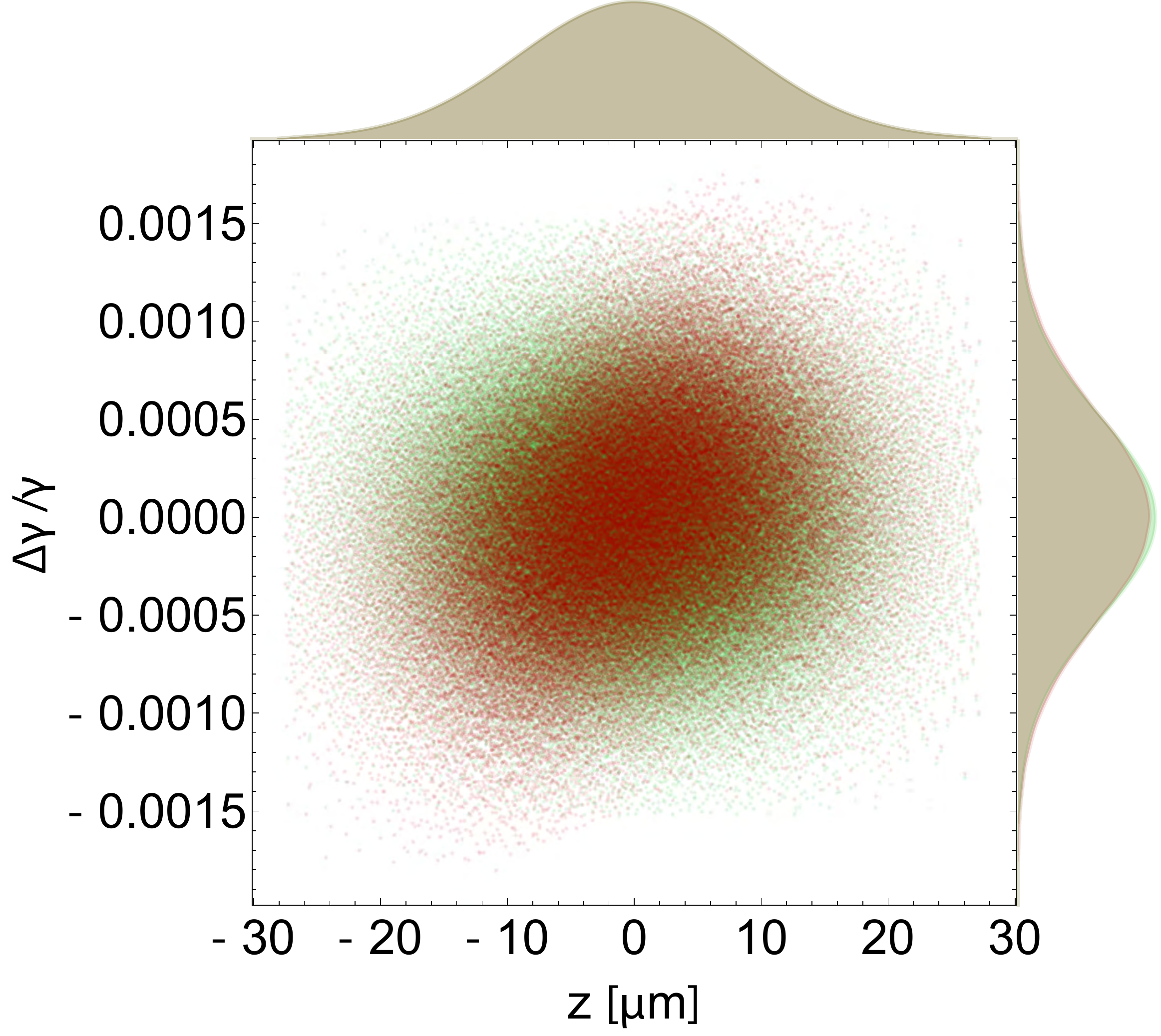} \
	\includegraphics[width=0.31\textwidth]{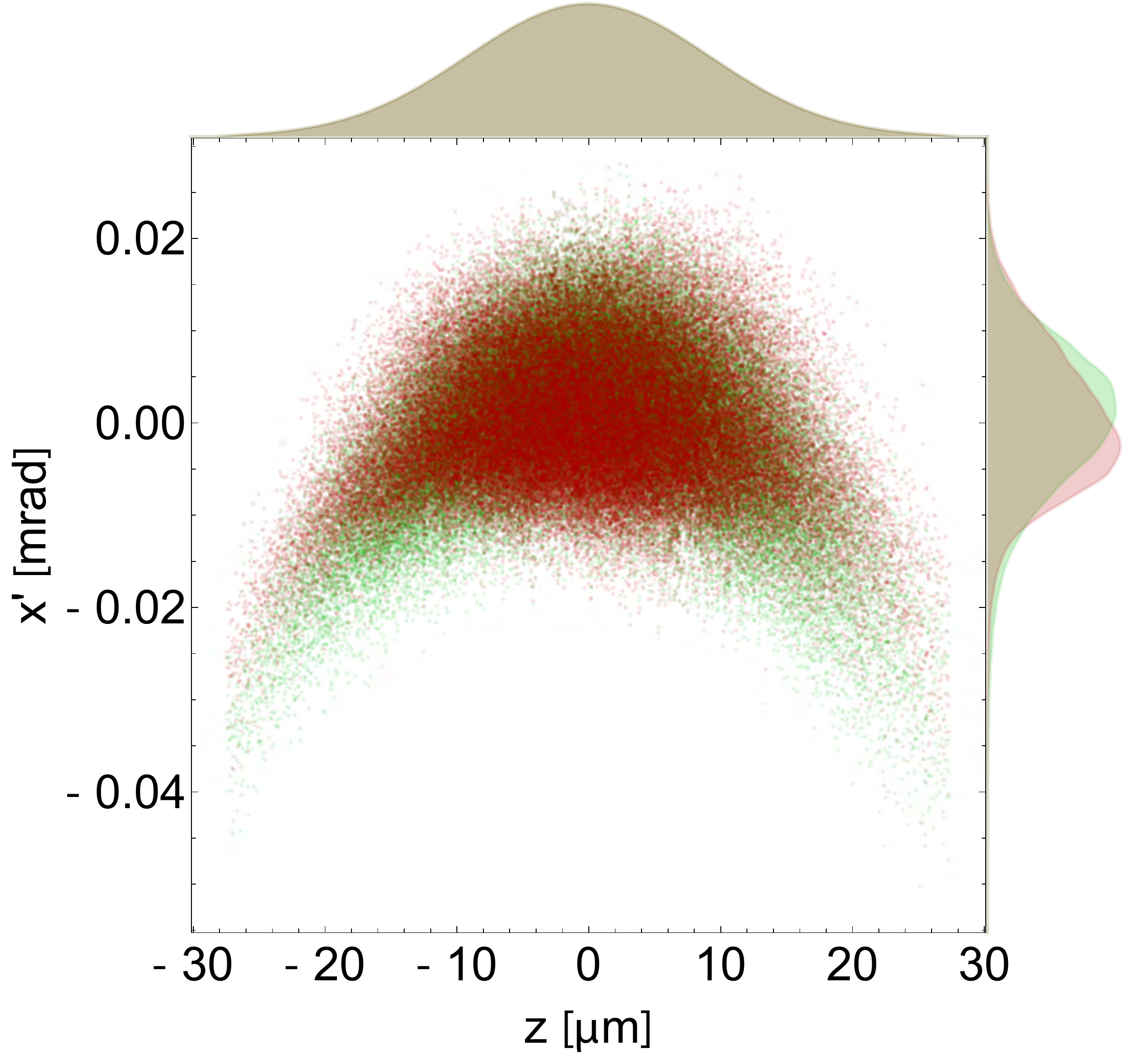} 
	\includegraphics[width=0.32\textwidth]{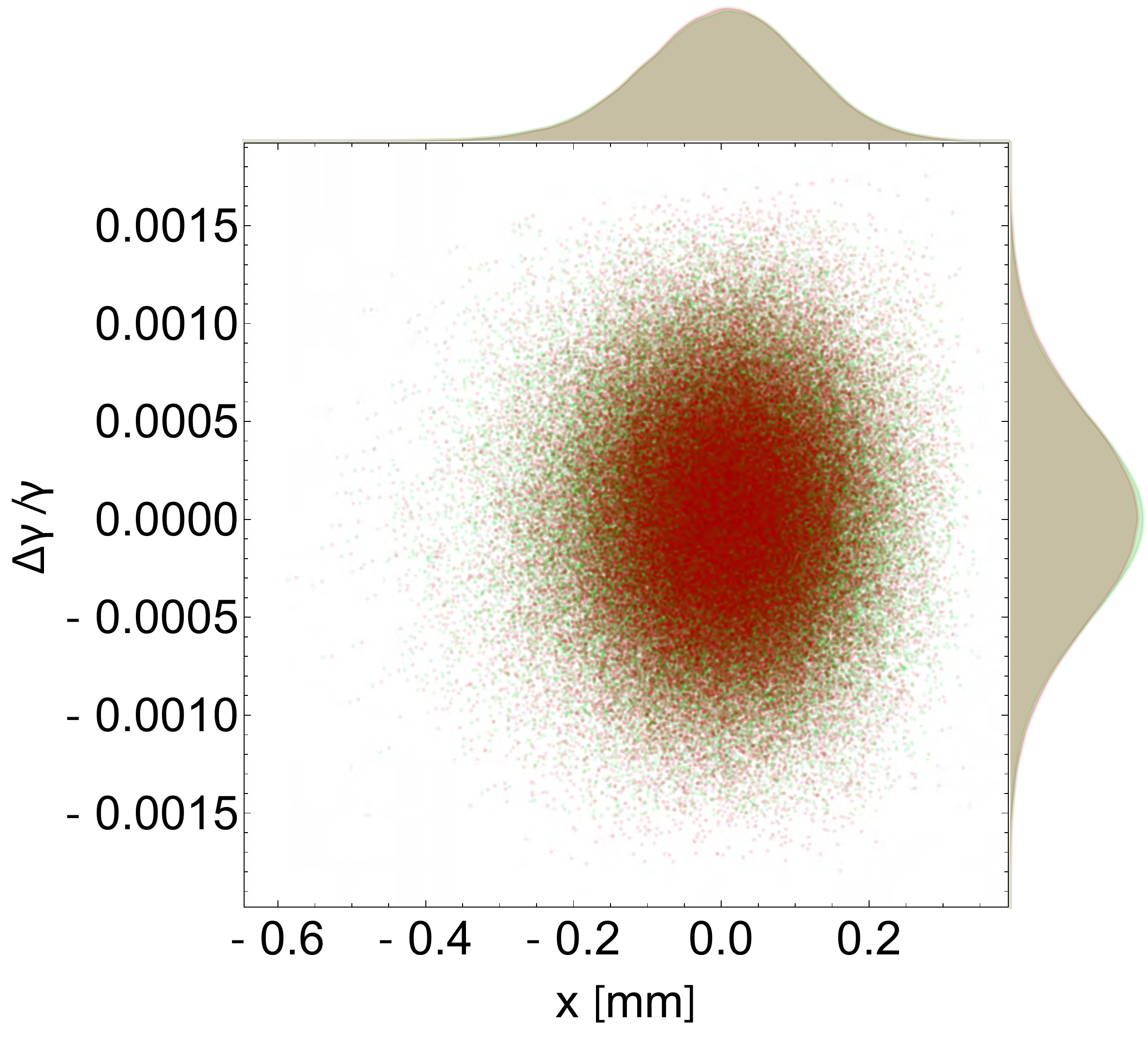} \
	\includegraphics[width=0.3\textwidth]{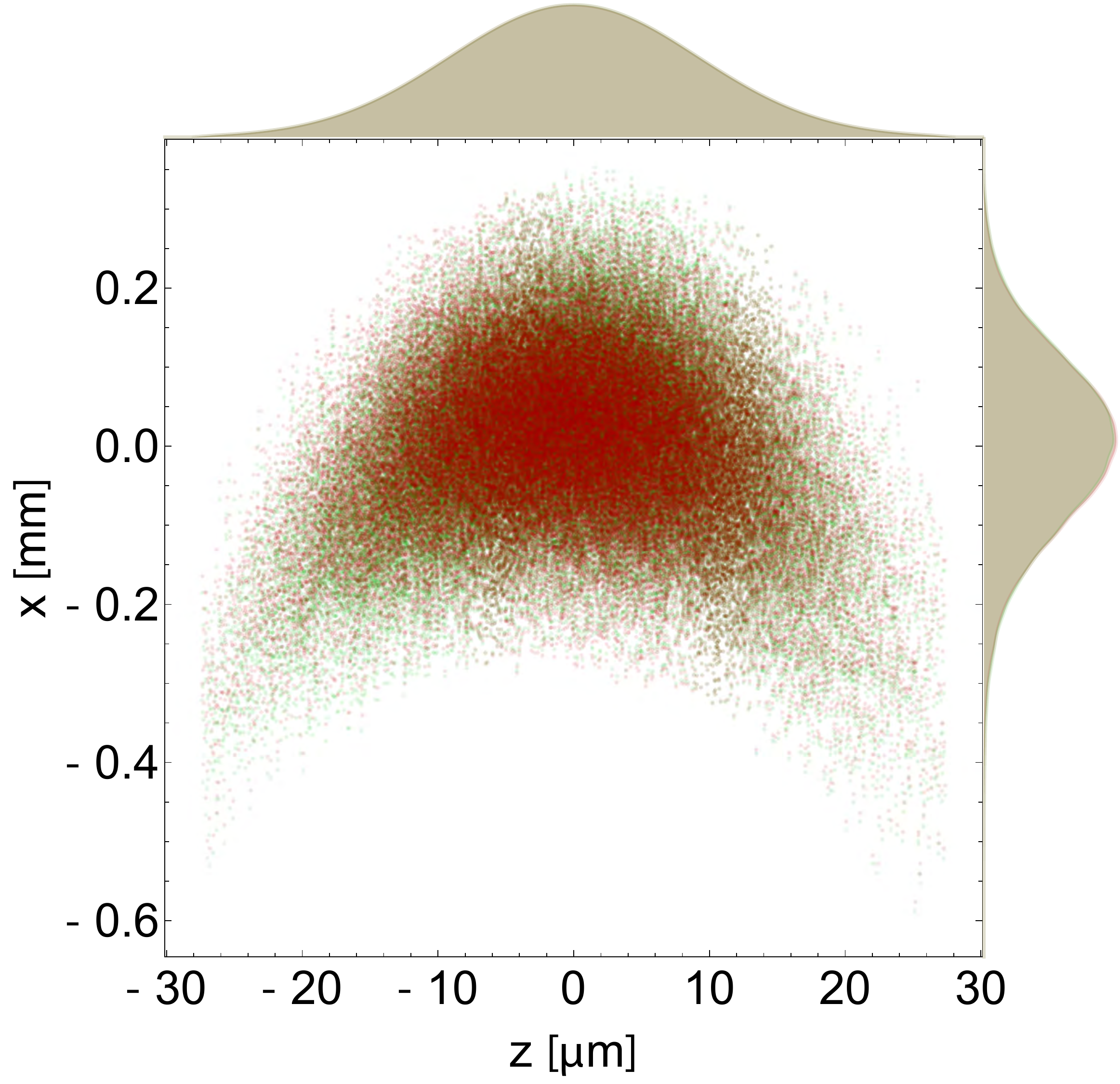} \
	\includegraphics[width=0.32\textwidth]{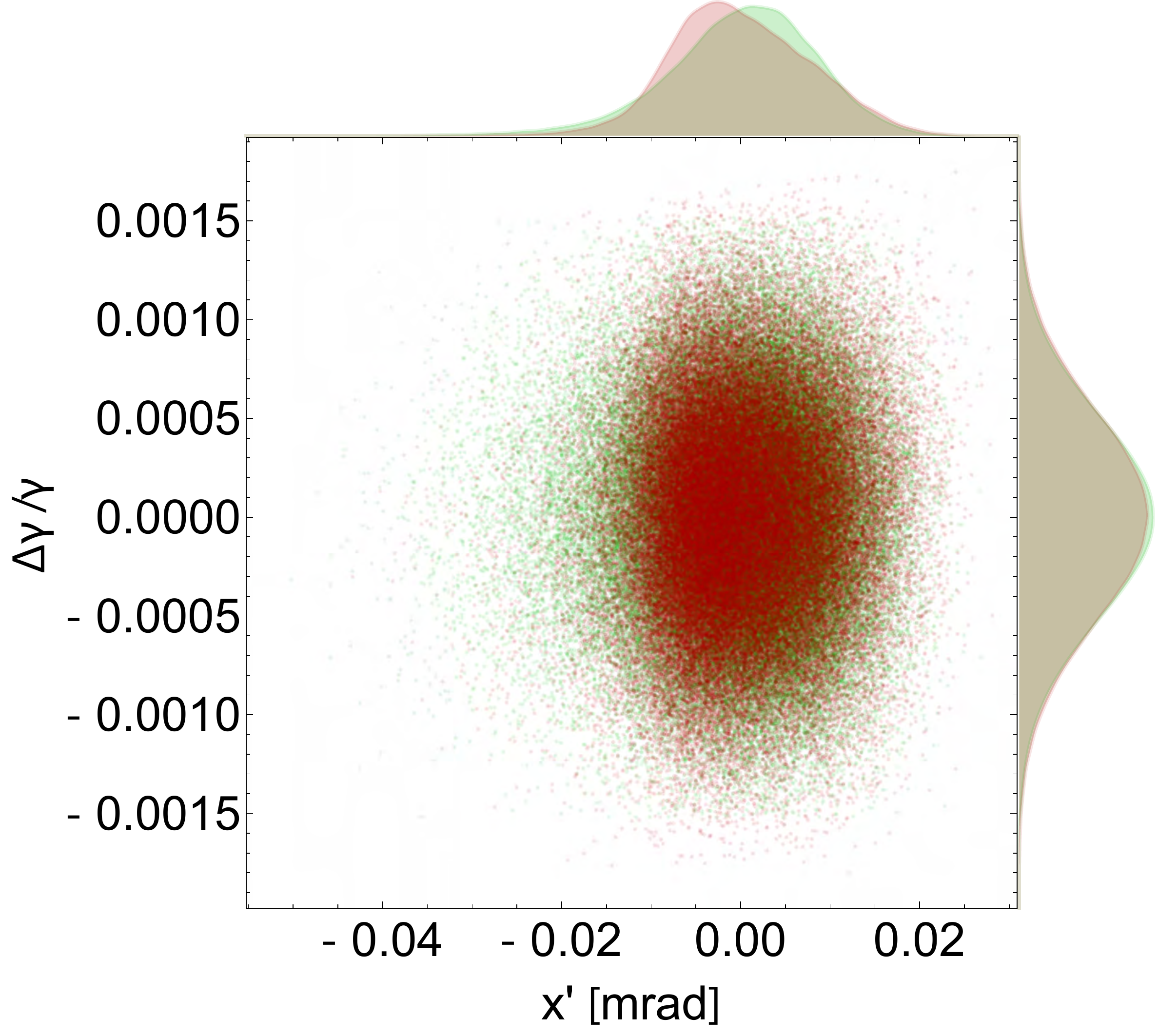} 
	\caption{Phase spaces at the exit of asymmetrical DEEX with mirrored telescope optimized in CSR regime and simulated in nonlinear (green) and CSR (red) regime.} 
	\label{fig:OptScheme-trick_NLvsCSR}
\end{figure*}

\begin{figure*}[ht]
	\includegraphics[width=0.470\textwidth]{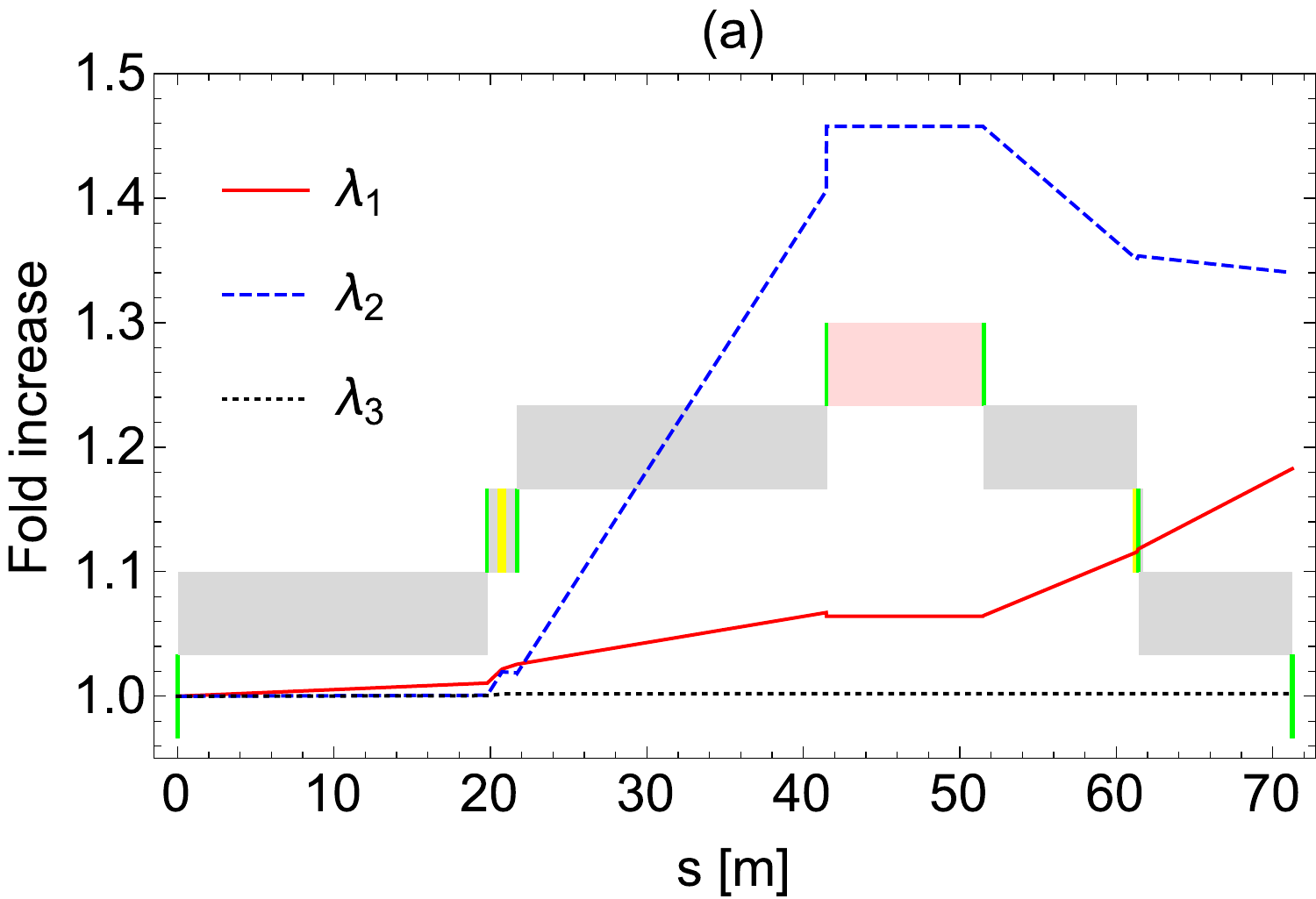}
	\includegraphics[width=0.470\textwidth]{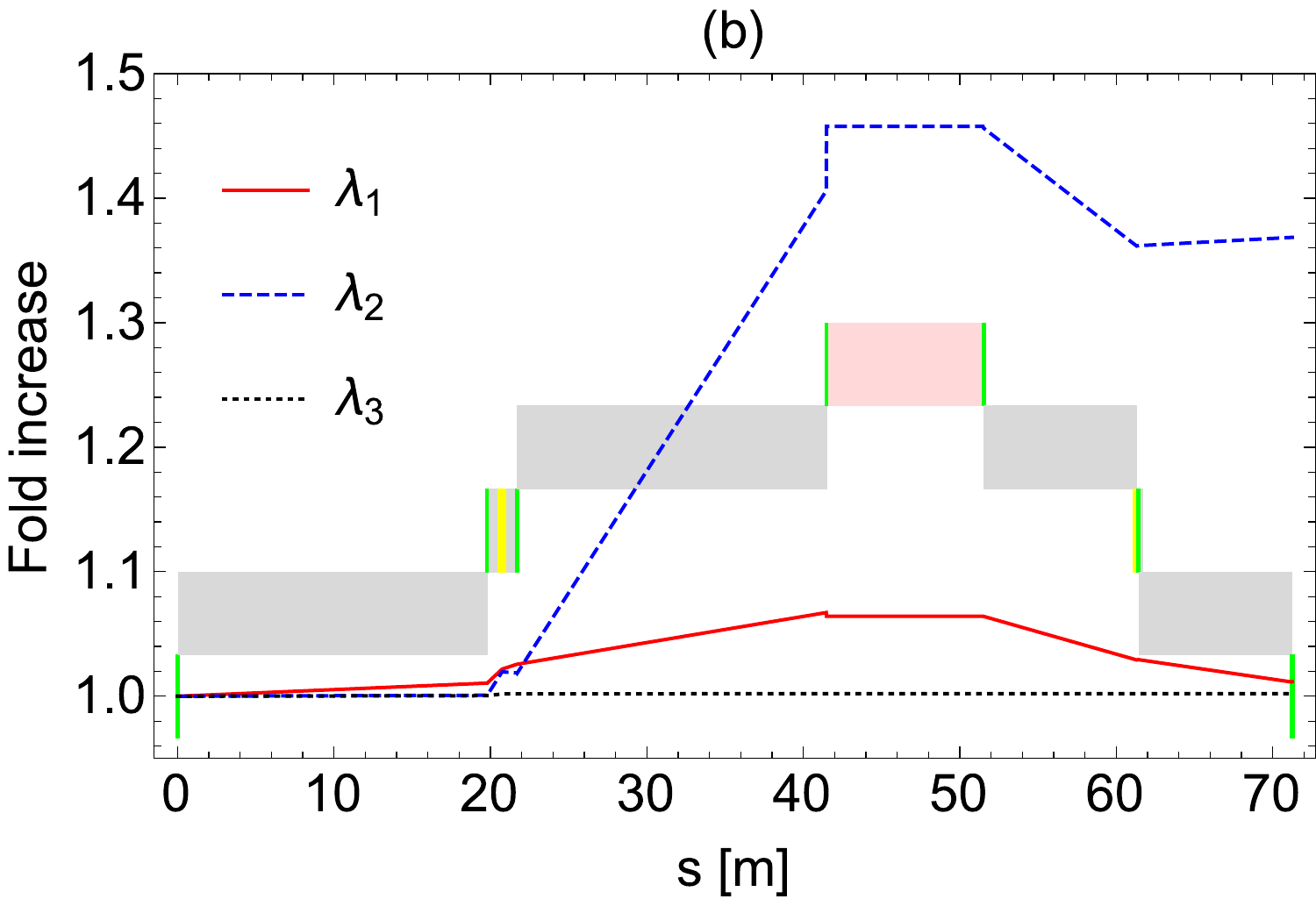}
	\caption{Fold increase of normalized eigen emittances along the beamline for direct (a) and mirrored (b) configuration in CSR regime.
	} 
	\label{fig:Emittance_evol_opt_Scheme}
\end{figure*}

In this section we demonstrate that emittance degradation can be significantly minimized by allowing for a unique configuration of each emittance exchanger, which was found by ES-driven optimization for the direct telescope scheme. These parameters demonstrated even better performance for the scheme with a mirrored telescope, while further adjustment of them with ES did not bring significant improvements. The parameters of the beamline elements are summarized in Table~\ref{tbl:beamline} and compared with that of the symmetric design. The final transverse emittance $\epsilon_{n_x}$ for the scheme with a direct telescope is 45.6\% bigger than its initial value, while the longitudinal emittance is enlarged by 17.9\%. In contrast, the corresponding values for the scheme with CSR-compensation experience 38.7\% growth of $\epsilon_{n_x}$ and only 1.2\% enlargement of $\epsilon_{n_z}$. The dynamics in the nonlinear regime through the asymmetric scheme are similar for two telescope configurations. Nonlinear effects practically do not affect the longitudinal phase space resulting in invariant longitudinal emittance. In contrast, both schemes demonstrate similar and worser (than in CSR regime) output beam quality in ($x,\;x'$) transverse phase space, which is characterized by a roughly 72\% enlargement of the corresponding emittance and depicted in green in Fig.~\ref{fig:OptScheme-notrick_NLvsCSR} for the mirrored configuration and Fig.~\ref{fig:OptScheme-trick_NLvsCSR} for the direct configuration. This result should not be surprising since the schemes were specifically optimized in the CSR-regime, which lead to the mutual compensation of the nonlinear and CSR effects. The fold-increase evolution of eigen emittances along the beamline is depicted in Fig.~\ref{fig:Emittance_evol_opt_Scheme} and demonstrates almost complete compensation of CSR effects' impact on the longitudinal dynamics for the scheme with mirrored telescope and partial compensation of the nonlinear and CSR effects for the transverse dynamics in both configurations.

\begin{figure*}[ht]
	\includegraphics[width=0.470\textwidth]{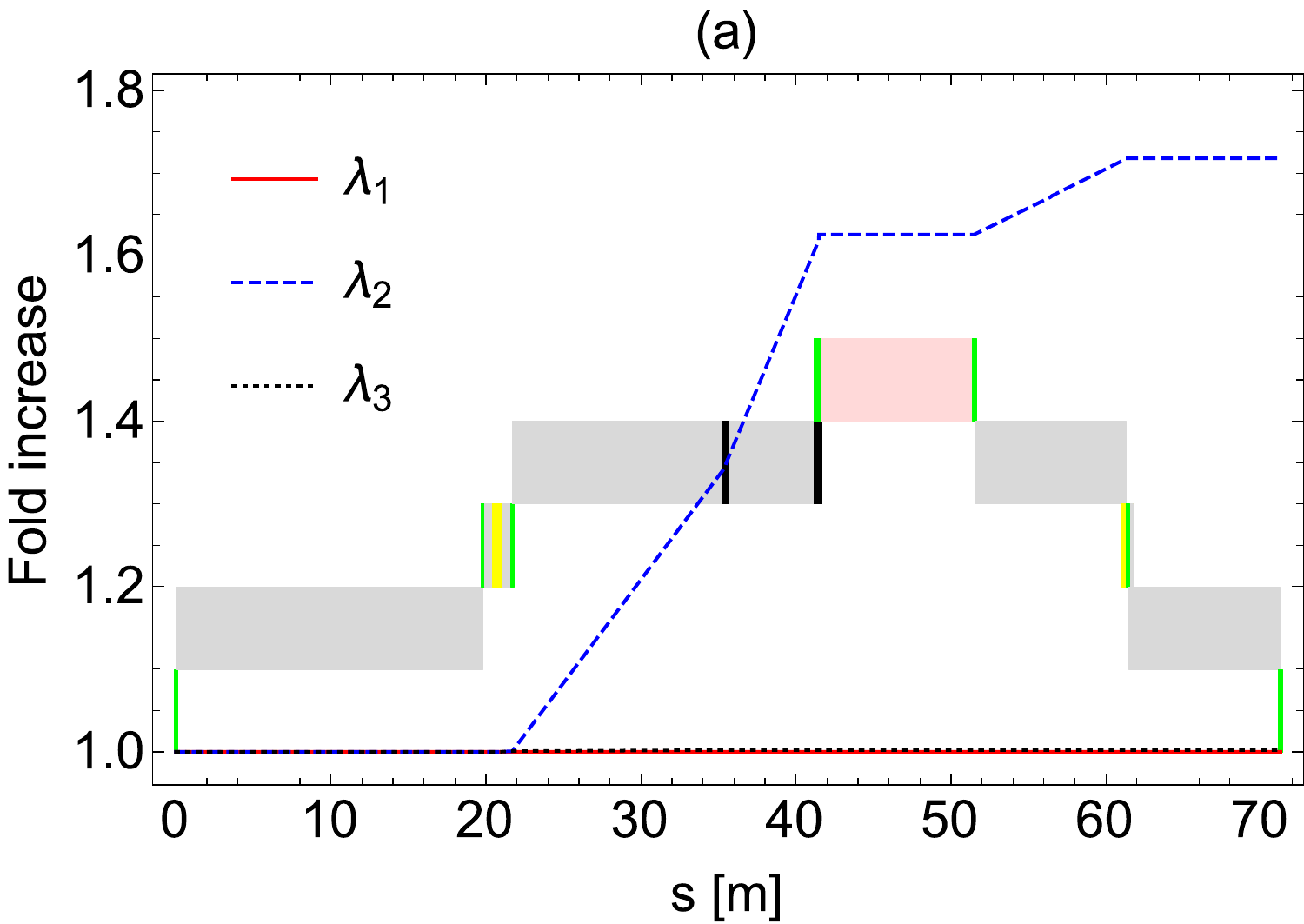}
	\includegraphics[width=0.470\textwidth]{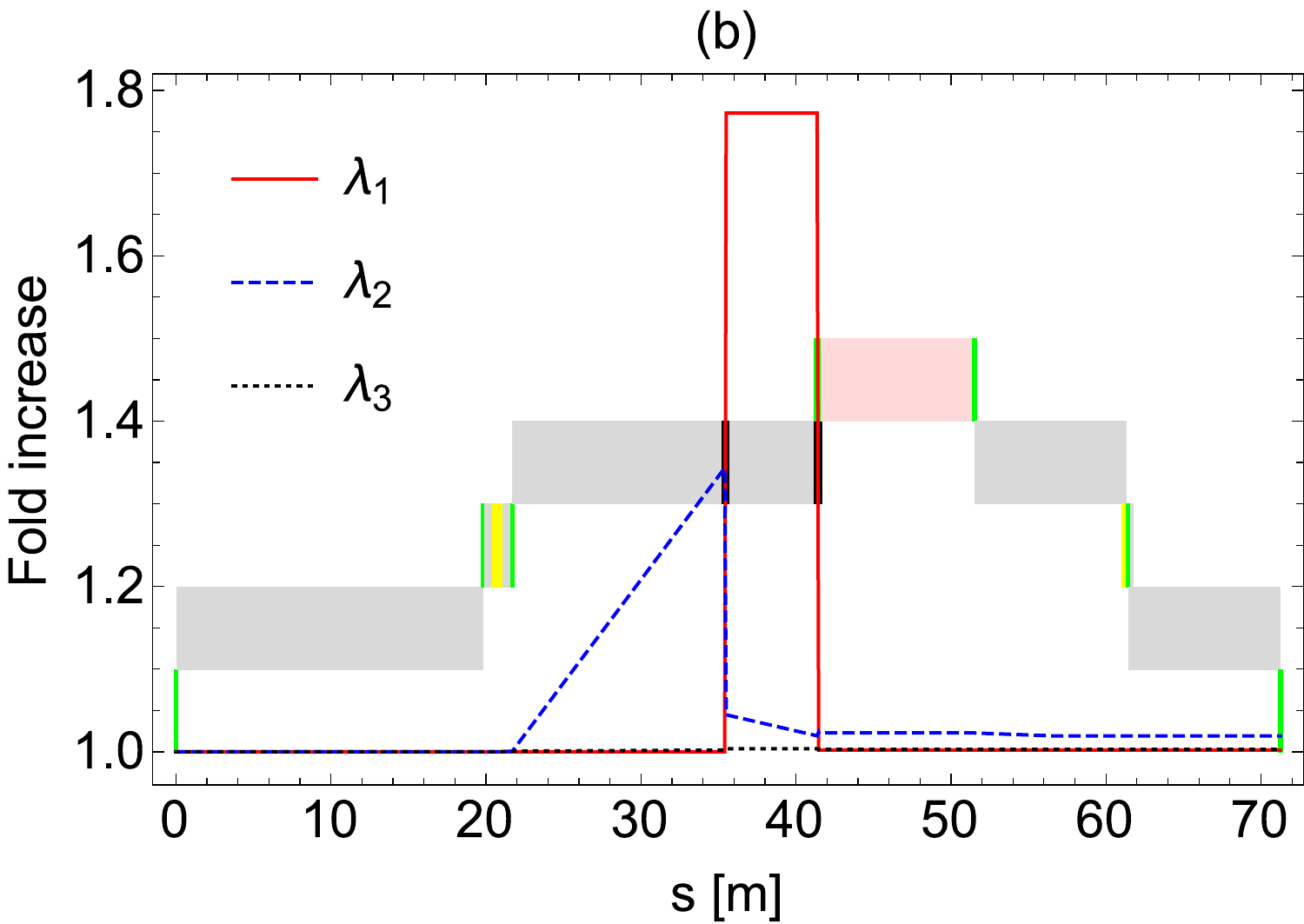}
	\caption{Fold increase of normalized eigen emittances along the beamline in nonlinear regime for mirrored configuration with turned off (a) and turned on (b) correcting sextupoles (marked black).
	} 
	\label{fig:Emittance_evol_opt_Scheme_Sext}
\end{figure*}

\emph{\bf Nonlinear optics}. 
Nonlinear effects in the obtained asymmetrical configuration are partially compensated by choosing appropriate linear optics and input Twiss parameters delivering the proper beam dynamics along the DEEX bunch compressor. The major enlargement of eigen emittance $\lambda_2$ happens in the drift section between the third and fourth bending magnets in the CSR regime (Fig.~\ref{fig:Emittance_evol_opt_Scheme}) and become even bigger in the nonlinear regime as demonstrated in Fig.~\ref{fig:Emittance_evol_opt_Scheme_Sext} (a). In the approximation of zero bunch charge (CSR effects are neglected) the nonlinear effects can be fully compensated by inserting two sextupole magnets in this drift section, while properly adjusting their relative location and strength. The first sextupole ($k_1=-4.33$~m$^{-2}$) is placed in the location dividing the drift section as 7 to 3 and aimed to turn around the evolution of eigen emittance $\lambda_2$ as depicted in Fig.~\ref{fig:Emittance_evol_opt_Scheme_Sext} (b). This sextupole also strongly affects the eigen emittance $\lambda_1$ since the transverse and longitudinal dynamics are highly coupled at this point of the beamline. A second sextupole ($k_2=1.55$~m$^{-2}$) is placed directly in front of the fourth bending magnet in order to compensate for the introduced enlargement of $\lambda_1$, which at this location is practically solely characterized by the transverse dynamics, thus eigen emittance $\lambda_2$ is practically unaffected by this sextupole. The sextupoles strengths and locations were tuned exclusively by analyzing the evolution of eigen emittances along the beamline and resulted in approximately 2\% enlargement of $\epsilon_{n_x}$ and 0.2\% enlargement of $\epsilon_{n_x}$ in nonlinear regime.

Turning on CSR effects in the approximation of 100~pC bunch destroys both the sextupole compensation of the nonlinear effects and CSR compensation, accomplished by the proper non-symmetry of the DEEX design with mirrored telescope, and results in 74\% enlargement of $\epsilon_{n_x}$ and 2.1\% enlargement of $\epsilon_{n_z}$ which are even worse than in the optimal asymmetrical scheme without sextupoles. A further adjustment of sextupole strengths and locations within the drift sections (two more sextupoles were additionally added between the fifth and sixth bending magnets) was accomplished by two strategies for ES-driven optimization. 

Initially, optimization was started from the optimal asymmetrical beamline configuration for 100 pC bunch with turned off sextupoles and converged very slowly. In contrast, the alternative strategy relied on optimization of all parameters starting from the optimal asymmetric scheme with sextupoles for 0 pC bunch by slowly increasing the bunch charge to the designed value of 100~pC. Thus, the output beamline configuration, optimized for 20 pC bunch charge was used for 40 pC bunch charge, etc. For the operating regime of LCLS-II assuming a bunch charge of 100 pC we found a configuration of the scheme with mirrored telescope resulting in as low as 26.6\% enlargement of transverse emittance $\epsilon_{n_x}$. 

\begin{figure*}[ht]
	\includegraphics[width=0.3\textwidth]{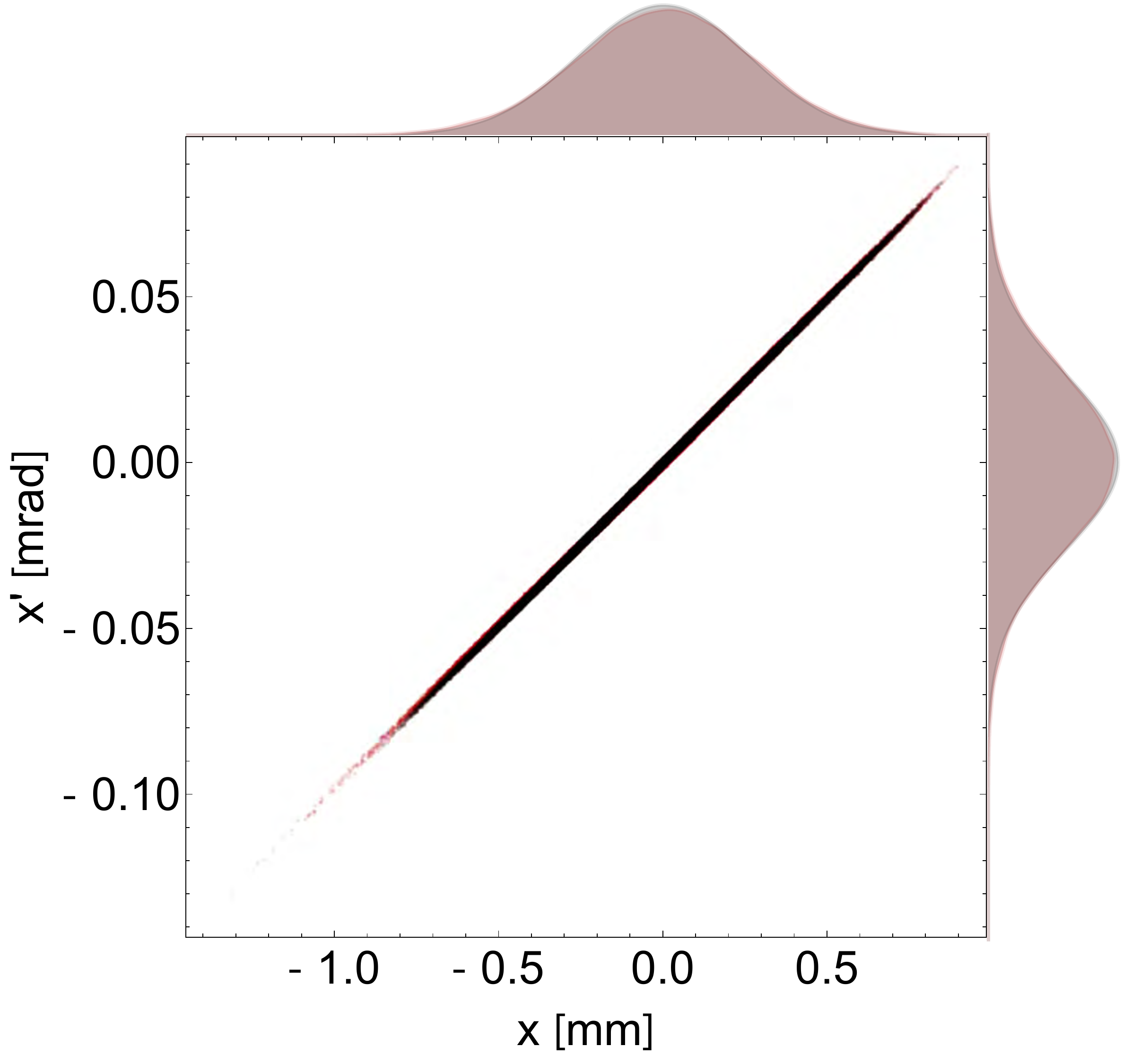} \
	\includegraphics[width=0.32\textwidth]{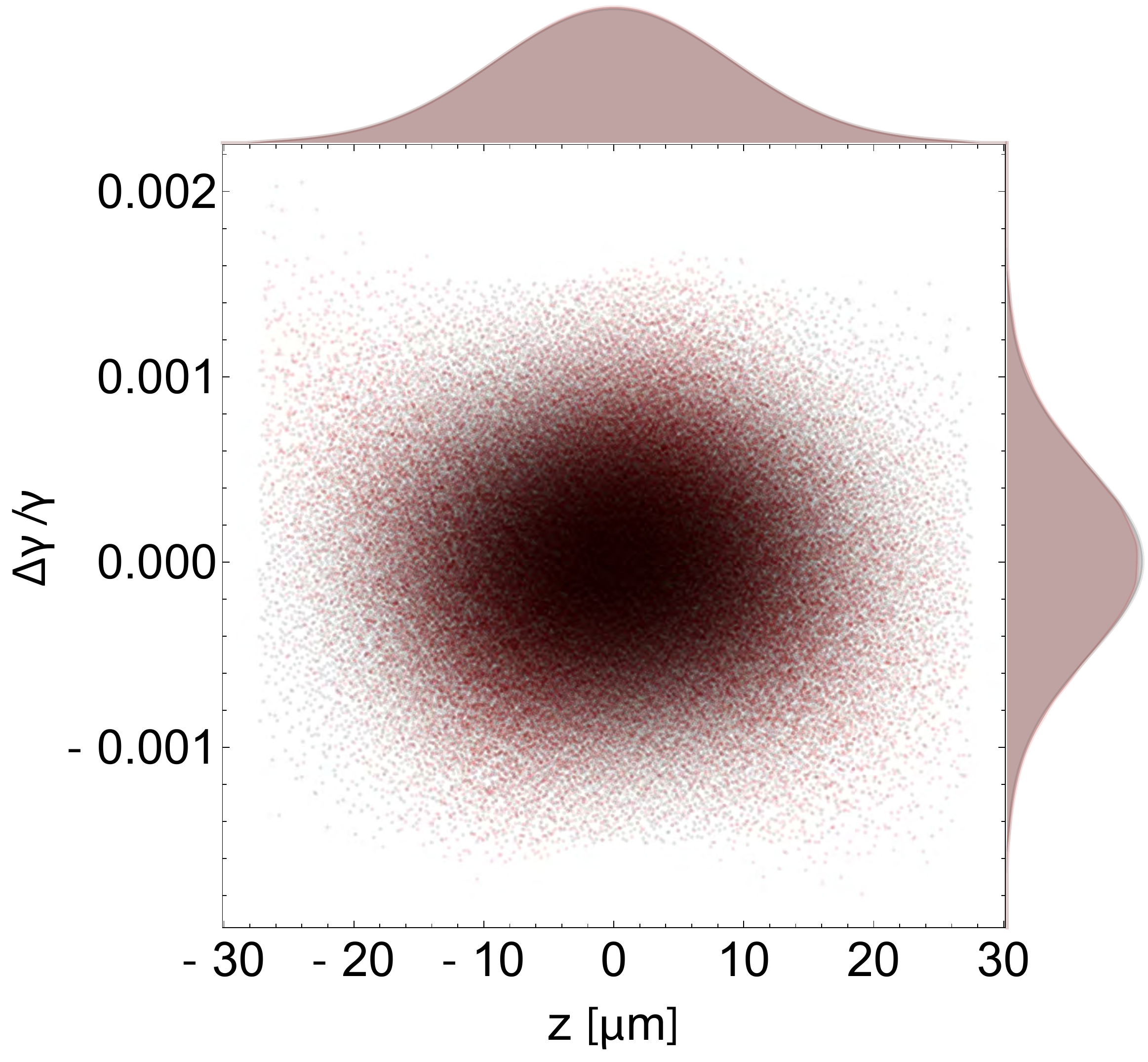} \
	\includegraphics[width=0.31\textwidth]{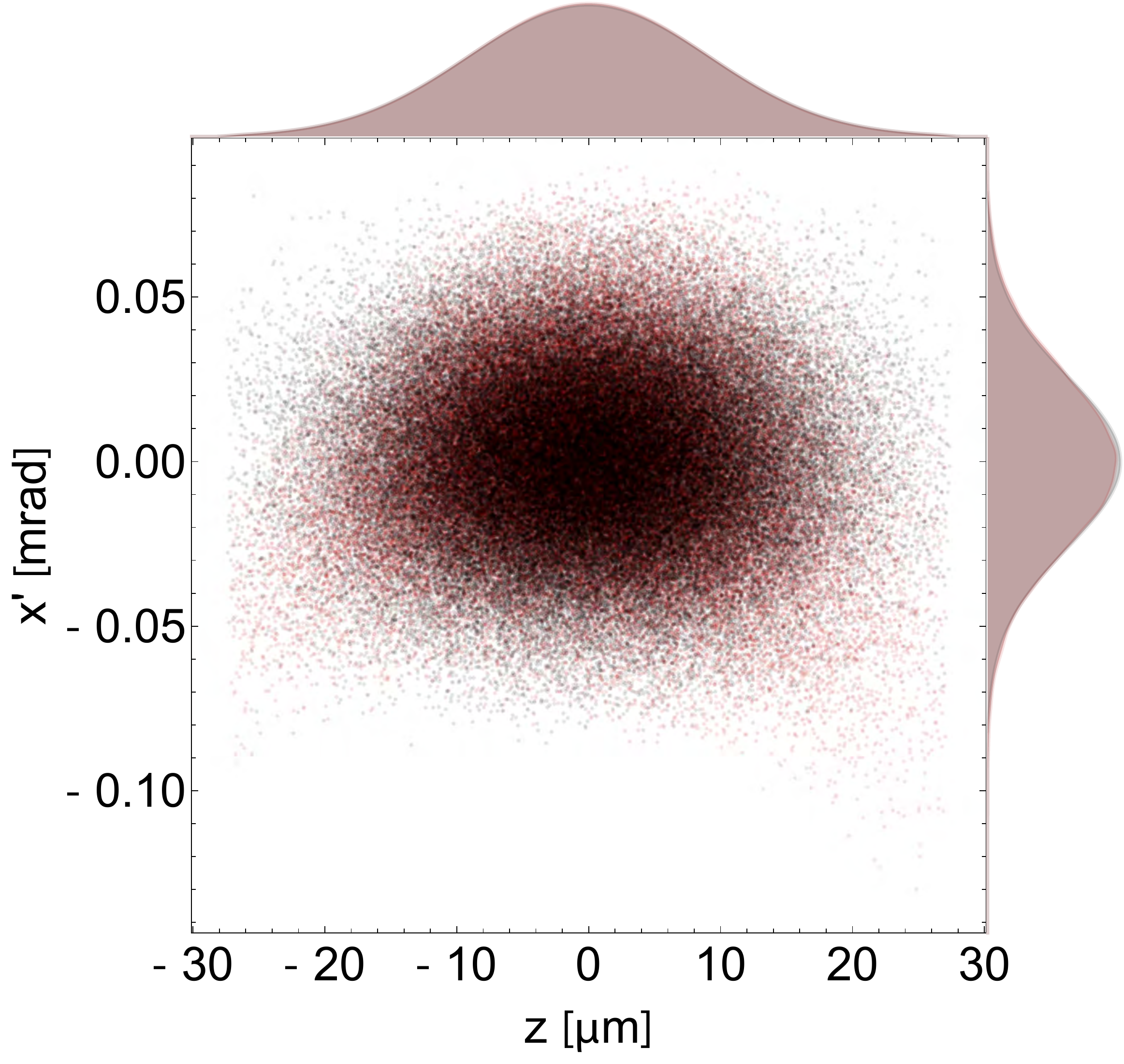} 
	\includegraphics[width=0.32\textwidth]{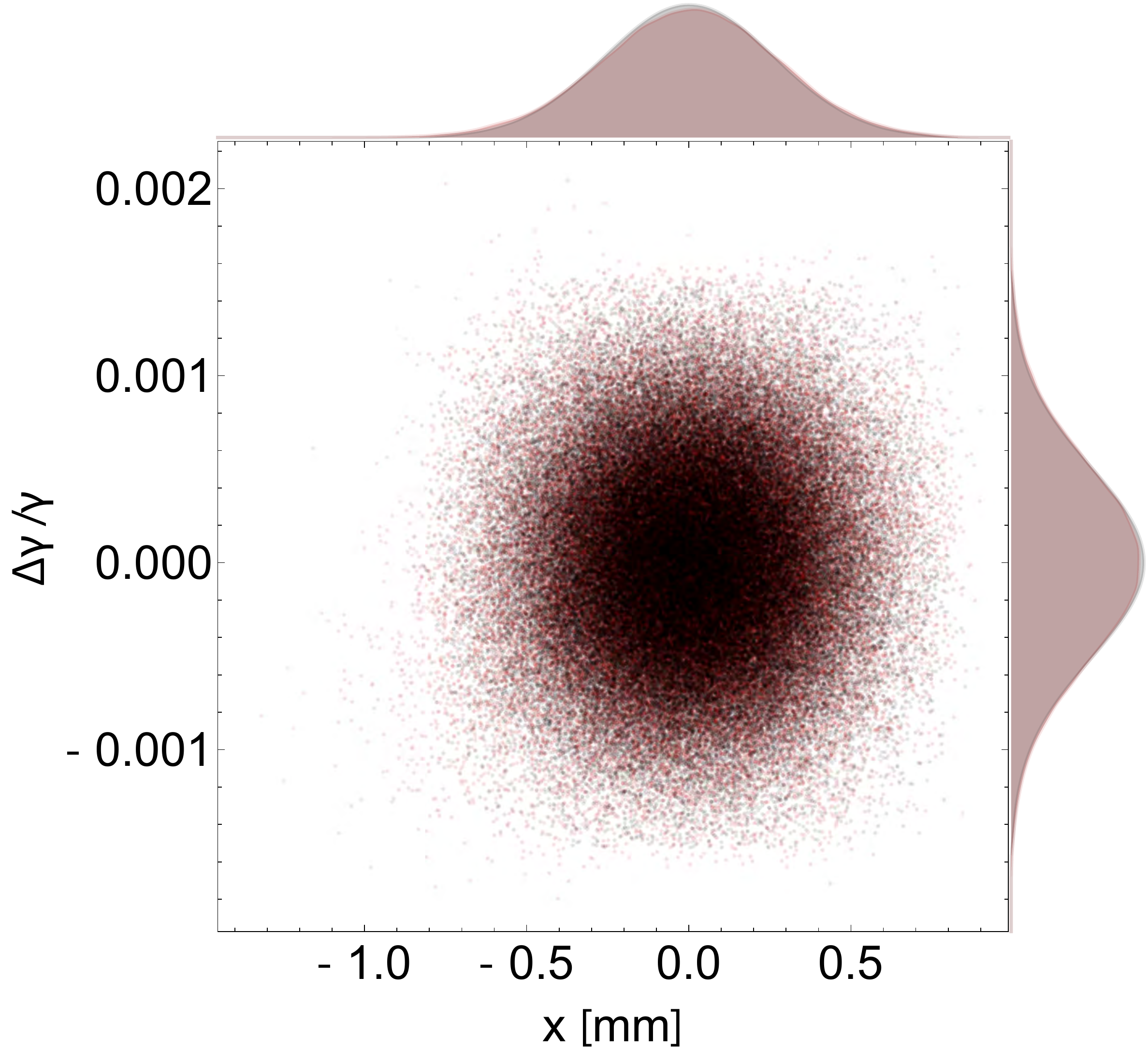} \
	\includegraphics[width=0.3\textwidth]{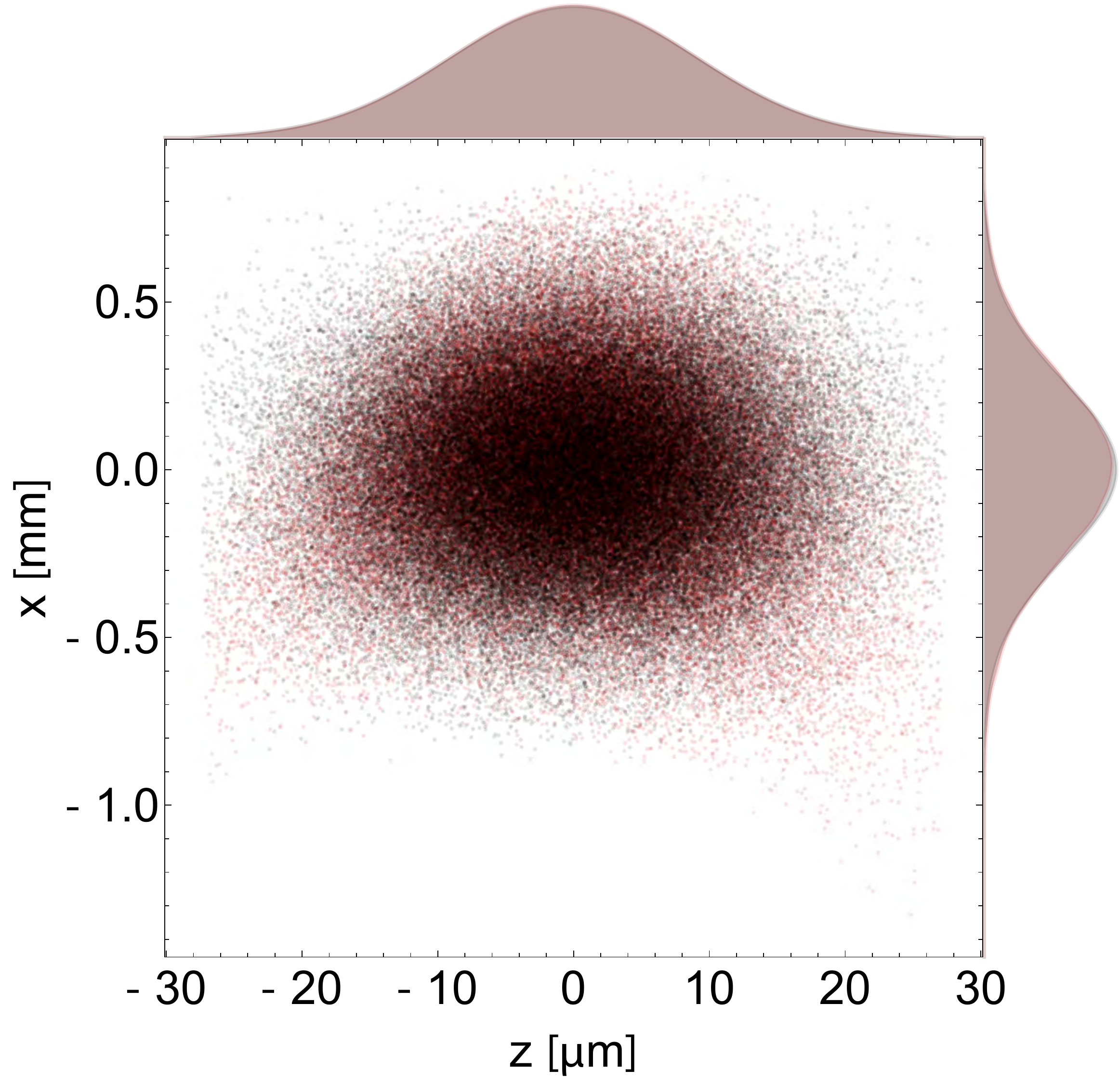} \
	\includegraphics[width=0.32\textwidth]{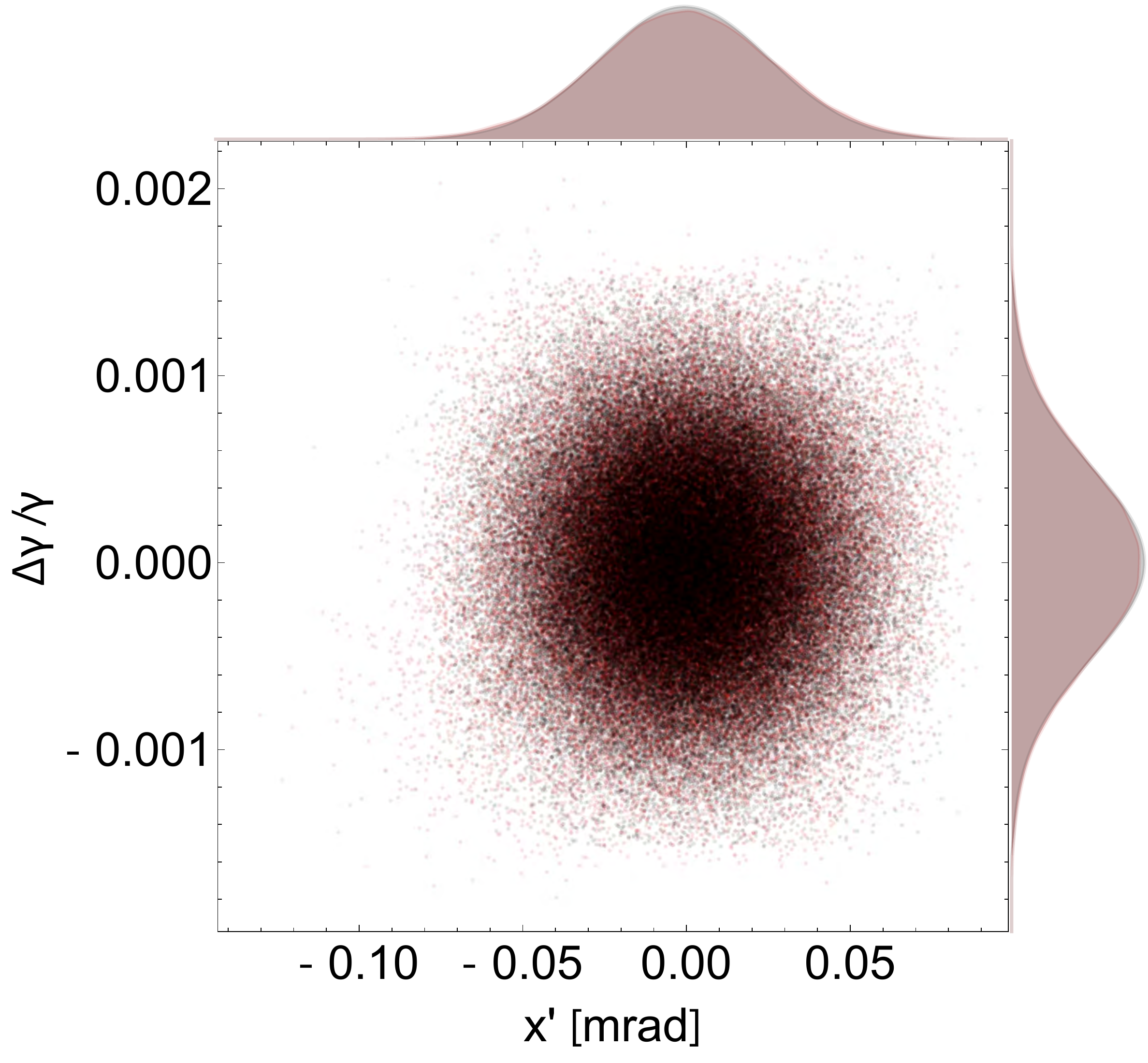} 
	\caption{Phase spaces at the exit of optimized asymmetrical DEEX with sextupoles in linear (black) and CSR (red) regime with chirping ($R_{56}=0.867$~m$^{-1}$) mirrored telescope demonstrating 26.6\% growth of $\epsilon_{n_x}$ and 2\% enlargement of $\epsilon_{n_z}$.} 
	\label{fig:OptSchemeSex-trick_LvsCSR_100_chirp-removed}
\end{figure*}

\begin{figure*}[ht]
	\includegraphics[width=0.3\textwidth]{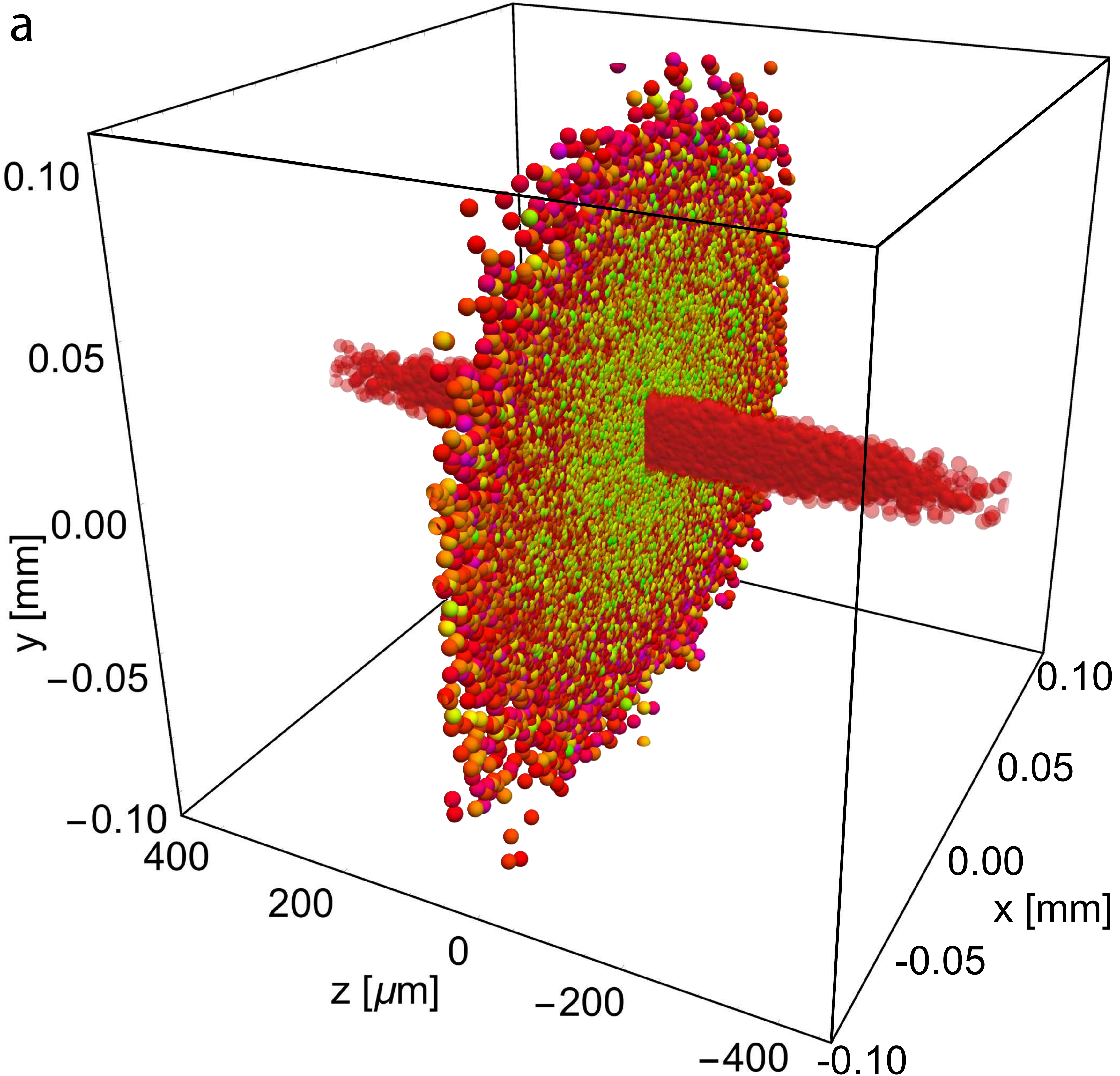} \
	\includegraphics[width=0.3\textwidth]{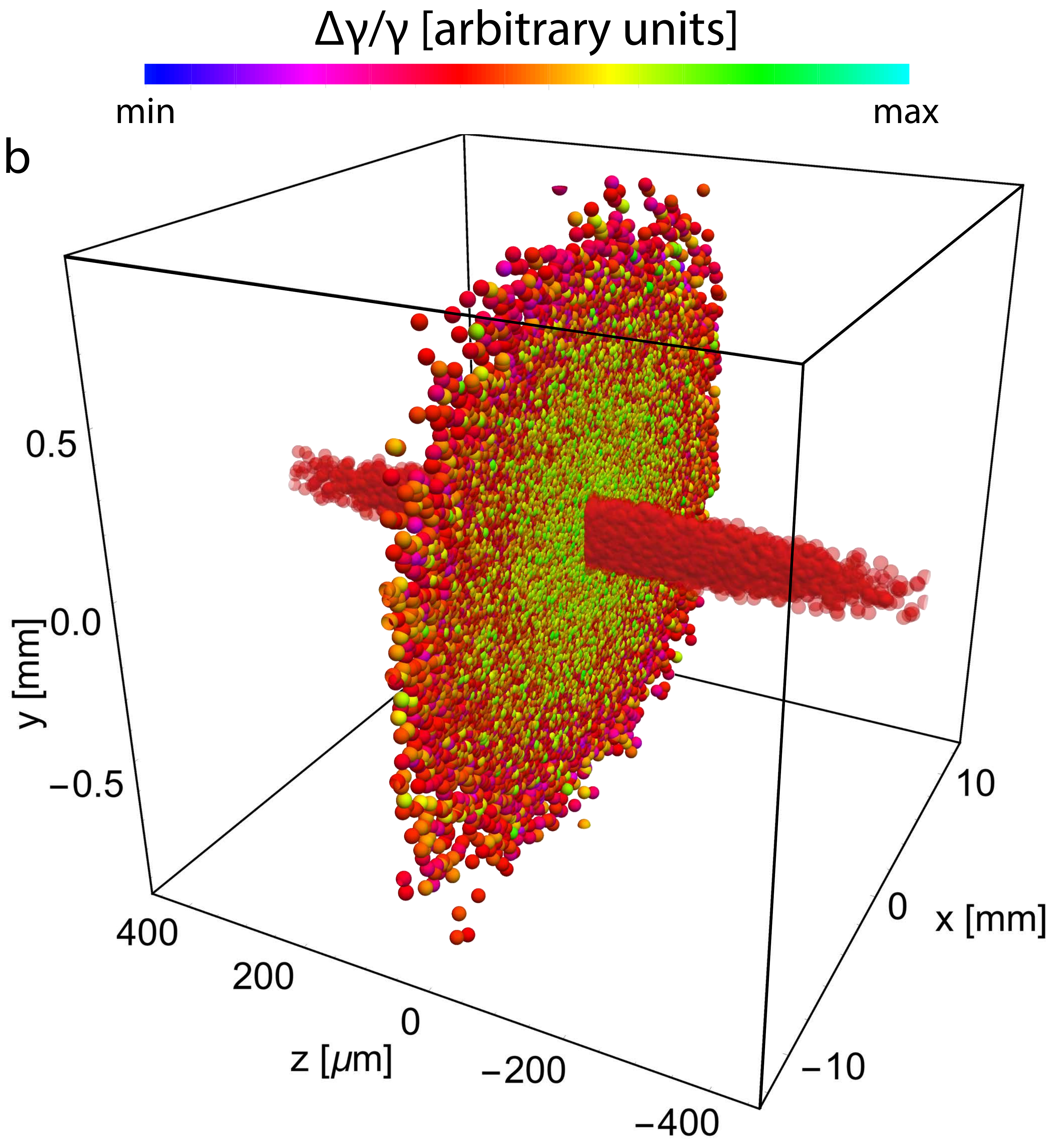} \
	\includegraphics[width=0.3\textwidth]{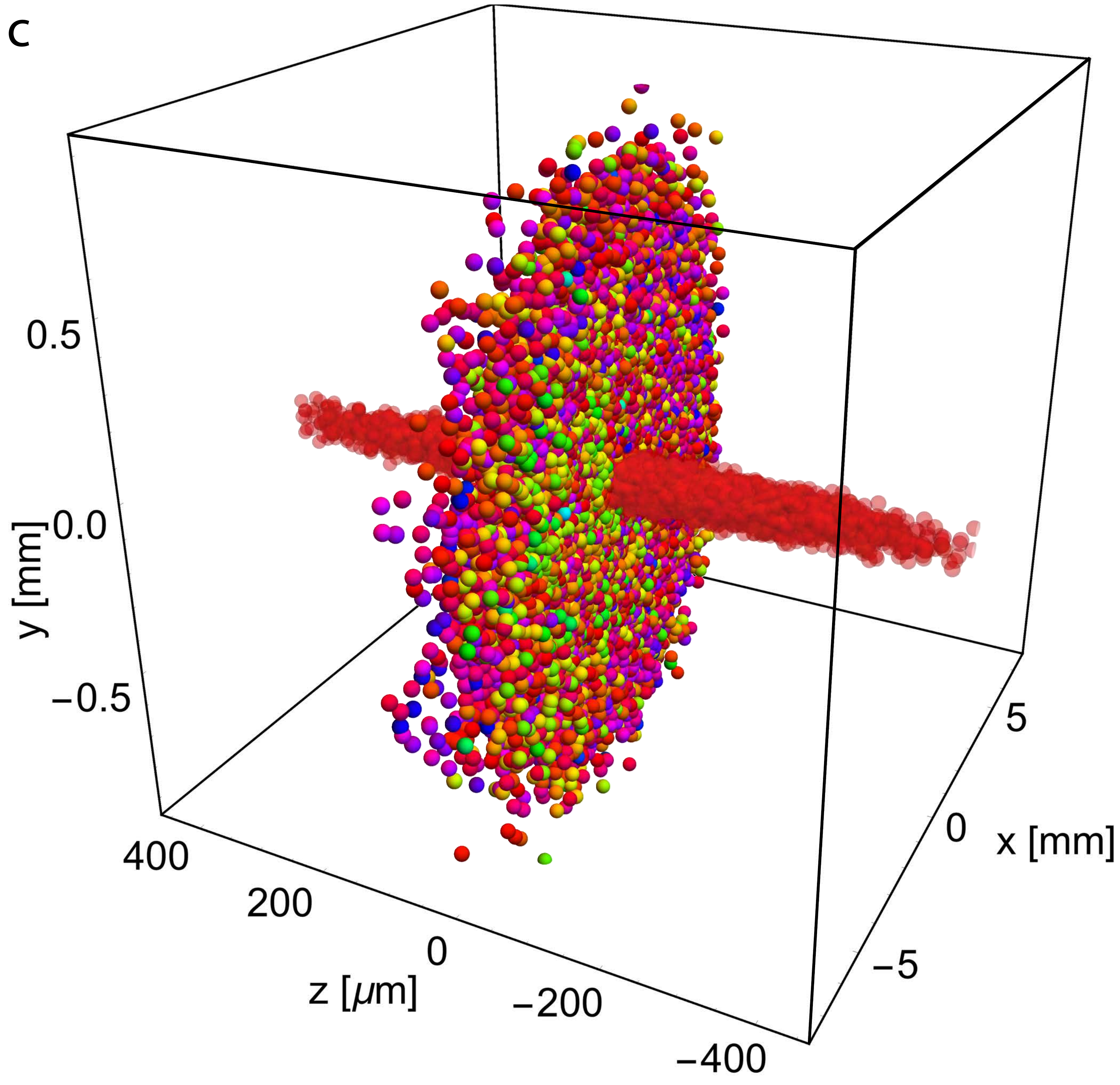} 
	\includegraphics[width=0.3\textwidth]{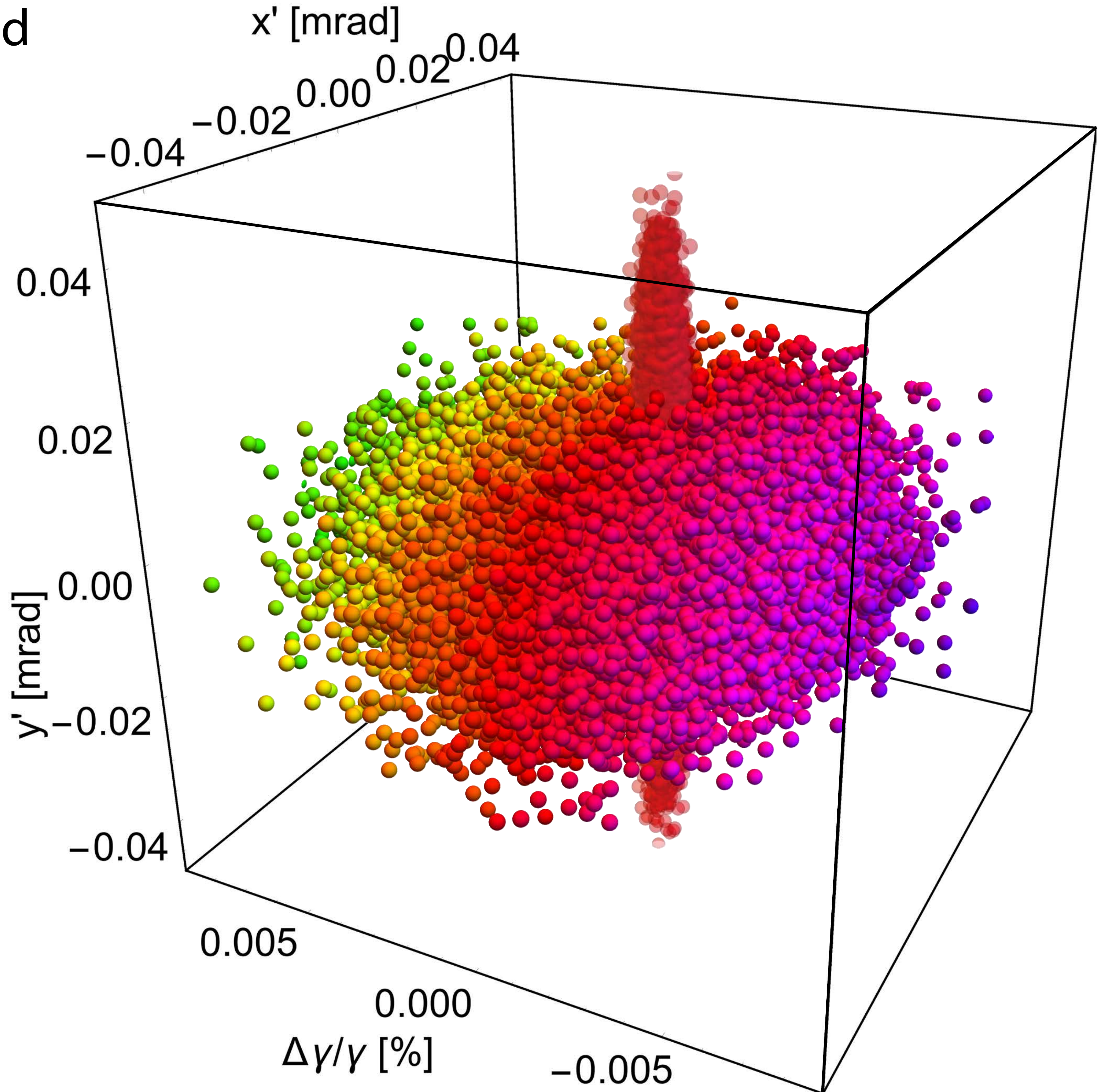} \
	\includegraphics[width=0.3\textwidth]{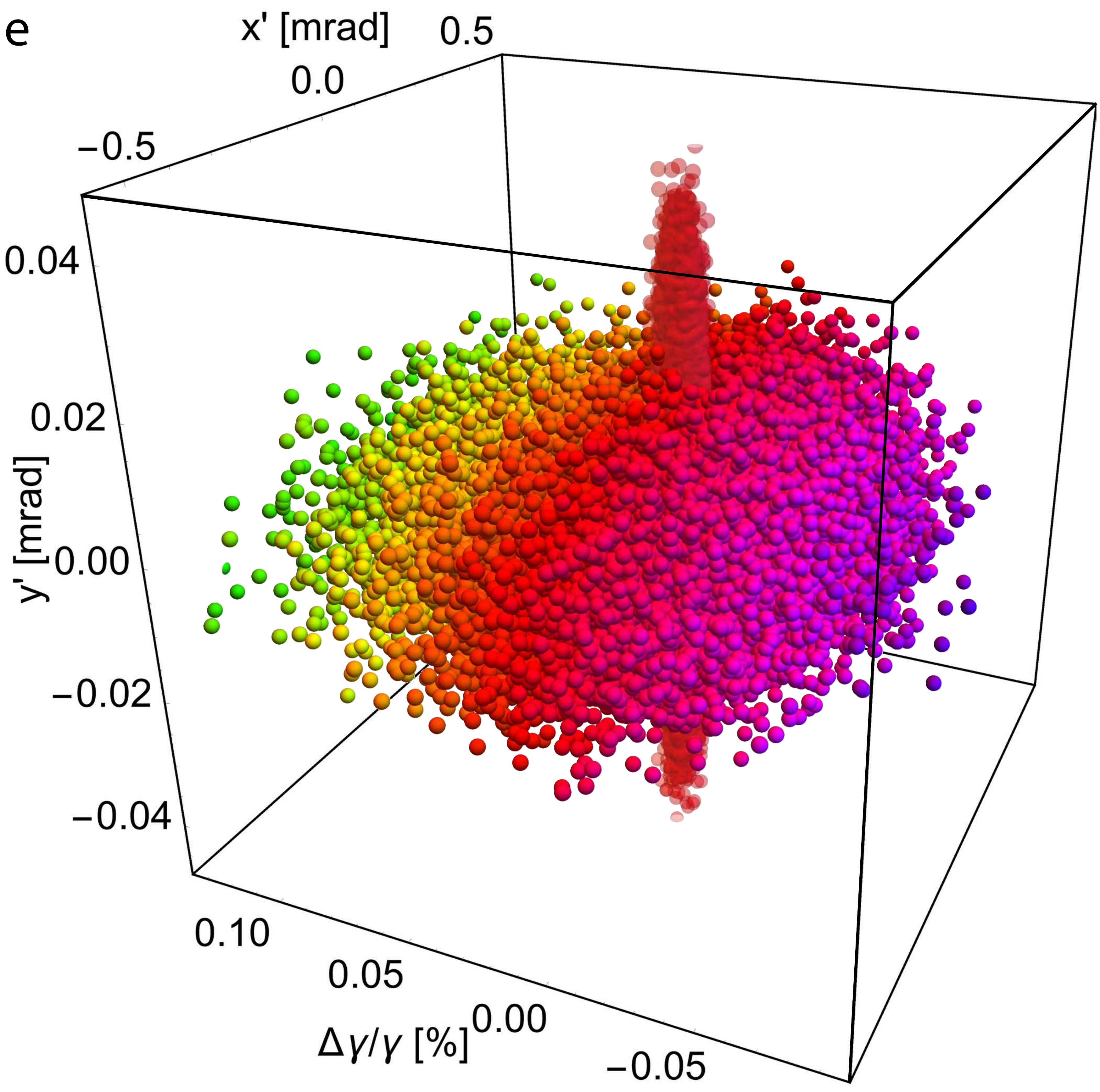} \
	\includegraphics[width=0.3\textwidth]{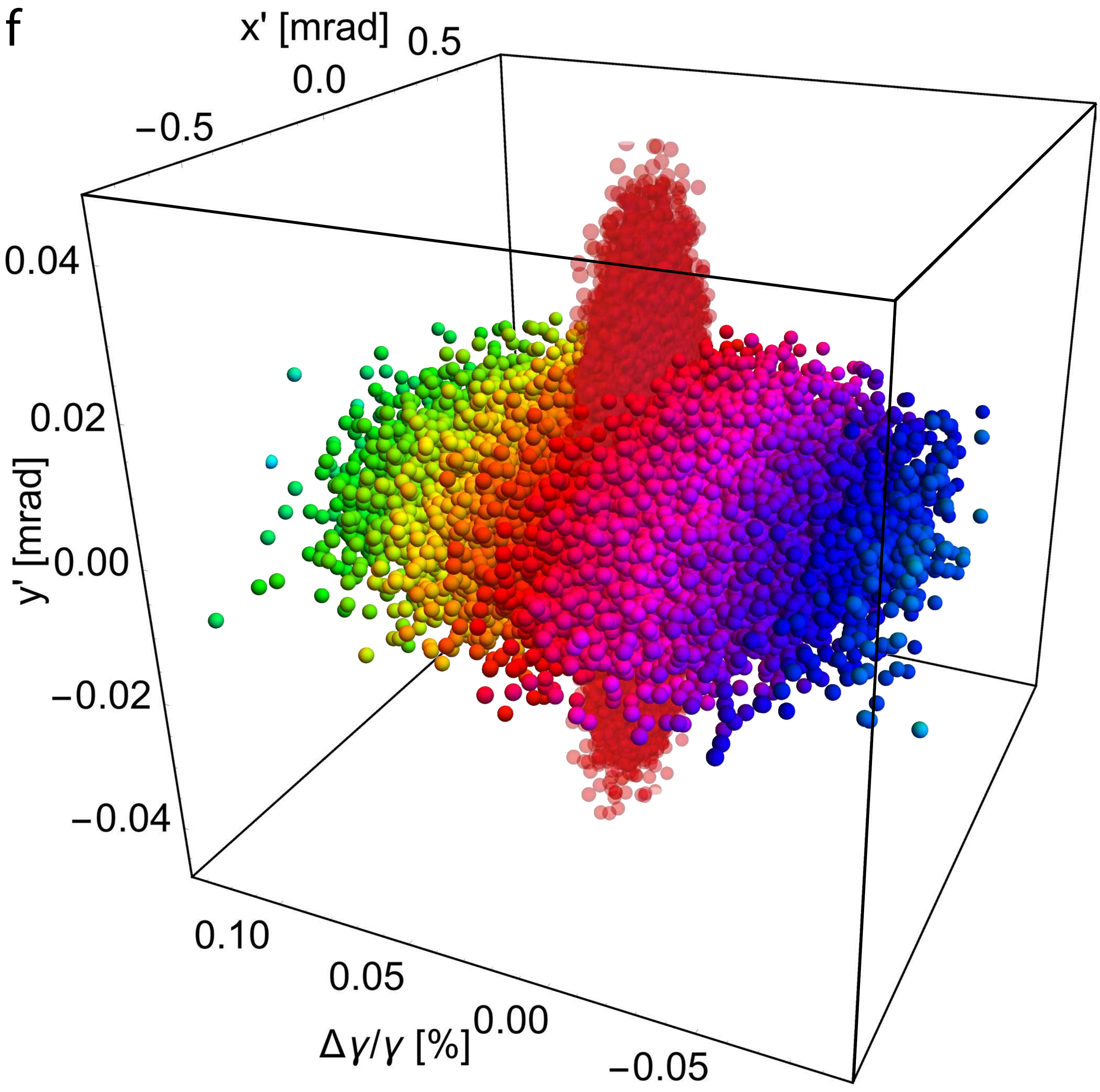} 
	\includegraphics[width=0.3\textwidth]{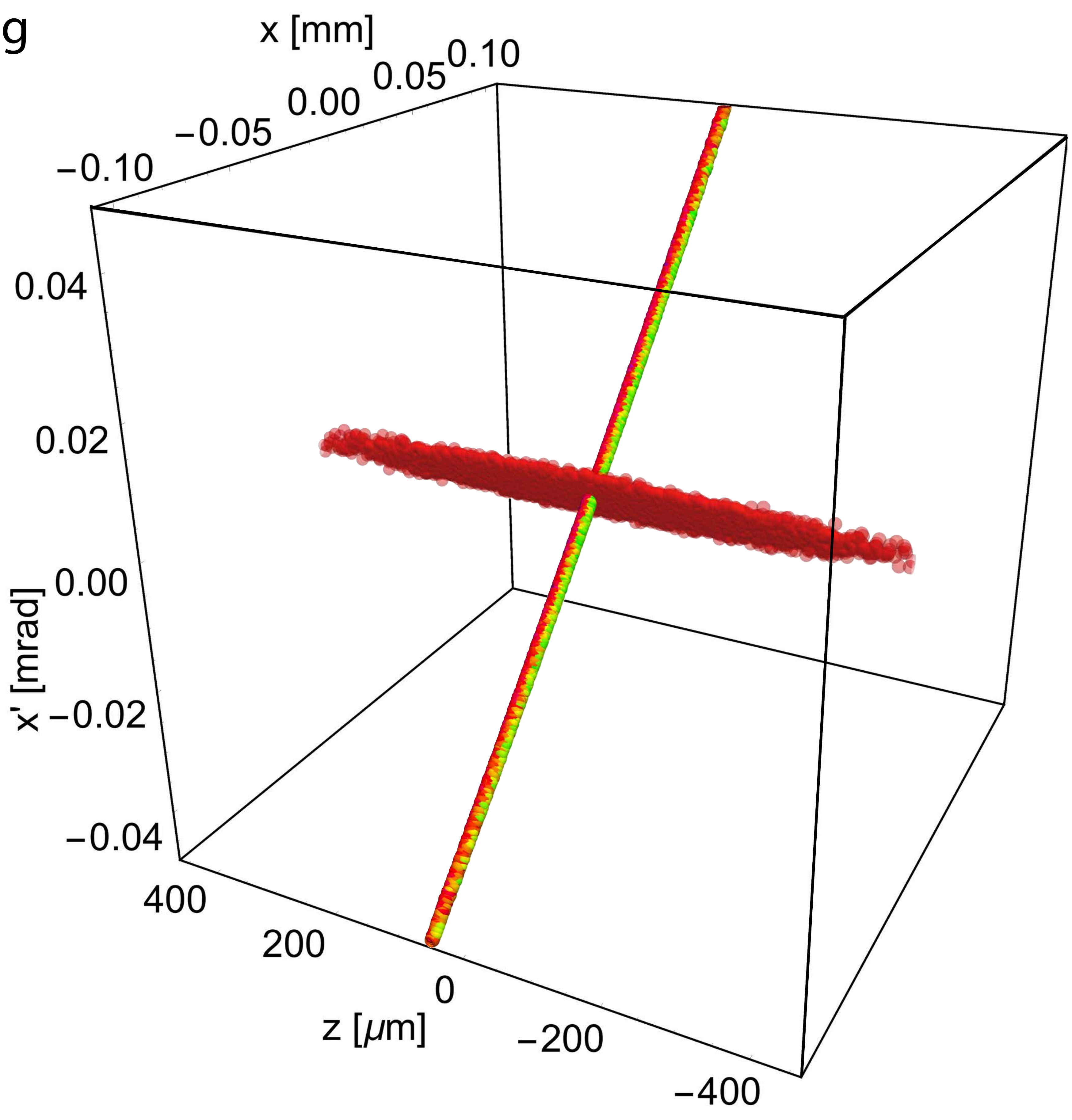} \
	\includegraphics[width=0.3\textwidth]{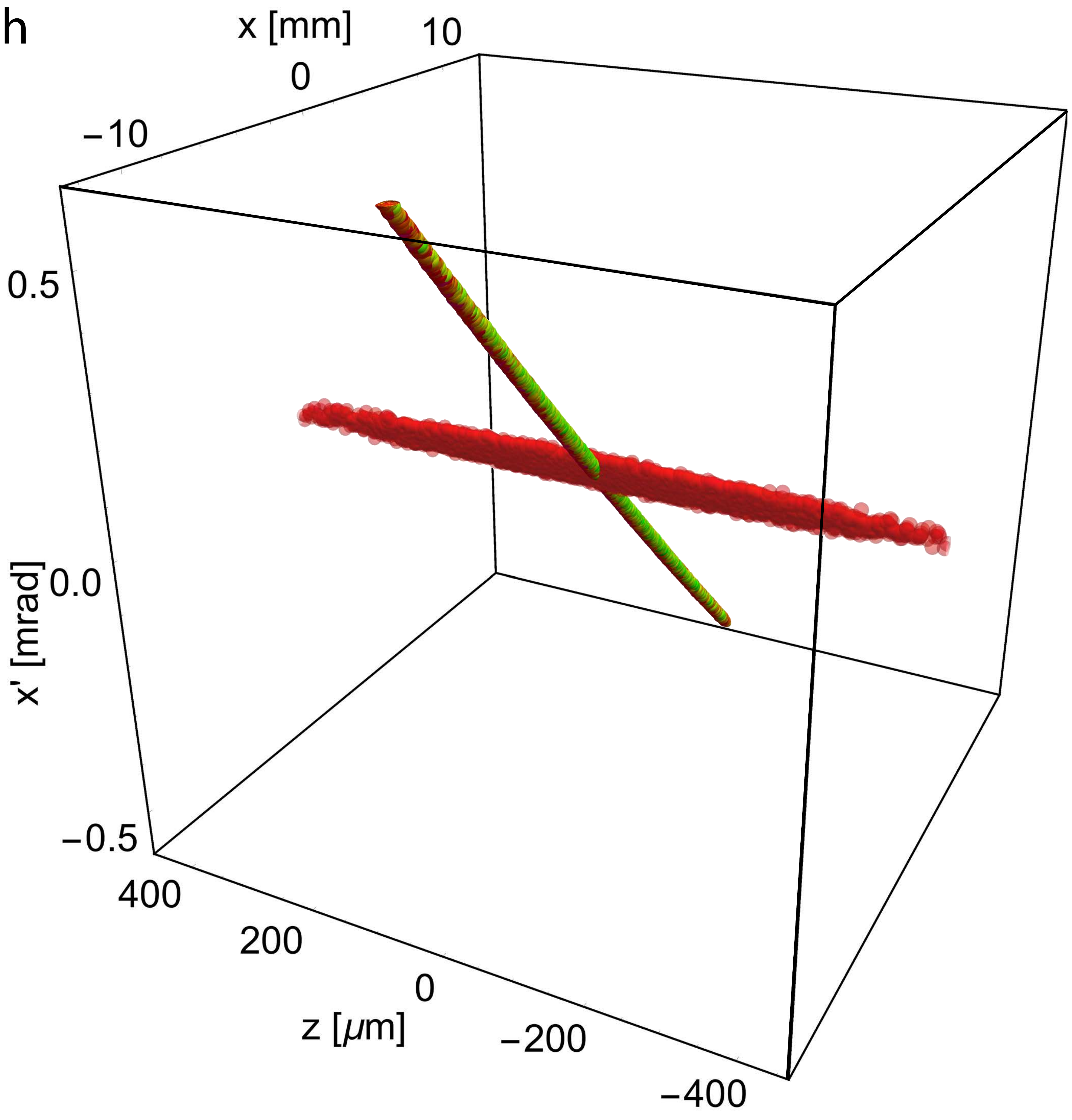} \
	\includegraphics[width=0.3\textwidth]{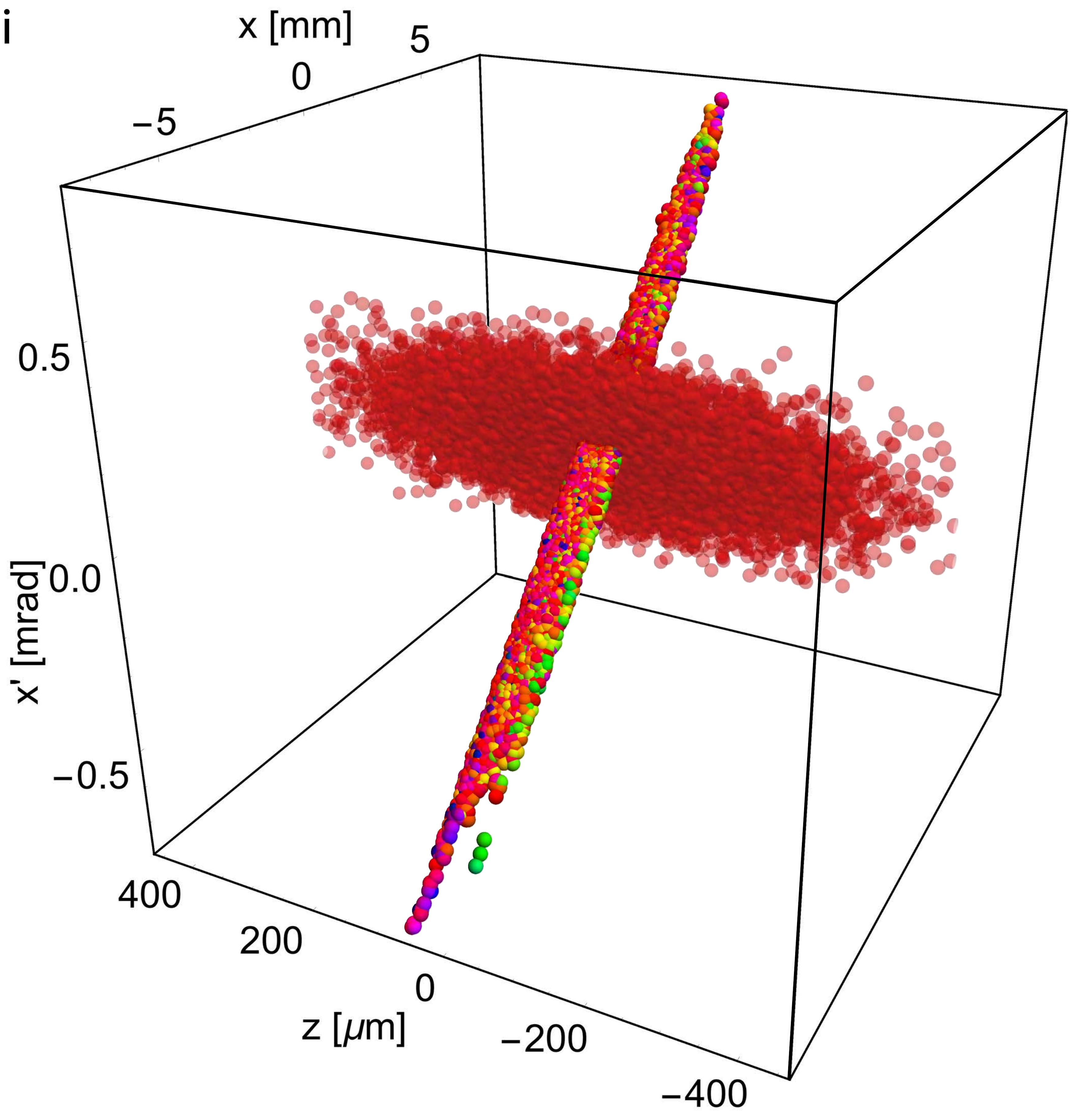} 
	\caption{The top row of subplots (a-c) demonstrates the evolution of the beam shape in real space $(x,y,z)$ while propagating through the optimized DEEX. Subplot (a) shows the beam at the exit of first EEX, (b) shows the beam after the mirrored transverse-optics, and (c) is at the exit of the beamline. The energy spread, $\Delta\gamma/\gamma$, is specified by the color map in arbitrary normalized units. All of the subplots also show the beam's phase space distribution at the entrance of the beamline for comparison. The initial phase spaces are all uniformly red because of very low initial energy spread, and are plotted with slight opacity to help differentiate from the subsequent phase spaces. The second row (d-f) shows the $(x',y',\Delta\gamma/\gamma)$ phase space at the same locations as the top row. The third row (g-i) depicts the corresponding evolution of the beam in 4D phase space $(x,x',z,\Delta\gamma/\gamma)$ at the same locations along the beamline, visualizing the phase space exchanges resulting in significant compression of the bunch at the cost of enlarging the energy spread.} 
	\label{fig:3D_phase_space}
\end{figure*}

Compensation of CSR effects by turning around the longitudinal phase space of the beam by mirrored telescope results in a small CSR-induced chirp (-14.7 m$^{-1}$) corresponding to the correlated energy spread of $-212$~keV, while the final uncorrelated energy spread is $817$~keV. This slightly enlarges the total energy spread of the beam ($844$~keV), but longitudinal emittance $\epsilon_{n_z}$ practically remains unchanged demonstrating only 2\% enlargement. The imposed chirp can be removed via a downstream off-crest acceleration, alternatively, by introducing an initial correlated energy spread $12.5$~keV, or even more elegantly, by slightly modifying the transverse-optics of the mirrored telescope. The later results in almost complete correspondence of the output phase space in linear (without chirp) and CSR regime as depicted in Fig.~\ref{fig:OptSchemeSex-trick_LvsCSR_100_chirp-removed}. 
The required modification of the telescope matrix in Eq.~(\ref{eq:telescope}) can be described by introducing a single coefficient $R_{65}$ resulting in the redefined matrix elements:
\begin{align}\label{eq:TELEparameters-AB_chirp}
R_{11}&=-\frac{-L_B\psi_A + m(m+R_{65}\psi_A) (L_B\psi_B +1) 
}{m\eta_A\eta_B},\\
R_{21}&=-\frac{\psi_A + m(m+R_{65}\psi_A) \psi_B}{m \eta_A \eta_B},\\
R_{22}&=\frac{L_A (\psi_A+ 
	m^2\psi_B)-\eta_A^2-mR_{65}\psi_B(\eta_A^2-L_A\psi_A)}{m\eta_A\eta_B},\\
\label{eq:TELEparameters-AB_chirp_F}
R_{12}&=\frac{R_{11}R_{22}-1}{R_{21}},
\end{align}
which correspond to the original matrix elements defined in Equations~(\ref{eq:TELEparameters-AB}-\ref{eq:TELEparameters-AB_F}), if $R_{65}=0$.
The new transverse-optics results in the modified longitudinal part of the final matrix of the DEEX BC which imposes a chirp on the beam in the linear regime while compressing it:
\begin{align}\label{DEEX-matrix-chirp}
R'_{DEEX_M}=\begin{pmatrix}
R'_{11} & R'_{12} & 0 & 0  \\ 
R'_{21} & R'_{22} & 0 & 0  \\
0 & 0 & -1/m & 0   \\
0 & 0 & R_{65} & -m   
\end{pmatrix}.
\end{align}
The transverse part of the matrix remains unchanged and its matrix elements are described by Equations~(\ref{eq:DEEX-matrix_parameters}-\ref{eq:DEEX-matrix_parametersF}).
One can completely compensate the CSR-induced chirp for the 100~pC bunch by choosing the proper value of $R_{65}=0.867$~m$^{-1}$ (Fig.~\ref{fig:OptSchemeSex-trick_LvsCSR_100_chirp-removed}). The input Twiss and beamline parameters for the DEEX BC relying on different treatments aimed to compensate for nonlinear and CSR effects for 100 pC bunch are summarized in Table~\ref{tbl:beamline}, while Table~\ref{tbl:results} provide with the input emittances and quantify their degradation in distinct scheme configurations. Finally, we visualize the evolution of the bunch through the optimized DEEX bunch compressor in CSR regime in Fig.~\ref{fig:3D_phase_space}. 
\begin{table}[ht]
	\begin{ruledtabular}
		\caption{The input Twiss and DEEX BC parameters for different beamline configurations: initial symmetrical design (IS), symmetrical design with optimized Twiss parameters (ST) and asymmetrical design (AS) for the mirrored/direct telescope; asymmetrical design with sextupoles for mirrored (ASX) and modified-mirrored (ASXC) transverse-optics.}
		\begin{tabular}{llllll}\label{tbl:beamline}
			{\bf Design }& IS & ST& AS&ASX& ASXC   \\ 
			$\beta_x$, m & 100 & 31.6/31 & 11.504 &8.857&8.857  \\
			$\alpha_x$, m & 0 & 7.9 & 4.254 &3.186&3.186  \\
			$\theta_A$, deg & 2.5 & 2.5 & 2.306 &2.158 &2.158  \\
			$\theta_B$, deg & 2.5 & 2.5 & 0.628 &0.200& 0.200 \\
			$r_A$, m & 1.14 & 1.14 & 0.502&0.500&0.500 \\
			$r_B$, m & 1.14 & 1.14 & 3.907 &3.932& 3.932 \\
			$S_{1A}$, m & 4.8 & 4.8  & 19.738  &20.199&20.199\\
			$S_{1B}$, m & 4.8 & 4.8  & 9.778 &9.699& 9.699\\
			$S_{2A}$, m &1.6 & 1.6 & 0.945 &0.930& 0.930\\
			$S_{2B}$, m &1.6 & 1.6 & 0.017 &0.017& 0.017\\
			$\kappa_A$, $\rm m^{-1}$ &-4.71 &-4.71&-1.255 &-1.311&-1.311\\
			$\kappa_B$, $\rm m^{-1}$ &4.71 &4.71&9.291 &29.495&29.495\\
			$k_1$, $\rm m^{-2}$ &0 &0&0 &-5.05&-5.047\\
			$k_2$, $\rm m^{-2}$ &0 &0&0 &-0.42&-0.422\\
			$k_3$, $\rm m^{-2}$ &0 &0&0 &-22.59&-22.588\\
			$k_4$, $\rm m^{-2}$ &0 &0&0 &12.13&12.134\\
			$R_{65}$, m$^{-1}$ & 0& 0 & 0 &0 &0.867\\
			$R_{11}$ & $\pm$5.768&$\pm$5.768 &$\pm$0.228 &0.638 & 0.638\\
			$R_{12}$, m & $\mp$37.186&$\mp$37.186 &$\mp$0.429 &-0.644 &-0.644\\
			$R_{21}$, m$^{-1}$ & $\mp$3.495 & $\mp$3.495 &$\mp$0.255 &-0.143 &-0.143\\
			$R_{22}$ &$\pm$22.709 &$\pm$22.709 &$\pm$4.869 &1.710 &1.711\\
			
		\end{tabular}
	\end{ruledtabular}
\end{table}
\begin{table}[ht]
	\begin{ruledtabular}
		\caption{Emittance degradation in CSR regime for 100~pC bunch for the different beamline configurations with mirrored/standard telescope (apart from ASX design where results are provided for the mirrored telescope only while are similar for the ASXC design with chirping transverse-optics).}
		\begin{tabular}{llllll}\label{tbl:results}
			{\bf Design}& Initial & IS & ST& AS& ASX \\ 
			$\epsilon_{n_x}$ & 0.45 $\mu$m & 1722/1707\% & 335/366\% & 39/46\%& 27\%  \\
			$\epsilon_{n_y}$ & 0.45 $\mu$m & 0\% & 0\% & 0\% & 0\%  \\
			$\epsilon_{n_z}$ & 14.09  $\mu$m & 5/15\% & 3/20\% & 1/18\%& 2\% \\
			$\epsilon_{n_{6D}}$ & 2.85 $\mu$m$^3$ & 1779/1730\% & 338/385\%&38/58\%&26\% \\
			$\mu$ & 1  & 1.04/1.28 & 1.03/1.20&1.04/1.18&1.05 \\

		\end{tabular}
	\end{ruledtabular}
	
\end{table}

\begin{table}[ht]
	\begin{ruledtabular}
		\caption{Emittance degradation in DEEX BC in CSR regime optimized for possible operating regimes of LCLS-II for different bunch charges and transverse emittances from~\cite{LCLS2-conceptual_design} while input $\epsilon_{n_z}$ = 14.09 $\mu$m for all regimes.}
		\begin{tabular}{llllll}\label{tbl:regimesLCLS2}
			{\bf Bunch charge}, pC& 20 & 50 & 100  & 200 & 300  \\ 
			Input $\epsilon_{n_x}$, $\mu$m & 0.2  & 0.3 & 0.45 & 0.55& 0.65  \\
			Enlargement $\epsilon_{n_x}$, \% & 6.3 & 16.0 & 26.5 & 59.6& 90.3 \\
			Enlargement $\epsilon_{n_z}$, \% & 0.17 & 0.58 & 1.95 & 6.7&13.6  \\
			Enlargement $\epsilon_{n_{6D}}$, \% & 5.85 & 14.8 & 26.0 & 66.2& 110 \\
			$\mu$ & 1.01 & 1.03 & 1.05&1.05&1.06 \\
			$E_c$, keV ($R_{65}=0$) & -26  & -89  & -212 & -452 & -705 \\
			$R_{65}$, m$^{-1}$ & 0.1  & 0.365 & 0.867&1.85&2.88 \\

		\end{tabular}
	\end{ruledtabular}
	
\end{table}

\emph{\bf Different operating regimes}.
Here we briefly provide optimization results for other possible operating regimes of LCLS-II which are summarized in Table~\ref{tbl:regimesLCLS2}. Assuming $\epsilon_{n_x,n_y}=0.20\;\mu$m for 20 pC charge, the scheme demonstrates almost negligible changes in the quality of the output beam in CSR regime in comparison to the linear regime. The transverse emittance experiences only 6.3\% growth, while longitudinal emittance is increased in less than 0.2\%. The CSR-induced chirp (-26~keV) is smaller than for 100 pC bunch and can be compensated by choosing $R_{65}=0.1$~m$^{-1}$. For the 200 pC operating and initial emittance $\epsilon_{n_x,n_y}=0.55\;\mu$m the scheme still demonstrates the acceptable performance: 59.6\% degradation of $\epsilon_{n_x}$ and 6.7\% enlargement of $\epsilon_{n_z}$. The CSR-induced chirp is larger than in 100~pC regimes but can be completely compensated by choosing $R_{65}=1.85$~m$^{-1}$.

%%%%%%%%%%%%%%%%%%%%%%%%%%%%%%%%%%%%%%%%%%%%%%%%%%%%%%%%%%%%%%%%%%%%%%%%%%%%%%%%%%%%%%%%
%%%%%%%%%%%%%%%%%%%%%%%%%%%%%%%%%%%%%%%%%%%%%%%%%%%%%%%%%%%%%%%%%%%%%%%%%%%%%%%%%%%%%%%%
\section{Discussion} \label{sec:diss}
%%%%%%%%%%%%%%%%%%%%%%%%%%%%%%%%%%%%%%%%%%%%%%%%%%%%%%%%%%%%%%%%%%%%%%%%%%%%%%%%%%%%%%%%
%%%%%%%%%%%%%%%%%%%%%%%%%%%%%%%%%%%%%%%%%%%%%%%%%%%%%%%%%%%%%%%%%%%%%%%%%%%%%%%%%%%%%%%%
Typical linear or circular accelerators have an extremely large number of beam elements and, therefore, thousands of parameters which must be tuned to achieve desired performance. Such an optimization problem is equivalent to the mathematical problem of finding an extremum of a highly coupled, nonlinear, multi-variable function. Optimization during design is very important because once an accelerator is built, the possibility of adjusting parameters of the accelerating elements and of the whole beamline performance becomes extremely limited. While designing an accelerator, the beam dynamics simulation is the most valuable technique to determine optimal beam elements and their boundary conditions. The initial design, taking into account linear dynamics propagation, can be evaluated quickly using matrix formalism and gives an understanding of which beam elements should be used and where. Once nonlinear effects and collective effects are introduced, special simulation tools are required, tracking codes, such as {\sc elegant}, {\sc parmela}~\cite{parmela} and others. For light sources such as LCLS, LCLS2, European XFEL, Spring-8 or the planned MaRIE XFEL, the design stage includes tracking of the electrons from injection at the cathode all way to the undulator, where X-ray photons are simulated using FEL simulation software such as {\sc genesis}~\cite{GENESIS}, \emph{etc}. 

A light source design problem can be represented as the simultaneous maximization of brightness and minimization of bandwidth of the produced radiation. In such an optimization problem, there are thousands of parameters that one can adjust, while even a single simulation run of the electron beam from the cathode to the photons at the exit of the undulator can take up to several days on a high performance computer cluster. The process can be subdivided into at least three stages: 1) cathode and electron gun simulations. 2) Beam transport through the accelerator. 3) Electron propagation and photon generation in the undulator. In this paper, we discussed beamline transport optimization concentrating on the realization of an important structural component, the bunch compressor. We demonstrated that a novel extremum seeking (ES) algorithm not only can dramatically improve the performance of the scheme (here, an ability to preserve the beam quality and compression factor), but, in addition, can help to discover or prove interesting physics behavior in such a system. It is our belief that this method can be useful for a wide range of accelerator scientists who face large, multi-dimensional, computationally expensive optimization problems.

In summary, all bunch compressors, especially advanced schemes, such as a double chicane or double EEX, have a large number of beam optics elements. In addition, each element is usually characterized by multiple parameters. While working well in the linear regime, all of these schemes requires multi-dimensional (over all set of parameters) optimization in the case of real dynamics (including nonlinear and collective effects) to minimize emittance growth and phase space deformations. Since every individual simulation run of a bunch compressor with nonlinear and CSR effects takes quite some time, we needed to apply fast adaptive multi-dimensional optimization algorithms.

%%%%%%%%%%%%%%%%%%%%%%%%%%%%%%%%%%%%%%%%%%%%%%%%%%%%%%%%%%%%%%%%%%%%%%%%%%%%%%%%%%%%%%%%
%%%%%%%%%%%%%%%%%%%%%%%%%%%%%%%%%%%%%%%%%%%%%%%%%%%%%%%%%%%%%%%%%%%%%%%%%%%%%%%%%%%%%%%%
\section{Conclusions and Future Work} \label{sec:conclusion}
%%%%%%%%%%%%%%%%%%%%%%%%%%%%%%%%%%%%%%%%%%%%%%%%%%%%%%%%%%%%%%%%%%%%%%%%%%%%%%%%%%%%%%%%
%%%%%%%%%%%%%%%%%%%%%%%%%%%%%%%%%%%%%%%%%%%%%%%%%%%%%%%%%%%%%%%%%%%%%%%%%%%%%%%%%%%%%%%%
In this paper, we demonstrated a new type of a bunch compressor, the asymmetrical double emittance exchanger, in what each emittance exchanger and transverse-optics is tuned to match the parameters of the propagating beam. In the symmetric configuration of the scheme, the CSR effects' impact on the degradation of the longitudinal phase space was compensated along the beamline by turning around the beam in the mirrored telescope followed by a second EEX. The unique asymmetric design was revealed by the model independent Extremum Seeking optimization effectively working in a high-dimensional space of parameters, while 6-dimensional emittance was used as a cost function for optimization. This asymmetrical configuration lead to partial compensation of CSR and nonlinear effects on the transverse dynamics and, as a result, significantly improved the quality of the output transverse phase space, minimizing enlargement of the corresponding emittance. Furthermore, nonlinear effects were completely suppressed in the approximation of zero bunch charge by incorporating several sextupoles and tuning them exclusively relying on the eigen-emittance analysis. This result was used as a starting point for the ES-driven optimization which then adjusted all parameters of the scheme while slowly increasing the bunch charge to the operational value of 100~pC. This lead to the great correspondence of the output beam phase spaces in the approximation of the idealized linear dynamics and real dynamics accounting for nonlinear and CSR effects resulting in 26.5\% enlargement of $\epsilon_{n_x}$ and 2\% enlargement of $\epsilon_{n_z}$ respectively to their initial values. Optimization results for other operational regimes of LCLS-II accounting for the corresponding emittance quantities for distinct bunch charges prove that DEEX-based BC can be effective across a wide operational range of the machine.    

The designed bunch compressor does not require energy-phase compensations and completely eliminates chicane-type optics, hence, has the potential to have less impact from LSC-induced micro-bunch instabilities. Detailed investigation of how the DEEX-based bunch compressor will quantitatively interact with microbunching instabilities imposed in the first bunch compressor and other upstream elements and if it is capable to substitute both compressors at once is the subject of future research, in which nonlinear effects associated with the transverse deflecting cavities in the thick-lens approximation and structural components of the transverse-optics, neglected here, should be accounted for. Apart from FEL-driving linacs, ultra-short low-emittance bunches are required for the Plasma-Wakefield Accelerators~\cite{PWA}, \emph{i.e.} FACET at SLAC, and for the high-energy particle colliders such as LHC at CERN, \emph{etc}. It is our belief that the designed DEEX-based bunch compressor can be of great interest for these machines alongside its feasibility for FELs.

%%%%%%%%%%%%%%%%%%%%%%%%%%%%%%%%%%%%%%%%%%%%%%%%%%%%%%%%%%%%%%%%%%%%%%%%%%%%%%%%%%%%%%%%
%%%%%%%%%%%%%%%%%%%%%%%%%%%%%%%%%%%%%%%%%%%%%%%%%%%%%%%%%%%%%%%%%%%%%%%%%%%%%%%%%%%%%%%%
\section{Acknowledgments}
%%%%%%%%%%%%%%%%%%%%%%%%%%%%%%%%%%%%%%%%%%%%%%%%%%%%%%%%%%%%%%%%%%%%%%%%%%%%%%%%%%%%%%%%
%%%%%%%%%%%%%%%%%%%%%%%%%%%%%%%%%%%%%%%%%%%%%%%%%%%%%%%%%%%%%%%%%%%%%%%%%%%%%%%%%%%%%%%%
We would like to thank Prof. Philippe Piot, Dr. Nikolai Yampolsky and Dr. Bruce Carlsten for useful scientific discussions.

%%%%%%%%%%%%%%%%%%%%%%%%%%%%%%%%%%%%%%%%%%%%%%%%%%%%%%%%%%%%%%%%%%%%%%%%%%%%%%%%%%%%%%%%
%%%%%%%%%%%%%%%%%%%%%%%%%%%%%%%%%%%%%%%%%%%%%%%%%%%%%%%%%%%%%%%%%%%%%%%%%%%%%%%%%%%%%%%%
\appendix
%%%%%%%%%%%%%%%%%%%%%%%%%%%%%%%%%%%%%%%%%%%%%%%%%%%%%%%%%%%%%%%%%%%%%%%%%%%%%%%%%%%%%%%%
%%%%%%%%%%%%%%%%%%%%%%%%%%%%%%%%%%%%%%%%%%%%%%%%%%%%%%%%%%%%%%%%%%%%%%%%%%%%%%%%%%%%%%%%

%%%%%%%%%%%%%%%%%%%%%%%%%%%%%%%%%%%%%%%%%%%%%%%%%%%%%%%%%%%%%%%%%%%%%%%%%%%%%%%%%%%%%%%%
%%%%%%%%%%%%%%%%%%%%%%%%%%%%%%%%%%%%%%%%%%%%%%%%%%%%%%%%%%%%%%%%%%%%%%%%%%%%%%%%%%%%%%%%
\section{Extremum Seeking}\label{apx:ES}
%%%%%%%%%%%%%%%%%%%%%%%%%%%%%%%%%%%%%%%%%%%%%%%%%%%%%%%%%%%%%%%%%%%%%%%%%%%%%%%%%%%%%%%%
%%%%%%%%%%%%%%%%%%%%%%%%%%%%%%%%%%%%%%%%%%%%%%%%%%%%%%%%%%%%%%%%%%%%%%%%%%%%%%%%%%%%%%%%

The ES feedback scheme is designed for n-dimensional dynamic systems of the form
\begin{equation}
	\frac{d\x}{dt} = \f(\x,\p,t),
\end{equation}
where $\f$ is an unknown function, $\x = \left ( x_1, \dots, x_n \right )$ are physical parameters such as beam properties throughout an accelerator, and $\mathbf{p}(t) = \left ( p_1(t), p_2 (t), \dots , p_m(t) \right )$ are tunable parameters which influence $\x$, such as magnet lengths or electromagnetic field amplitudes. ES can be used to minimize an analytically unknown, time-varying, user-defined ``cost function," $C(\x,\p,t)$, whose minimization corresponds to optimization of certain system properties, such as beam loss or a measure of how much phase space deviates from a desired distribution in a particular location. The ES adaptation only needs access to a possibly noise corrupted measurement of $C(\x,\p,t)$ of the form:
\begin{equation}
	\hat{C}(\x,\p,t) = C(\x,\p,t) + \underbrace{\nu(t)}_{\mathrm{noise}}.
\end{equation}
In order to minimize the analytically unknown $C$, the ES dynamics are given by
\begin{equation}
	\frac{dp_j}{dt} = \sqrt{\alpha \omega_j} \cos \left ( \omega_j t + k \hat{C} \right ). \label{dlaw_dynamic}
\end{equation}
If $\omega_j = \omega r_j$ such that $r_j \neq r_i$ for $i \neq j$, then in the limit as $\omega \rightarrow \infty$, the dynamics (\ref{dlaw_dynamic}) are approximated by the average dynamics
\begin{equation}
	\frac{d\p}{dt} = -\frac{k\alpha}{2}\nabla_\p C, \label{ave_dynamics}
\end{equation}
a gradient descent of the actual, unknown function $C$ with respect to $\p$ \cite{ref-SchSch,ref-SchBk}. In (\ref{dlaw_dynamic}) the term $\alpha$ controls the size of the dither amplitude of each parameter and can be increased to escape local minima or to speed up convergence towards the minimum. The term $k>0$ serves as a feedback gain which also speeds up convergence. For the average dynamics approximation (\ref{ave_dynamics}) to hold, $\omega$ must be very large relative to the natural time variation of the system and may have to be increased as the values of $k$ or $\alpha$ are increased.

The $\cos(\cdot)$ terms in (\ref{dlaw_dynamic}) may be replaced by $\sin(\cdot)$ functions, or the two can be mixed together. Convergence of the scheme requires that the perturbing functions are orthogonal in $L^2[0,T]$, which means that the inner products of the functions uniformly converge to zero:
\begin{equation} 
	  \lim_{\omega \rightarrow \infty} \int_{0}^{T} \sin(\omega r_1 t) \sin(\omega r_2 t) dt = 0. \nonumber
\end{equation}

For complex physical systems, such as particle accelerators, the values of the parameters $\p$ may vary over several orders of magnitude and may have various bounds which should not be violated. Therefore it is usually best to normalize all parameters within a given range such as $\p_\norm \in [-1,1]$. Normalization is performed by first defining upper and lower bounds on all parameters:
\begin{eqnarray}
	\p_{\mathrm{max}} &=& \left ( p_{1,\mathrm{max}}, \dots, p_{m,\mathrm{max}} \right ), \\
	\p_{\mathrm{min}} &=& \left ( p_{1,\mathrm{min}}, \dots, p_{m,\mathrm{min}} \right ).
\end{eqnarray}
We then define
\begin{equation}
	\Delta \p_+ = \frac{ \p_{\mathrm{max}} + \p_{\mathrm{min}} }{2}, \quad \Delta \p_- = \frac{ \p_{\mathrm{max}} - \p_{\mathrm{min}} }{2},
\end{equation}
and normalize all parameters via
\begin{equation}
	\p \ \longrightarrow \ \frac{\p - \Delta \p_+ }{\Delta \p_-} = \p_\norm \in [-1, 1], \label{normalize}
\end{equation}
and un-normalize normalized values back to their physical ranges via
\begin{equation}
	\p_\norm \ \longrightarrow \  \p_\norm \Delta \p_- + \Delta \p_+ = \p \in [\p_{\mathrm{min}},\p_{\mathrm{max}}]. \label{un_normalize}
\end{equation}

The iterative ES procedure is:
\\
{\bf 1).} Choose physics-based estimates for initial parameter values, $\p(1)$. 
\\
{\bf 2).} Perform a simulation or run an experiment with parameter settings $\p(1)$ and make the desired measurement of the analytically unknown cost function
\begin{equation}
	\hat{C}(1) = C(\x(\p(1)),\p(1),t_1) + \nu(t_1).
\end{equation}
{\bf 3).} Calculate normalized parameter values $\p_{\norm}(1)$ according to (\ref{normalize}). 
\\
{\bf 4).} Calculate new normalized parameter values $\p_\norm(2)$ according to the ES update rule:
\begin{equation}
	p_{\norm,j}(2) = p_{\norm,j}(1) + \Delta \sqrt{\alpha_j \omega_j} \cos \left ( \omega_j \Delta + k \hat{C}(1) \right ). \label{dlaw}
\end{equation}
{\bf 5).} Force normalized parameters values to remain within the bounds $[-1,1]$ via the simple check:
\begin{equation}
	| p_{\norm,j}(2) | > 1 \ \Longrightarrow \ p_{\norm,j}(2) = \frac{\mathrm{sign}(p_{\norm,j}(2))}{|p_{\norm,j}(2)|}. \label{bound_check}
\end{equation}
{\bf 6).} Un-normalize parameters to get new physical values $\p(2)$, according to (\ref{un_normalize}) and repeat the simulation or experiment to record $\hat{C}(2)$. 
\\
{\bf 7).} The procedure is continued iteratively with updates at step $n$ of the form:
\begin{equation}
	p_{\norm,j}(n+1) = p_{\norm,j}(n) + \Delta \sqrt{\alpha_j \omega_j} \cos \left ( \omega_j n \Delta  + k \hat{C}(n) \right ), \label{dlawn}
\end{equation}
which, for $\Delta \ll \frac{1}{\max{\omega_j}} \ll 1$ is a finite difference approximation of the dynamics (\ref{dlaw_dynamic}) and results in minimization of $C$.

Several useful features of this feedback scheme are:
\\
{\bf 1).} The average gradient descent, (\ref{ave_dynamics}) takes place relative to the actual, unknown function $C$, despite being based only on its noise-corrupted measurement $\hat{C}$.
\\
{\bf 2).} Unlike standard gradient-descent approaches, convergence time does not grow exponentially with the number of parameters and can handle noisy, time-varying systems.
\\
{\bf 3).} Despite operating on an analytically unknown dynamic system and minimizing an analytically unknown function, $\hat{C}$, this feedback has analytically known bounds on parameter variation and update rates, because the unknown function enters the parameter dynamics as the argument of a known, bounded function:
\begin{eqnarray}
	\left | p_j(n+1) - p_j(n) \right | & = & \left |  \Delta \sqrt{\alpha \omega_j} \cos \left ( \omega_j \Delta n + k \hat{C}(n) \right ) \right | \nonumber \\
	& \leq &  \left |  \Delta \sqrt{\alpha \omega_j}  \right | ,
\end{eqnarray}
which makes this method safe and useful for in-hardware and in-software implementation.

%%%%%%%%%%%%%%%%%%%%%%%%%%%%%%%%%%%%%%%%%%%%%%%%%%%%%%%%%%%%%%%%%%%%%%%%%%%%%%%%%%%%%%%%
%%%%%%%%%%%%%%%%%%%%%%%%%%%%%%%%%%%%%%%%%%%%%%%%%%%%%%%%%%%%%%%%%%%%%%%%%%%%%%%%%%%%%%%%
\section{Matrices of the beamline elements}\label{apx:elements}
%%%%%%%%%%%%%%%%%%%%%%%%%%%%%%%%%%%%%%%%%%%%%%%%%%%%%%%%%%%%%%%%%%%%%%%%%%%%%%%%%%%%%%%%
%%%%%%%%%%%%%%%%%%%%%%%%%%%%%%%%%%%%%%%%%%%%%%%%%%%%%%%%%%%%%%%%%%%%%%%%%%%%%%%%%%%%%%%%
Linear matrices of the beamline elements used in the paper are summarized in this section assuming electron coordinates  $\boldsymbol{\xi}=$($x,\;x',\;y,\;y',\;z=s-ct,\;z'=\delta\gamma/\gamma_0$) in 6D phase space.
The matrix of a free space propagation (drift) of length $L$ for highly relativistic beam ($\gamma\gg1$):
\begin{equation}\label{eq:drift}
R_{Drift}(L)=\begin{pmatrix}
1 & L &0 &0& 0 &0  \\ 
0 & 1 & 0 &0& 0 &0\\ 
0 & 0 & 1 &L& 0 &0\\
0 & 0 & 0 &1& 0 &0\\
0 & 0 & 0&0&1 &0\\
0 & 0 & 0&0&0 &1\\
\end{pmatrix}
.
\end{equation}
A bending magnet can be represented by the transfer matrix under an assumption that the reference particle trajectory is orthogonal to edge surfaces of the magnet~\cite{book_Intro_part_Accel_Conte}:
\begin{multline}\label{eq:bend}
R_{bend}(r,\theta)=\\=\begin{pmatrix}
\cos\theta &r\sin\theta&0 &0& 0 &r(1-\cos\theta) \\ 
r^{-1}\sin\theta & \cos\theta & 0 &0& 0 &\sin\theta\\ 
0 & 0 & 1 &r\theta& 0 &0\\
0 & 0 & 0 &1& 0 &0\\
-\sin\theta & -r(1-\cos\theta) & 0&0&1 &r(\sin\theta-\theta)\\
0 & 0 & 0&0&0 &1\\
\end{pmatrix}
,
\end{multline}
where $\theta$ is the bending angle and $r$ is the bending radius, and $D=r\theta$ is the length along the beam path. 
Focusing or defocussing effects by the edge of the magnet appear if the beam arrives to/leaves it at an angle $\alpha$~\cite{book_Intro_part_Accel_Conte}:
\begin{equation}\label{eq:edge}
R_{edge}(r,\alpha)=\begin{pmatrix}
1 & 0 &0 &0& 0 &0  \\ 
r^{-1}\tan\alpha & 1 & 0 &0& 0 &0\\ 
0 & 0 & 1 &0& 0 &0\\
0 & 0 & -r^{-1}\tan\alpha &1& 0 &0\\
0 & 0 & 0&0&1 &0\\
0 & 0 & 0&0&0 &1\\
\end{pmatrix}
.
\end{equation}
The total transfer matrix of a bending magnet with equal edge angles can be found as:
\begin{equation}\label{eq:bend_edge}
R_{bend'}(r,\theta,\alpha)=R_{edge}\left(r,\alpha\right)\cdot R_{bend}(r,\theta)\cdot R_{edge}\left(r,\alpha\right).
\end{equation}
A dogleg is a sequence of two identical but oppositely oriented bending magnets separated by drift. This beamline shifts the beam in the bending plane and imposes transverse-longitudinal correlations. 
If $\alpha=\theta/2$ for each bending magnet, the combination of two bending magnets accounting for the edge focusing effects separated by drift $L_1$ can be found as a multiplication product of its components:
\begin{multline}\label{eq:dogleg_line}
R_{dogleg}(r,\theta,L_1)=\\=R_{bend'}(r,\theta,\frac{\theta}{2})\cdot R_{Drift}(L_1)\cdot R_{bend'}(-r,-\theta,-\frac{\theta}{2}),
\end{multline}
and results in the following transfer matrix for a dogleg shifting the trajectory up:
\begin{equation}\label{eq:dogleg}
R_{dogleg}(\eta)=\begin{pmatrix}
1 & L_{eff} &0 &0& 0 &\eta  \\ 
0 & 1 & 0 &0& 0 &0\\ 
0 & 0 & R'_{33} &R'_{34}& 0 &0\\
0 & 0 & R'_{43} &R'_{44}& 0 &0\\
0 & \eta & 0&0&1 &\psi\\
0 & 0 & 0&0&0 &1\\
\end{pmatrix}
.
\end{equation}
The only difference between matrices of dogleg up and dogleg down is in the sign of dispersion $\pm\eta$.
Parameters of the dogleg can be found as:
\begin{align}\label{eq:dogleg_eta}
&\eta=-2 (L_1 + r \sin\theta) \tan\left(\frac{\theta}{2}\right),\\
&\psi=4 L_1 \tan^2\left(\frac{\theta}{2}\right)-2 r (\theta + \sin\theta) + 8 r \tan\left(\frac{\theta}{2}\right) 
,\\
&L_{eff}=L_1+2r\sin\theta,
\end{align}
and matrix elements characterizing horizontal dynamics:

\begin{align}\label{eq:dogleg_L_eff_y}
&R'_{33}=R'_{44}=\\&=1 - \tan\frac{\theta}
{2} \left(\theta \tan\frac{\theta}{2}-2\right) \left(\frac{L_1}{r}
\left( \theta \tan\frac{\theta}{2} - 1\right) - 2  \theta\right),\\
&R'_{34}=\left( \theta \tan\frac{\theta}{2}-1\right) \left( 
L_1 \left(\theta \tan\frac{\theta}{2}-1\right) - 2 r \theta \right),\\
&R'_{43}=\frac{{R'_{33}}^2-1}{R'_{34}},
\end{align}
A chicane with four bending magnets or C-chicane combined with two doglegs is described by the corresponding transfer matrix:
\begin{equation}\label{eq:chicane}
R_{chicane}=\begin{pmatrix}
1 & L_{x} &0 &0& 0 &0  \\ 
0 & 1 & 0 &0& 0 &0\\ 
0 & 0 & R_{33} &R_{34}& 0 &0\\
0 & 0 & R_{43} &R_{44}& 0 &0\\
0 & 0 & 0&0&1 &R_{56}\\
0 & 0 & 0&0&0 &1\\
\end{pmatrix}
.
\end{equation}
The chicane strength is equal to:
\begin{equation}\label{eq:chic_R56}
R_{56}=2\psi,
\end{equation}
and the effective drift length in the vertical plane is ($x,\;x'$):
\begin{equation}\label{eq:chic_Lx}
L_{x}=2L_{eff}+L_2,
\end{equation}
where $L_2$ is the distance between two doglegs and matrix elements describing ($y,\;y'$) dynamics: 
\begin{align}\label{eq:chic_Ly}
R_{33}&=R_{44}={R'_{33}}^2 + L_2 R'_{33} R'_{43} + R'_{34}R'_{43},\\
R_{34}&=R'_{33} (L_2 R'_{33} + 2 R'_{34}),\\
R_{43}&=\frac{R_{33}^2-1}{R_{34}}.
\end{align}
Elements imposing/removing the correlated energy spread along the bunch is described by the matrix:
\begin{equation}\label{eq:chirper-6D}
R_{chirp}=\begin{pmatrix}
1 & L_{c} &0 &0& 0 &0  \\ 
0 & 1 & 0 &0& 0 &0\\ 
0 & 0 & 1 &L_{c}& 0 &0\\
0 & 0 & 0 &1& 0 &0\\
0 & 0 & 0&0&1 &0\\
0 & 0 & 0&0&R_{65}&1\\
\end{pmatrix}
,
\end{equation}
where $R_{65}<0$ is for chirper and vice verse. The effective vertical and horizontal lengths are chosen to be identical for simplicity but differ for chirper ($L_c$) and dechirper ($L_d$).
The overall transfer matrix of the chicane-based bunch compressor can be found as:
\begin{equation}\label{eq:chicane_BC-6D}
R_{BC}=\begin{pmatrix}
1 & L_{BC_x} &0 &0& 0 &0  \\ 
0 & 1 & 0 &0& 0 &0\\ 
0 & 0 & R''_{33} &R''_{34}& 0 &0\\
0 & 0 & R_{43} &R''_{44}& 0 &0\\
0 & 0 & 0&0&1+R_{56}R_{65} &R_{56}\\
0 & 0 & 0&0&R''_{65}&R''_{66}\\
\end{pmatrix}
,
\end{equation}
where $L_{BC_{x}}=L_{c}+L_{x}+L_{d}$, and matrix elements describing transverse dynamics in ($y,\;y'$) phase space are:
\begin{align}\label{eq:chicane_R33}
R''_{33}&=R_{33} + L_d R_{43},\\
R''_{44}&=R_{33} + L_c R_{43},\\
R''_{34}&=(L_c+L_d)R_{33} + R_{34} + L_c L_d R_{43},
\end{align}
while matrix elements characterizing longitudinal dynamics defined according to the following equations:
\begin{align}\label{eq:chicane_R65}
R''_{65}&=\frac{R_{56}\sigma_{z'_0}^2}{(1+R_{56}R_{65})^2\sigma_{z_0}^2+R_{56}^2\sigma_{z'_0}^2},\\
R''_{66}&=\frac{(1+R_{56}R_{65})\sigma_{z_0}^2}{(1+R_{56}R_{65})^2\sigma_{z_0}^2+R_{56}^2\sigma_{z'_0}^2}.
\end{align}

A TDC-cavity used between two doglegs in an emittance exchanger can be represented by the following transfer matrix in the thin-lens approximation:
\begin{equation}\label{eq:TDC_4D}
R_{TDC}(\kappa)=\begin{pmatrix}
1 & 0 &  0 & 0 &  0 & 0   \\ 
0 & 1 &  0 & 0 & \kappa &0\\ 
0 & 0 & 1 &  0 & 0 & 0\\
0 & 0 & 0 &  1 & 0 & 0\\
0 & 0 &0 & 0 &1 &0\\
\kappa&0&0 & 0 &0 &1\\
\end{pmatrix}
.
\end{equation}
The beam dynamics in the ($y,y'$) plane of the DEEX BC is uncoupled and the associated transfer matrix can be found as a combination of its structural components:
\begin{equation}\label{eq:2EEX-Y-comb}
R_{BC_{yy'}}=R_{EEX_{yy'}^-}\cdot R_{T_{yy'}}\cdot R_{EEX_{yy'}^+}.
\end{equation}
where the transfer matrix through each EEX:
\begin{equation}\label{eq:EEX-Y-comb}
R_{EEX_{yy'}}=R_{dogleg_{yy'}}\cdot R_{Drift}(S_2)\cdot R_{dogleg_{yy'}}.
\end{equation}
The corresponding transfer matrix through each dogleg (Eq.~\ref{eq:dogleg}) is invariant of its orientation (up or down). Matrix elements for a dogleg with asymmetrical edge angles of first magnet ($\alpha_1=0$ and $\alpha_2=\theta$) and second magnet ($\alpha_1=\theta$ and $\alpha_2=0$) used in each EEX can be found as:
\begin{align}
R'_{33}&=1 + \tan\theta \left(-2\theta + 
\frac{S_1}{D} \tan\theta (-1 + \theta \tan\theta)\right),\\
R'_{43}&=\frac{\sin\theta \tan\theta (-2 D + S_1 \tan\theta^2)}{D^2}.
\end{align} 
The transfer matrix for ($y,\;y'$) dynamics through a telescope is equal to: 
\begin{equation}
R_{T_{yy'}}=\begin{pmatrix}
-1 &-L_{T_y}  \\ 
0 & -1  
\end{pmatrix},
\end{equation}
where effective drift length along the $y-$direction:
\begin{equation}
L_{T_y}=\frac{1}{3} L_{eff} \left(2 - \frac{(1 + m)^2}{(1 + m^2) q}\right).
\end{equation}

%%%%%%%%%%%%%%%%%%%%%%%%%%%%%%%%%%%%%%%%%%%%%%%%%%%%%%%%%%%%%%%%%%%%%%%%%%%%%%%%%%%%%%%%
%%%%%%%%%%%%%%%%%%%%%%%%%%%%%%%%%%%%%%%%%%%%%%%%%%%%%%%%%%%%%%%%%%%%%%%%%%%%%%%%%%%%%%%%
\section{Beam Invariants}\label{apx:beam}
%%%%%%%%%%%%%%%%%%%%%%%%%%%%%%%%%%%%%%%%%%%%%%%%%%%%%%%%%%%%%%%%%%%%%%%%%%%%%%%%%%%%%%%%
%%%%%%%%%%%%%%%%%%%%%%%%%%%%%%%%%%%%%%%%%%%%%%%%%%%%%%%%%%%%%%%%%%%%%%%%%%%%%%%%%%%%%%%%
The distribution of the electron beam in 6D phase space is described by the $\Sigma-$matrix, a symmetric matrix of the second order momenta:
%\begin{widetext}
\begin{equation}\label{eq:Sigma-matrix}
\Sigma=\begin{pmatrix}
\mean{x^2} & \mean{xx'} & \mean{xy} &\mean{xy'}& \mean{xz} &\mean{xz'}  \\ 
\mean{xx'} & \mean{x'^2} & \mean{x'y} &\mean{x'y'}& \mean{x'z} &\mean{x'z'}\\ 
\mean{xy} & \mean{x'y} & \mean{y^2} &\mean{yy'}& \mean{yz} &\mean{yz'}\\
\mean{xy'} & \mean{x'y'} & \mean{yy} &\mean{y'^2}& \mean{y'z} &\mean{y'z'}\\
\mean{xz} & \mean{x'z} & \mean{yz} &\mean{y'z}& \mean{z^2} &\mean{zz'}\\
\mean{xz'} & \mean{x'z'} & \mean{yz'} &\mean{y'z'}& \mean{zz'} &\mean{z'^2}\\
\end{pmatrix},
\end{equation}
where $\mean{...}$ denotes averaging over the ensemble.
Dynamics uncoupled between different degrees of freedom are characterized by normalized emittances~\cite{Lapostolle}:
\begin{align}\label{eq:emitt_x}
\epsilon_{n_x}&=\gamma\beta\sqrt{\mean{x^2}\mean{x'^2}- \mean{xx'}^2},\\
\label{eq:emitt_y}
\epsilon_{n_y}&=\gamma\beta\sqrt{\mean{y^2}\mean{y'^2}- \mean{yy'}^2},\\
\epsilon_{n_z}&=\gamma\beta\sqrt{\mean{z^2}\mean{z'^2}- \mean{zz'}^2}.
\end{align}
An advanced analysis is needed when dynamics between different degrees of freedom are coupled. For instance, this is necessary to optimize a complicated beamline such as an EEX, where transverse and longitudinal phase spaces are coupled resulting in multiple correlations in the $\Sigma$-matrix elements off the diagonal blocks in the process of exchange. The electron beam matrix is block-diagonal at the entrance and at the exit of the beamline in the approximation of linear single-particle dynamics. However, nonlinear and collective effects may result in additional correlations in the $\Sigma-$matrix off the diagonal blocks. The Hamiltonian motion of a beam has three invariants, which can be chosen as the quantities known as the eigen emittances $\widetilde{\lambda}_j$, introduced by Dragt~\cite{Dragt, Dragt1}:
\begin{equation}
det\left(J\Sigma-i\widetilde{\lambda}_jI\right)=0,
\end{equation} 
where $J$ is the unit block-diagonal antisymmetric symplectic matrix satisfying $J^2=-I$, where $I$ is the $6\times 6$ identity matrix. Normalized eigen-emittances are defined as: $\lambda_j=\gamma\beta\widetilde{\lambda}_j$. Eigen emittances make the understanding of a ``magical'' EEX straightforward. Beam optics in the process of the ideal exchange simply flips the projection directions of eigen emittances to those of regular emittances.

%%%%%%%%%%%%%%%%%%%%%%%%%%%%%%%%%%%%%%%%%%%%%%%%%%%%%%%%%%%%%%%%%%%%%%%%%%%%%%%%%%%%%%%%
%%%%%%%%%%%%%%%%%%%%%%%%%%%%%%%%%%%%%%%%%%%%%%%%%%%%%%%%%%%%%%%%%%%%%%%%%%%%%%%%%%%%%%%%


\begin{thebibliography}{99}
\bibitem{Energy-spread} Compressing a beam in the longitudinal direction by the conventional scheme leads to enlarging the energy spread. Decreasing both quantities at the same time can be realized via a single emittance exchange coming at the cost of the transverse phase space enlargement.
\bibitem{universal-chicane-BC} At least two chicanes, chirper and dechirper are required to make a beam-independent chicane-based BC.
\end{thebibliography}
\end{document}